\documentclass[structabstract]{aa}
\usepackage{graphicx,lscape}
\usepackage{txfonts,natbib}
\bibpunct{(}{)}{;}{a}{}{,}
\begin{document}
   \title{Temperature, gravity and bolometric correction scales\\ for non-supergiant OB stars\thanks{Based on observations collected at the
   Centro Astron\'omico His\-pano Alem\'an (CAHA) at Calar Alto, operated
   jointly by the Max-
   Planck Institut f\"ur Astronomie and the Instituto de Astrof\'isica
   de Andaluc\'ia (CSIC), proposals H2001-2.2-011
   and H2005-2.2-016.}\fnmsep
   \thanks{Based on observations collected at the European Southern
   Obser\-vatory, Chile, ESO 074.B-0455(A) and from the ESO Archive.}\fnmsep
   \thanks{Based on spectral data retrieved from the ELODIE archive at
   Ob\-servatoire de Haute-Provence (OHP).}
}
 
   \author{Mar\'ia-Fernanda Nieva\inst{1,2}
                    }
  \institute{Dr. Karl Remeis-Sternwarte \& ECAP, Universit\"at Erlangen-N\"urnberg, Sternwartstr. 7, D-96049 Bamberg, Germany
         \and
             Max-Planck-Institut f\"ur Astrophysik, Karl-Schwarzschild-Str. 1, D-85741 Garching, Germany            }

   \date{}

 
  \abstract
   {Precise and accurate determinations of the atmospheric parameters effective temperature and surface gravity are mandatory to derive reliable chemical abundances in OB stars. Furthermore, fundamental parameters like distances, masses, radii, luminosities can also be derived from the temperature and gravity of the stars.}
   {Atmospheric parameters recently determined at high precision with several independent spectroscopic indicators in NLTE, with typical uncertainties of $\sim$300\,K for temperature and of $\sim$0.05\,dex for gravity, are employed to calibrate 
photometric relationships. This is in order to investigate whether a faster tool to estimate atmospheric parameters can be provided.}
   {Temperatures and gravities of 30 calibrators, i.e. well-studied OB main sequence to giant stars in the solar neighbourhood, are compared to reddening-independent quantities of the Johnson and Str\"omgren photometric systems, assuming normal reddening. In addition, we examine the spectral and luminosity classification of the star sample and compute bolometric corrections.}
   {Calibrations of temperatures and gravities are proposed for various photometric indices and spectral types. Once the luminosity of the stars is well known, effective temperatures can be determined at a precision 
of $\sim$400\,K for luminosity classes III/IV and $\sim$800\,K for luminosity class V. Furthermore, surface gravities can reach internal~ uncertainties as low as $\sim$0.08\,dex when using 
our calibration to the Johnson $Q$-parameter. Similar precision is achieved for gravities derived from the $\beta$-index and the precision is lower for both atmospheric parameters when using the Str\"omgren indices $[c1]$ and $[u-b]$. In contrast, external~uncertainties are larger for the Johnson than for the Str\"omgren calibrations.
Our uncertainties are smaller than typical differences among~ other methods in the literature, reaching values up to $\pm$2000\,K for temperature and $\pm$0.25\,dex for gravity, and in extreme cases, $+$6000\,K and $\pm$0.4\,dex, respectively.
A parameter calibration for sub-spectral types is also proposed. 
Moreover, we present a new bolometric correction relation to temperature based on our empirical data, rather than on synthetic grids.}
{The photometric calibrations presented here are useful tools to estimate effective temperatures and surface gravities of non-supergiant OB stars in a fast manner. This is also applicable to some single-line spectroscopic binaries, but caution has to be taken for undetected double-lined spectroscopic binaries and single objects with anomalous reddening-law, dubious~photometric quantities and/or luminosity classes, for which the systematic uncertainties may increase significantly. We recommend to use these calibrations only as a first step of the parameter estimation, with subsequent refinements based on spectroscopy. {A larger sample covering more  uniformly the parameter space under consideration will allow refinements to the present calibrations}.}

 \authorrunning{M. F. Nieva}
   \titlerunning{Temperature, gravity and bolometric correction scales}

   \maketitle
%

\begin{table*}[t!]
 \centering
\caption[]{
Star id, spectral type, photometric quantities$^1$, and spectroscopic stellar parameters.\\[-6mm] \label{photo}}
 \setlength{\tabcolsep}{.1cm}
 \begin{tabular}{r@{\hspace{1mm}}rrclcc@{\hspace{1mm}}rrrrrrrrrrrrrrr}
 \noalign{}
\hline
\hline
  & HD  & HR  &Name & Sp.\,T$^{2}$ && $V$ & $B-V$ & $U-B$ & $Q$ & $b-y$& $m1$&  $c1$& $\beta$ & $T_\mathrm{eff}$& $\log g$  &~~~$\xi$ &$v\sin\,i$&
~~~ $\zeta$\\
  &     &     &     &                       && mag &  mag  &  mag  &  mag& mag
& mag & mag & mag&K~~&dex &&km\,s$^{-1}$&\\
\hline\\[-3mm]
  1 & 36591 & 1861&               &B1\,V   &        &  5.339 & $-$0.194  & $-$0.911  & $-$0.771 &$-$0.077  &   0.074  &  -0.002&   2.612& 27000 & 4.12 &  3 & 12 &\ldots\\
    &       &     &               &        &  $\pm$ & \ldots & $ $\ldots & $ $\ldots & $ $\ldots& \ldots &   \ldots   &  \ldots & \ldots  &  300 & 0.05 &  1 &  1 &\ldots\\
  2 & 61068 & 2928&PT\,Pup        &B1\,V$^{3,4}$&   &  5.711 & $-$0.176  & $-$0.868  & $-$0.741 &$-$0.068  &   0.077  &   0.050&   2.617& 26300 & 4.15 &  3 & 14 & 20   \\
    &       &     &               &        &  $\pm$ &  0.015 & $ $0.010  & $ $0.015  & $ $0.017 & 0.003	&   0.001  &   0.002&	0.003&     300 & 0.05 &  1 &  2 &  1   \\
  3 & 63922 & 3055&               &B0.2\,III$^{4}$ &        &  4.106 & $-$0.185  & $-$1.010  & $-$0.877 &$-$0.051  &   0.043  &  -0.078&   2.591& 31200 & 3.95 &  8 & 29 & 37   \\
    &       &     &               &        &  $\pm$ &  0.005 & $ $0.007  & $ $0.000  & $ $0.005 & 0.006	&   0.004  &   0.002&	0.005&     300 & 0.05 &  1 &  4 &  8   \\
  4 & 74575 & 3468&$\alpha$\,Pyx &B1.5\,III&        &  3.679 & $-$0.183  & $-$0.875  & $-$0.743 &$-$0.063  &   0.066  &   0.050&   2.604& 22900 & 3.60 &  5 & 11 & 20   \\
    &       &     &               &        &  $\pm$ &  0.006 & $ $0.003  & $ $0.007  & $ $0.007 & 0.002	&   0.002  &   0.007&	0.002&     300 & 0.05 &  1 &  2 &  1   \\
  5 &122980 & 5285&$\chi$\,Cen    &B2\,V$^{4}$&   &  4.353 & $-$0.195  & $-$0.774  & $-$0.634 &$-$0.094  &   0.096  &   0.168&   2.656& 20800 & 4.22 &  3 & 18 &\ldots\\
    &       &     &               &        &  $\pm$ &  0.007 & $ $0.006  & $ $0.013  & $ $0.014 & 0.003	&   0.006  &   0.006&	0.004&     300 & 0.05 &  1 &  1 &\ldots\\
  6 &149438 & 6165&$\tau$\,Sco    &B0.2\,V &        &  2.825 & $-$0.252  & $-$1.023  & $-$0.842 &$-$0.093  &   0.039  &  -0.090&   2.601& 32000 & 4.30 &  5 &  4 &  4   \\
    &       &     &               &        &  $\pm$ &  0.009 & $ $0.007  & $ $0.013  & $ $0.014 & 0.008	&   0.015  &   0.023&	0.007&     300 & 0.05 &  1 &  1 &  1   \\
  7 &   886 &  39 &$\gamma$\,Peg  & B2\,IV &        &  2.834 & $-$0.226  & $-$0.862  & $-$0.699 &$-$0.107  &   0.093  &   0.116&   2.627& 22000 & 3.95 &  2 &  9 &  8   \\
    &       &     &               &        &  $\pm$ &  0.015 & $ $0.012  & $ $0.012  & $ $0.015 & 0.003	&   0.001  &   0.000&	0.003&     400 & 0.05 &  1 &  2 &  2   \\
  8 & 29248 & 1463&$\nu\,$Eri     &B1.5\,IV$^{3,4}$ &   &  3.930 & $-$0.210  & $-$0.879  & $-$0.728 &$-$0.076  &   0.068  &   0.072&   2.610& 22000 & 3.85 &  6 & 26 & 15   \\
    &       &     &               &        &  $\pm$ &  0.023 & $ $0.009  & $ $0.007  & $ $0.010 & 0.000	&   0.004  &   0.010&	0.002&     250 & 0.05 &  1 &  2 &  5   \\
  9 & 35299 & 1781&               &B1.5\,V &        &  5.694 & $-$0.210  & $-$0.874  & $-$0.723 &$-$0.094  &   0.088  &   0.057&   2.631& 23500 & 4.20 &  0 &  8 &\ldots\\
    &       &     &               &        &  $\pm$ &  0.010 & $ $0.007  & $ $0.011  & $ $0.012 & 0.002	&   0.004  &   0.005&	0.003&     300 & 0.05 &  1 &  1 &\ldots\\
 10 & 35708 & 1810&$o$\,Tau       &B2\,V$^{3,4}$&   &  4.875 & $-$0.145  & $-$0.768  & $-$0.664 &$-$0.058  &   0.090  &   0.151&   2.642& 20700 & 4.15 &  2 & 25 & 17   \\
    &       &     &               &        &  $\pm$ &  0.012 & $ $0.006  & $ $0.005  & $ $0.007 & \ldots&   \ldots    &   \ldots  &   \ldots  &  200 & 0.07 &  1 &  2 &  5 \\
 11 & 36512 & 1855&$\upsilon$\,Ori&B0\,V   &        &  4.618 & $-$0.264  & $-$1.068  & $-$0.878 &$-$0.112  &   0.061  &  -0.095&   2.597& 33400 & 4.30 &  4 & 20 & 10   \\
    &       &     &               &        &  $\pm$ &  0.013 & $ $0.007  & $ $0.008  & $ $0.009 & 0.005	&   0.000  &   0.004&	0.003&     200 & 0.05 &  1 &  2 &  5   \\
 12 & 36822 & 1876&$\phi^1$\,Ori  &B0.5\,III$^{4}$ &      &  4.408 & $-$0.162  & $-$0.965  & $-$0.848 &$-$0.037  &   0.026  &  -0.062&   2.602& 30000 & 4.05 &  8 & 28 & 18   \\
    &       &     &               &        &  $\pm$ &  0.006 & $ $0.013  & $ $0.013  & $ $0.016 & \ldots&   \ldots    &  \ldots & \ldots  &     300 & 0.10 &  1 &  2 &  5   \\
 13 & 36960 & 1887&               &B0.7\,V$^{4}$ & &  4.785 & $-$0.250  & $-$1.015  & $-$0.830 &$-$0.106  &   0.072  &  -0.056&   2.600& 29000 & 4.10 &  4 & 28 & 20   \\
    &       &     &               &        &  $\pm$ &  0.007 & $ $0.010  & $ $0.016  & $ $0.018 & 0.004	&   0.002  &   0.003& 	0.003&     300 & 0.07 &  1 &  3 &  7   \\
 14 &205021 & 8238&$\beta$\,Cep   &B1\,IV  &        &  3.233 & $-$0.222  & $-$0.952  & $-$0.792 &$-$0.092  &   0.066  &   0.010&   2.605& 27000 & 4.05 &  4 & 28 & 20   \\
    &       &     &               &        &  $\pm$ &  0.014 & $ $0.006  & $ $0.011  & $ $0.012 & 0.002	&   0.003  &   0.006& 	0.003&     450 & 0.05 &  1 &  3 &  7   \\
 15 &209008 & 8385&18\,Peg        &B3\,III &        &  5.995 & $-$0.120  & $-$0.568  & $-$0.482 &$-$0.035  &   0.081  &   0.411&   2.660& 15800 & 3.75 &  4 & 15 & 10   \\
    &       &     &               &        &  $\pm$ &  0.008 & $ $0.014  & $ $0.013  & $ $0.016 & 0.000	&   0.003  &   0.002&	0.003&     200 & 0.05 &  1 &  3 &  3   \\
 16 &216916 & 8725&EN\,Lac        &B1.5\,IV$^{4}$ & &  5.587 & $-$0.144  & $-$0.836  & $-$0.732 &$-$0.047  &   0.066  &   0.092&   2.629& 23000 & 3.95 &  0 & 12 &\ldots\\
    &       &     &               &        &  $\pm$ &  0.015 & $ $0.008  & $ $0.007  & $ $0.009 & 0.004	&   0.003  &   0.007&	0.007&     200 & 0.05 &  1 &  1 &\ldots\\
 17 &  3360 &  153&$\zeta$\,Cas   & B2\,IV &  &        3.661 & $-$0.196  & $-$0.849  & $-$0.708 &$-$0.090  &   0.087  &   0.134&   2.627& 20750 & 3.80 &  2 & 20 & 12   \\
    &       &     &               &        &  $\pm$ &  0.017 & $ $0.006  & $ $0.024  & $ $0.024 & 0.000	&   0.000  &   0.000&	0.005&     200 & 0.05 &  1 &  2 &  5   \\
 18 & 16582 &  779&$\delta$\,Cet  &B2\,IV  &        &  4.067 & $-$0.219  & $-$0.850  & $-$0.692 &$-$0.099  &   0.091  &   0.102&   2.616& 21250 & 3.80 &  2 & 15 & 10   \\
    &       &     &               &        &  $\pm$ &  0.007 & $ $0.008  & $ $0.014  & $ $0.015 & 0.001	&   0.002  &   0.008&	0.004&     400 & 0.05 &  1 &  2 &  5   \\
 19 & 34816 & 1756&$\lambda$\,Lep &B0.5\,V$^{3}$&   &  4.286 & $-$0.273  & $-$1.010  & $-$0.813 &$-$0.110  &   0.073  &  -0.061&   2.602& 30400 & 4.30 &  4 & 30 & 20   \\
    &       &     &               &        &  $\pm$ &  0.005 & $ $0.015  & $ $0.005  & $ $0.012 & 0.001	&   0.004  &   0.003&	0.002&     300 & 0.05 &  1 &  2 &  7   \\
 20 &160762 & 6588&$\iota$\,Her   &B3\,IV  &        &  3.800 & $-$0.179  & $-$0.702  & $-$0.573 &$-$0.065  &   0.079  &   0.292&   2.661& 17500 & 3.80 &  1 &  6 &\ldots\\
    &       &     &               &        &  $\pm$ &  0.000 & $ $0.003  & $ $0.010  & $ $0.010 & 0.002	&   0.004  &   0.004&	0.001&     200 & 0.05 &  1 &  1 &\ldots\\
 21 & 37020$^{5}$ & 1893&$\theta$1 Ori A&B0.5\,V &        &  6.720 & $ $0.000  & $-$0.860  & $-$0.860 & 0.092  &   0.029  &  -0.048&   2.573& 30700 & 4.30 &  0 & 45 & 20   \\
    &       &     &               &        &  $\pm$ & \ldots & \ldots    & \ldots    & \ldots   & 0.002	&   0.003  &   0.008& 	0.022&     300 & 0.07 &  1 &  6 & 10   \\
 22 & 37042$^{5}$ &     &$\theta$2 Ori B&B0.5\,V$^{4}$ &        &  6.380 & $-$0.090  & $-$0.920  & $-$0.855 &$-$0.004  &   0.049  &  -0.080&   2.599& 29300 & 4.30 &  2 & 30 & 10   \\
    &       &     &               &        &  $\pm$ & \ldots & \ldots    & \ldots    & \ldots   & \ldots&   \ldots   &  \ldots  & \ldots  &     300 & 0.09 &  1 &  4 &  5   \\
 23 & 36959 & 1886&               &B1.5\,V &   &  5.670 & $-$0.240  & $-$0.910  & $-$0.737 &$-$0.092  &   0.085  &   0.039&   2.629& 26100 & 4.25 &  0 & 12 &  5   \\
    &       &     &               &        &  $\pm$ & \ldots & \ldots    & \ldots    & \ldots   & 0.005	&   0.004  &   0.005&	0.004&     200 & 0.07 &  1 &  2 &  2   \\
 24 & 37744 & 1950&               &B1.5\,V &        &  6.213 & $-$0.208  & $-$0.900  & $-$0.750 &$-$0.081  &   0.079  &   0.043&   2.629& 24000 & 4.10 &  0 & 33 & 15   \\
    &       &     &               &        &  $\pm$ &  0.011 & $ $0.006  & $ $0.017  & $ $0.018 & 0.005	&   0.010  &   0.009&	0.006&     400 & 0.10 &  1 &  4 &  7   \\
 25 & 36285 & 1840&               &B2\,V   &        &  6.315 & $-$0.195  & $-$0.823  & $-$0.683 &$-$0.086  &   0.093  &   0.115&   2.646& 21700 & 4.25 &  0 & 11 &  8   \\
    &       &     &               &        &  $\pm$ &  0.008 & $ $0.007  & $ $0.008  & $ $0.011 & 0.007 &   0.005  &   0.008&	0.008&     300 & 0.08 &  1 &  3 &  4   \\
 26 & 35039 &1765& $o$\,Ori       &B2\,IV$^{3}$& & 4.731 & $-$0.169  & $-$0.790  & $-$0.668&$-$0.069  &   0.083  &   0.173&   2.625 & 19600 & 3.56 &  4 & 12 &  7   \\
    &       &     &               &        &  $\pm$ &  0.011 & $ $0.009  & $ $0.006  & $ $0.011 & 0.002	&   0.001  &   0.004&	0.002&     200 & 0.07 &  1 &  3 &  4   \\
 27 & 36629$^{5}$ &     &               &B2\,V   &&  7.650 & $ $0.020  & $-$0.660  & $-$0.674 & 0.067  &   0.067  &   0.149&   2.661& 20300 & 4.15 &  2 & 11 &  5   \\
    &       &     &               &        &  $\pm$ &  \ldots& \ldots    & \ldots    & \ldots   & 0.005	&   0.006  &   0.015&	0.004&     400 & 0.10 &  1 &  3 &  2   \\
 28 & 36430 & 1848&               &B2.5\,V$^{4}$   &  &  6.217 & $-$0.180  & $-$0.739  & $-$0.609 &$-$0.085  &   0.111  &   0.202&   2.676& 19300 & 4.14 &  0 & 20 & 10   \\
    &       &     &               &        &  $\pm$ &  0.009 & $ $0.008  & $ $0.010  & $ $0.013 & 0.001	&   0.001  &   0.001&	0.005&     200 & 0.05 &  1 &  2 &  5   \\
 29 & 35912 & 1820&               &B2.5\,V $^{4}$  &        &  6.408 & $-$0.177  & $-$0.743  & $-$0.615&$-$0.080  &   0.102  &   0.211&   2.664 & 19000 & 4.00 &  2 & 15 &  8   \\
    &       &     &               &        &  $\pm$ &  0.020 & $ $0.008  & $ $0.007  & $ $0.011 & 0.002	&   0.002  &   0.003&	0.004&     300 & 0.10 &  1 &  4 &  4   \\
 30 & 46202 &     &               &O9\,V   &        &  8.186 & $ $0.178  & $-$0.737  & $-$0.865 & 0.207  &  -0.022  &  -0.036&   2.616& 34100 & 4.18 &  6 & 25 & 15   \\
    &       &     &               &        &  $\pm$ &  0.015 & $ $0.011  & $ $0.013  & $ $0.017 & 0.002	&   0.000  &   0.005&	0.001&     400 & 0.05 &  2 &  7 &  7   \\
\hline\\[-6mm]
 \end{tabular}
\begin{list}{}{}
\item[$^{1}$] Johnson \citet{mer91}; \citet{mor78} for stars \#1, 13, 21, 22, 23 and 27. Str\"omgren: \citet{hm98}.
\item[$^{2}$] \citet{hj82}; \citet{gc09} for stars \#6, 7, 11, 19 and 23; \citet{conti71} for star \#30.
\item[$^{3}$] Luminosity class corrected (see~App.~\ref{sect_st}).  Uncorrected classes are, for star \#2: III, \#8: III, \#10: IV, \#19: IV, \#26: V.
\item[$^{4}$] Spectral type corrected (see~App.~\ref{sect_st}). Uncorrected types are, for stars \#2, 8, 16, 28 and 29: B2, \#5, 10: B2.5, \#3, 12: B0, \#13: B0.5,\#22: B0.7.
\item[$^{5}$] Stars with anomalous reddening law. R$_V$ values are: 5.5 for stars \#21 and 22 and 4.2 for star \#27.
\end{list}
\end{table*}


\section{Introduction}
Stars with spectral types between $\sim$O9 and B2 belong to the 'natural' spectroscopic group 'OB', presenting strong H and \ion{He}{i} lines \citep{jj87}. 
The lower boundary of this group (B2) corresponds to lower mass and upper age limits of $\sim$8 M$_{\odot}$ and 30\,Myr, respectively, and the occurrence of core collapse supernovae (CCSN). 
OB stars delineate the spiral structure of our Milky Way and play an important role in cosmic evolution. They dominate the light of entire spiral and irregular galaxies and
influence the interstellar medium, star formation and galactic evolution through 
their ultraviolet radiation, stellar winds and supernovae explosions.   
Furthermore, they are contributors to nucleosynthesis and chemical evolution, 
since the CCSN are the primary sources of several chemical elements, e.g. oxygen. 
As they are very luminous ($\sim$10$^4$-10$^6$\,L$_{\odot}$), this group of stars became 
an important subject of astronomical research in the past decades. 

New developments allow to obtain better quality observations and more realistic physical models to analyse data, than those available in the past decades. Photographic spectrograms previously used have been replaced by very high-resolution
digital spectra. Furthermore, larger telescopes allow us to achieve higher signal-to-noise ratios.
And the advent of more complex computers permit to incorporate a more realistic physical 
background into our calculations of synthetic spectra that are used to  interpret observations.
These improvements have not only a significant effect on the determination of basic properties of the stars, such as those in the present work, i.e. effective temperatures and surface gravities, but also on other quantities that strongly depend on them. Examples are
the chemical abundances, distances, luminosities, radii, masses, and ages.
Improved determinations of these quantities allow us to better understand the physics of stars (their interior, nucleosynthesis, evolution) and the structure and evolution of galaxies.

In the present work, we mainly focus on the temperature and gravity scales of non-supergiant OB stars, i.e. main sequence to giant stars (core H-burning). And we also evaluate bolometric corrections in function of temperature.
We investigate whether our recent improvements made in the acquisition of observational data,  calculation of more realistic theoretical spectra and in the spectral analysis methodologies, affect the overall effective temperature and surface gravity scales of those objects. 
A quantitative spectroscopic analysis for a representative sample of 30 stars in the solar neighbourhood has been performed in \citet[hereafter Paper~I]{np12}, \citet[Paper~II]{ns11} and \citet[Paper~III]{b11}, where multiple independent ionization equilibria were evaluated iteratively together with fits to Balmer lines in NLTE in order to reproduce the whole observed spectrum in the optical. The analysis resulted in a unique set of atmospheric parameters at high precision and accuracy for each star. Based on these stellar parameter determination, we provide here 
new calibrations to determine temperatures and gravities of OB stars from photometric quantities and their spectral classification.

The paper is structured as follows. Section~\ref{sect_obs} presents the star sample and observations, Sect.~\ref{sect_param} briefly addresses the techniques used for stellar parameter determinations in previous papers, Sect.~\ref{sect_tgscales} discusses the temperature and gravity scales to photometric data.
 In Sect.~\ref{sect_comp}, our photometric scales and spectroscopic parameters are compared to other work from the literature,
Sect.~\ref{sect_stcali} focus on the spectral type calibration to temperature and Sect.~\ref{sect_bc} on the bolometric corrections as a function of temperature. Section~\ref{sect_summ} presents a summary and conclusions. Furthermore, App.~\ref{sect_st} addresses the spectral type and luminosity class examination and App.~\ref{sect_indicators} the spectroscopic indicators for temperature and gravity determinations.

\section{The star sample and observations}\label{sect_obs}

Our star sample covers a slightly broader temperature range than the standard definition for OB stars, with spectral types from B3 to O9. In the context of stellar evolution, the stars are located between the Zero Age 
(ZAMS) and the Terminal Age Main Sequence (TAMS). Their temperatures range from $\sim$16\,000 to 34\,000\,K, the surface gravities from $\sim$3.6 to 4.3 in logarithmic scale, and the masses from $\sim$5 to 20 M$_{\odot}$.
The sample is listed in Table~\ref{photo}, where the spectral type, photometric quantities and atmospheric parameters derived in the Papers~I-III are given. 
Object identifications include HD and HR numbers and names of the stars and our own numbering scheme to facilitate easy identification in the figures and
other tables (i.e. stars analyzed in Paper~I: \#1-20, Paper~II: \#21-29, Paper~III:\#30).

For the whole sample, high-resolution \'echelle spectra at very high signal-to-noise ($S/N$) ratio   -- ranging from 250 up to over 800 in $B$ -- and wide wavelength coverage were obtained, either by own observations, or from archives. Details on the observations and data reduction can be found in Papers I and III and for stars \#21-29 in \citet{s10}. 
Spectra for stars \#1--6 were obtained with {\sc Feros} 
on the ESO 2.2m telescope in La Silla/Chile, with a resolving
power of $R$\,$=$\,$\lambda/\Delta\lambda$\,$\approx$\,48\,000. 
Stars \#7--16 were observed with {\sc Foces} 
on the 2.2\,m telescope at Calar Alto/Spain at a resolution of $R$\,$\approx$\,40\,000.
Spectra of stars \#17--20 were extracted from the archive of the {\sc Elodie} \'echelle
spectrograph mounted on the 1.93\,m telescope of the Observatoire de
Haute-Provence/France at $R$\,$\approx$\,42\,000.
Spectra for stars \#21-29 were obtained with {\sc Fies} mounted at the 2.5\,m Nordic Optical Telescope (NOT) in El Roque de los Muchachos observatory on La Palma (Canary Islands, Spain). 
A spectrum obtained with {\sc Feros} for star \#30 was extracted from the ESO Archive.

We  aimed at selecting single stars,
as second light from a companion of a similar spectral class distorts the ratio of line- to
continuum-fluxes. Thus, objects in SB1 systems with much fainter companions (e.g. stars \#21 and 26) can therefore still qualify as targets, because the contribution of the secondary star in optical light (UBV, uvby) is marginal. Individual components in a visual binary are also considered as targets (e.g. star \#13 ).
The star sample contains normal, $\beta$Cephei and slowly pulsating B-stars\footnote{Pulsation periods of $\beta$Cephei stars vary between $\sim$3 to 8 hours, while those in SPB stars vary between $\sim$0.5 to 3 days.}.
Exposures taken during a small fraction of the same night were co-added. The analysis will thus yield atmospheric parameters like effective temperature and surface gravity valid only for the moment of observation. For this sample, we have adopted reliable photometric data from the literature, these however correspond to a different moment of observation than for the spectroscopic data, which might be of significance for some variable stars.

\begin{figure}[t!]
\resizebox{0.99\hsize}{!}{\includegraphics[width=12cm]{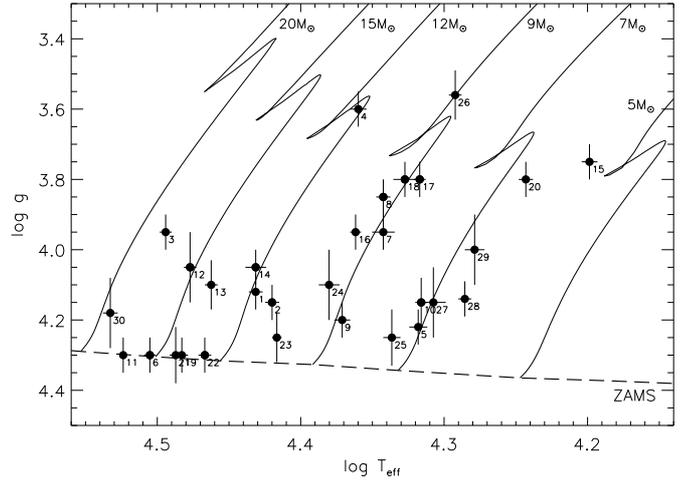}}
\caption[]{Location of the sample stars in the $T_\mathrm{eff}$-$\log g$ diagram. Atmospheric parameters and star identification are according to Table~\ref{photo}. Evolutionary tracks for non-rotating stars of metallicity $Z=0.014$ are from \citet{e12}. The location of the ZAMS is indicated.}
\label{f0}
\end{figure}

\section{Stellar parameter determination}\label{sect_param}
Atmospheric parameters were determined simultaneously via line-fitting of synthetic spectra computed in NLTE to the observed data, as described in Papers I--III. 
Atmospheric parameters are 
the effective temperature $T_\mathrm{eff}$, surface gravity $\log g$, 
microturbulence $\xi$, projected rotational velocity $v\,\sin\,i$ and macroturbulence $\zeta$.
Details on the synthetic spectra and the quantitative analysis are summarized in the following.

\subsection{Spectrum synthesis in NLTE}\label{sect_the}

The analysis was based on a grid of synthetic spectra 
in NLTE computed with the codes {\sc Detail} and {\sc Surface} \citep{gid81,but_gid85} on prescribed LTE atmospheric structures calculated with {\sc Atlas9} \citep{kur93b}.
Model atoms for H, He, C, N, O, Ne, Mg, Si and Fe for different ionization stages were employed. 
Opacities due to hydrogen and helium and the individual metals were considered in NLTE, and line blocking was accounted for in LTE via Kurucz' Opacity Distribution Functions \citep{kur93a}.
This approach provides an efficient way to compute realistic synthetic
spectra when the atmosphere is close to LTE. In particular, it is equivalent to full NLTE model atmosphere calculations for late O-type to early B-type stars with weak winds \citep{np07,pnb11}.

\begin{figure*}[t!]
\resizebox{0.47\hsize}{!}{\includegraphics[width=12cm]{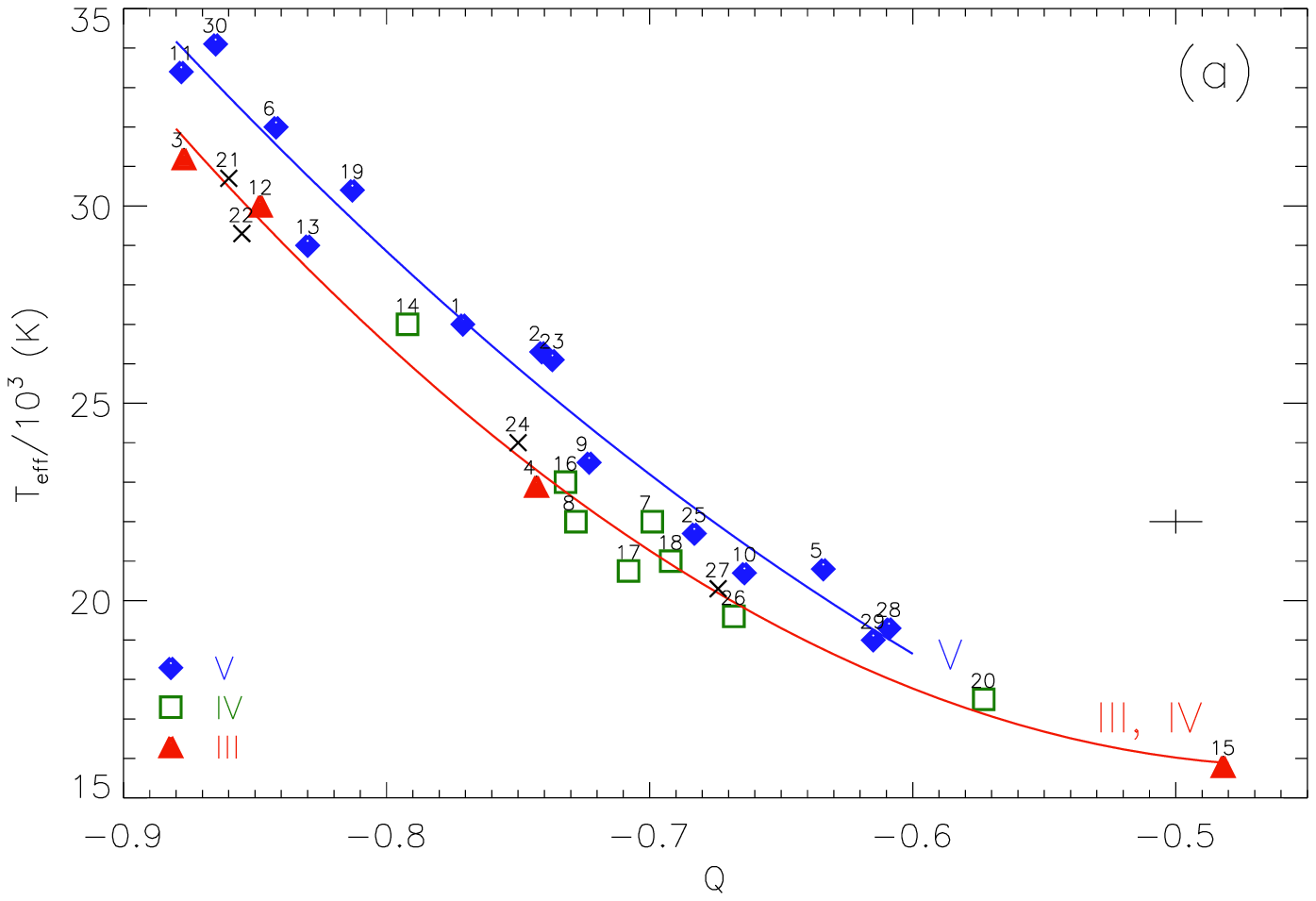}}
\resizebox{0.47\hsize}{!}{\includegraphics[width=12cm]{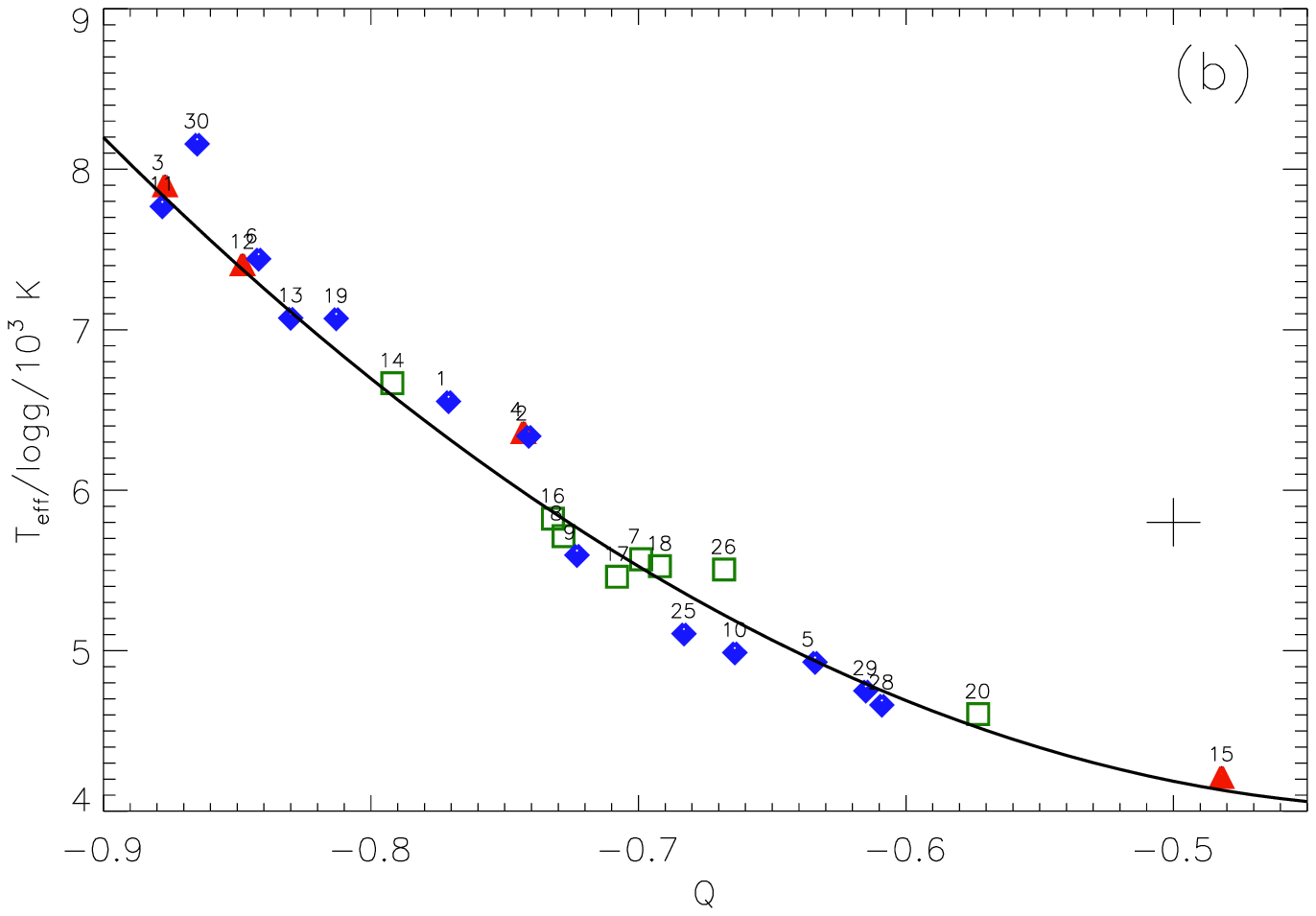}}
\resizebox{0.47\hsize}{!}{\includegraphics[width=12cm]{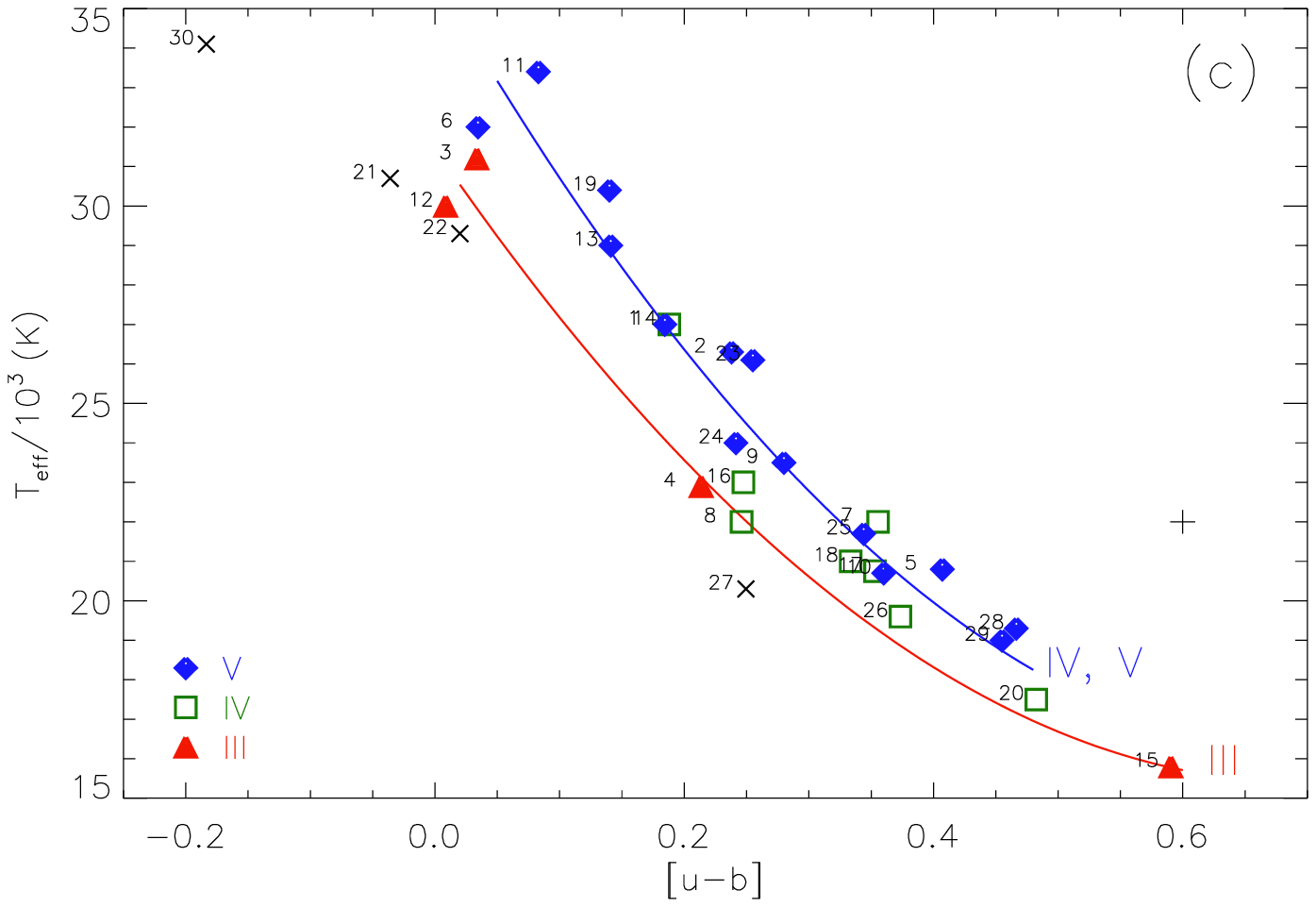}}
\resizebox{0.47\hsize}{!}{\includegraphics[width=12cm]{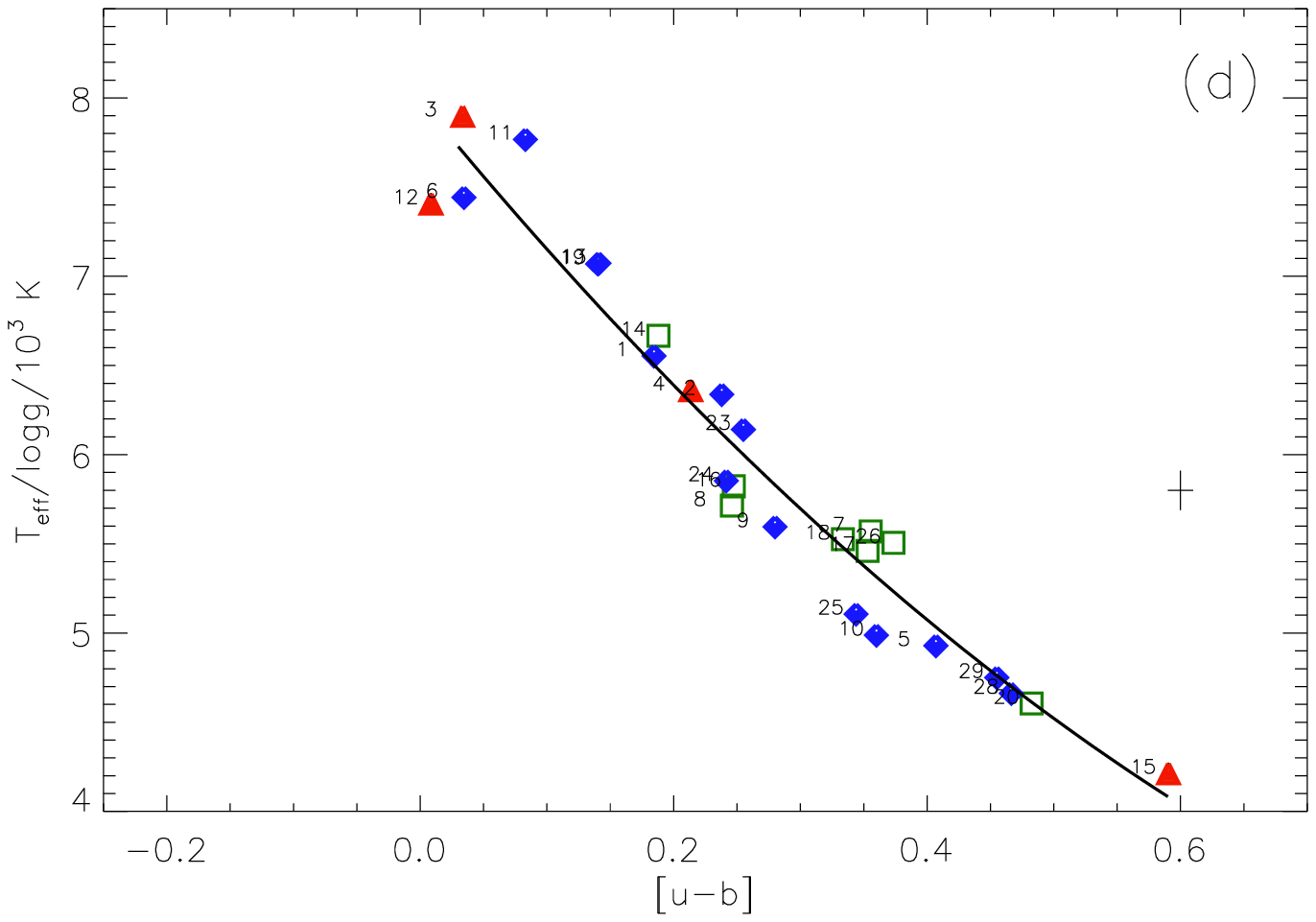}}
\resizebox{0.47\hsize}{!}{\includegraphics[width=12cm]{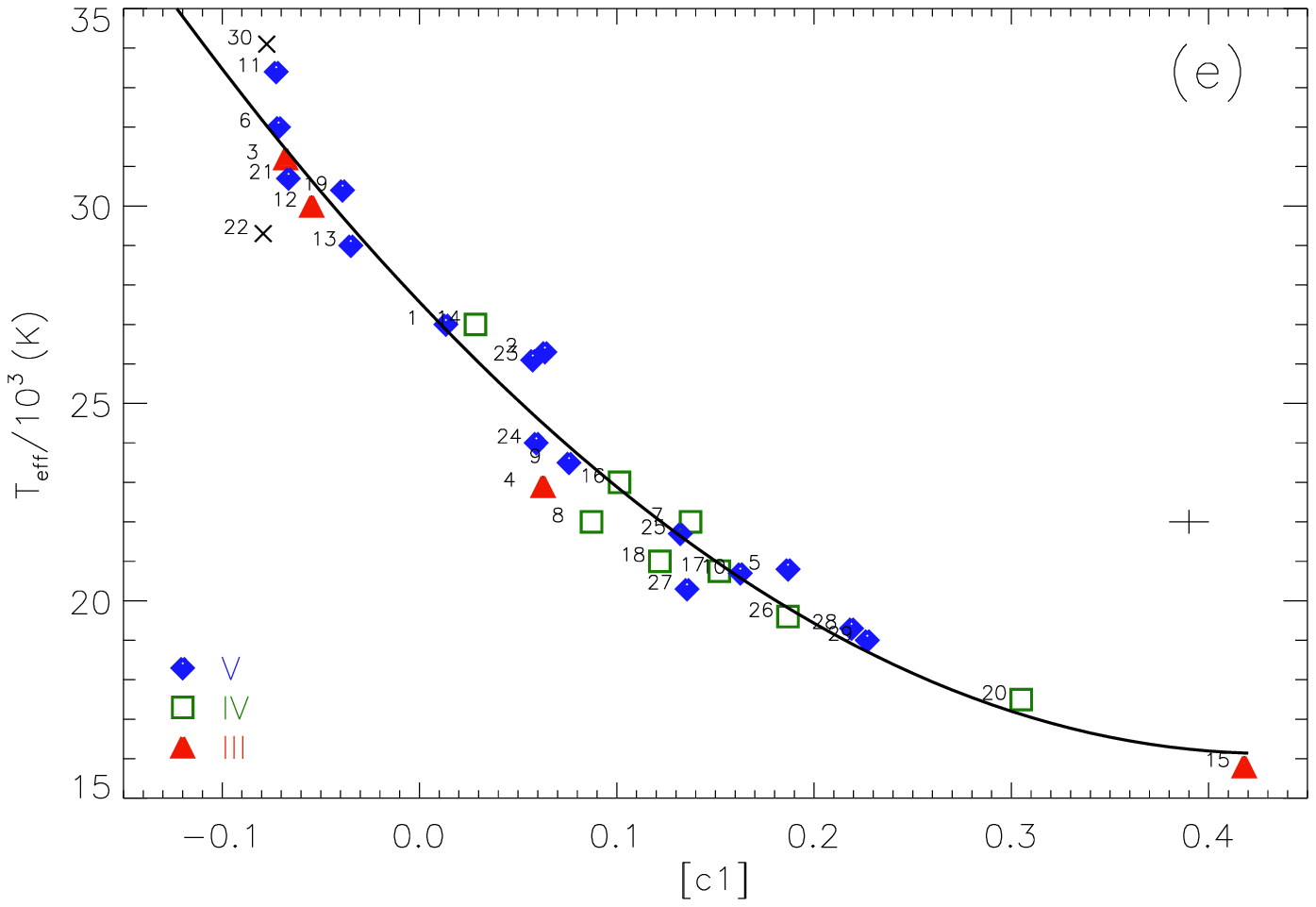}}
\hspace{1cm}
\resizebox{0.47\hsize}{!}{\includegraphics[width=12cm]{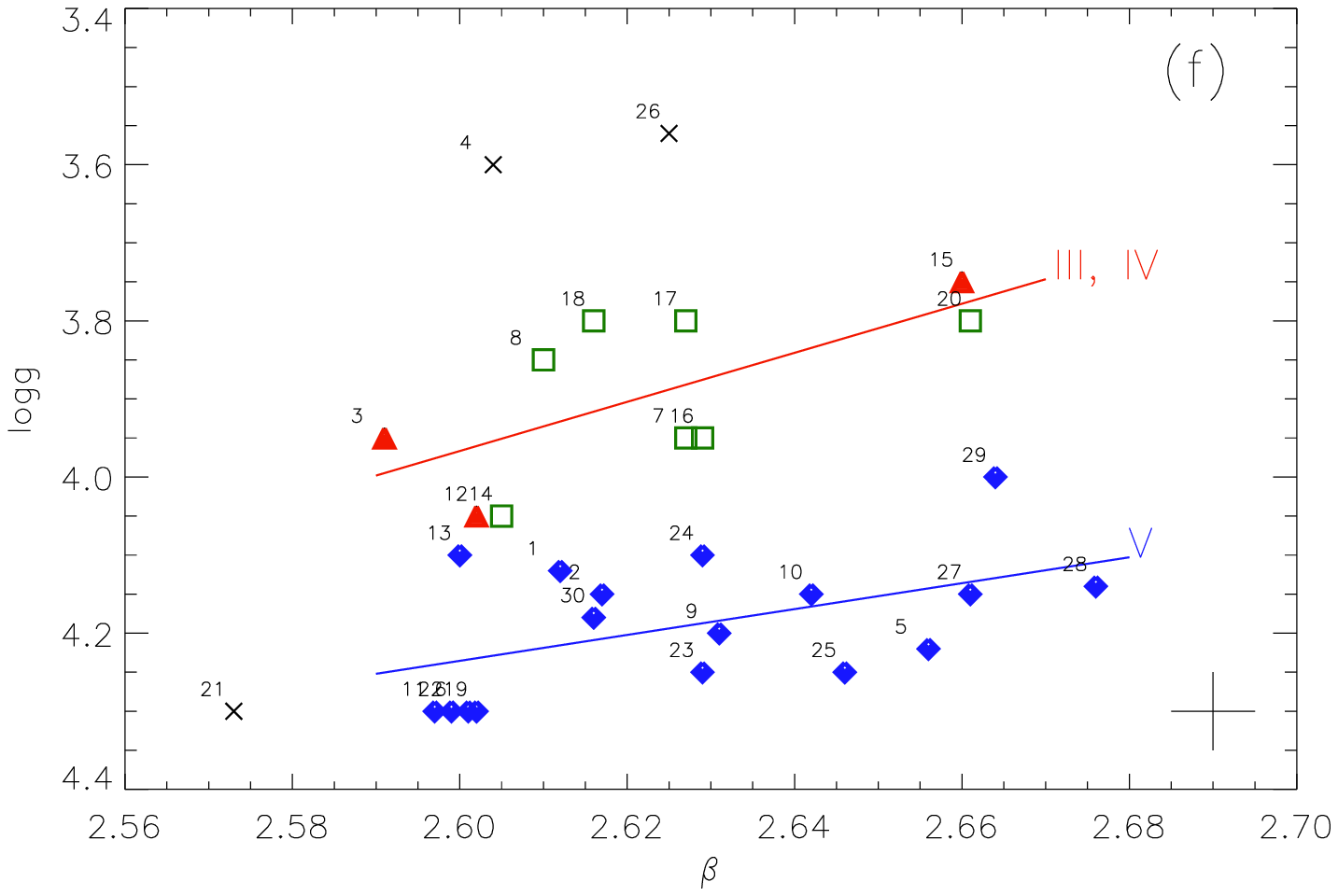}}
\caption[]{Functional relationships for the dependence of $T_\mathrm{eff}$ (left panels) and $\log g$  (right panels) on the photometric parameters $Q$ (a,b), $[u-b]$ (c,d), $[c1]$ (e) and $\beta$ (f). 
Symbols correspond to the sample stars and curves are polynomial fits to the data for different luminosity classes, encoded according to the legend. A few stars have been excluded from the fits (crosses). Typical error bars are indicated. See the text for details.}
\label{f1}
\end{figure*}

\subsection{Quantitative spectral analysis}\label{sect_an}

The spectral analysis follows the methodology introduced by \citet{n07} and applied for H/He and C by \citet{np07,np08} and for other elements by \citet{pnb08}. 
Stellar parameters are determined through ionization equilibria for different species and Balmer lines fits. 
Spectroscopic indicators for effective temperature and surface gravity of the star sample are summarized in Table~\ref{indicators}. 
Fitted spectra for stars \#1,6,7 and 15 (Paper~I) and \#9 (Paper~II), illustrate  examples of global fits for different spectral classes.
Numerous examples of fits to stellar spectral energy distributions (SEDs) can also be found in Papers~I and III. 
The reliability of our temperature and gravity determinations and the minimisation of systematic uncertainties are confirmed by simultaneous agreement of several independent indicators: ionization equilibria of different atomic species, fits to Balmer lines and fits to the spectral energy distributions from the UV to the IR. Furthermore, our surface gravities are also confirmed by simultaneous match of ionization equilibria, fits to Balmer lines and the agreement of spectroscopic distances with Hipparcos distances (see Paper~I). Note that every model atom for NLTE calculations is independent from each other, hence multiple ionization equilibria only render 4-5 independent indicators for temperature and gravity determinations.

The location of the (apparently slowly-rotating) sample stars in the $T_\mathrm{eff}$-$\log g$ plane
is visualized in Fig.~\ref{f0}, where also evolutionary tracks for non-rotating stellar models 
\citep{e12} are shown for comparison. Both, the models and the stars have the same metallicity of $Z$= 0.014,  the stars' metallicities being determined in Papers I-III. The star sample provides us with a good coverage of the main-sequence band, ranging from $\sim$6 to 20 M$_{\odot}$.

\section{Temperature and gravity scales}\label{sect_tgscales}

Useful temperature scales for early-type stars have been presented by \citet{napi93}, \citet{d99} and \citet{l02},
where the temperature can be estimated from reddening-independent photometric quantities (assuming a valid standard reddening-law).
The first work scales the temperature to the $[u-b]$\footnote{$[u-b]= (u-b)-1.5(b-y)= c_1 + 2m_1 + 0.5\,(b-y)$}-parameter from the narrow-band Str\"omgren $ubvy$-system. 
In contrast, the other two employ the $Q$\footnote{$Q=(U-B)-0.72\,(B-V)$}-index from the 
broad-band Johnson $UBV$-system.
In all three cases, the temperature determination is based on other independent photometric calibrations or on fits to spectral energy distributions. \citet{l02} proposed, in addition, temperature scales for two significantly different values of gravity.

Motivated by the aforementioned studies, we investigate whether new temperature and gravity scales to $Q$, $[u-b]$, the Str\"omgren $[c1]$\footnote{$[c1]=c1-0.2\,(b-y)$} and $\beta$\footnote{$\beta$ is a measure of the stellar H$\beta$ line intensity} can be established based on our previously derived spectroscopic parameters. Our atmospheric parameters derived simultaneously from several independent indicators and high quality spectral models and observations offer a unique 
set of input data for building such scales: uncertainties are as low as $\sim$1--2\% for effective temperatures and $\sim$10\% for surface gravities. This is in contrast to uncertainties of $\sim$ 5-10\% and $\sim$25\%, respectively, using standard methods. Typical uncertainties for effective temperature
given in the literature are $\le$3-5\%, however these do not take into account the large
discrepancies -- up to $\sim$6000\,K -- that exist between different temperature determination methods. Error estimates of up to 40-60\% in surface gravity are also found in the literature (see Sect.~\ref{sect_comp} for more details). 
Likewise, the most accurate photometric data available in the literature are adopted for this study. The compilation of photometric data is as homogeneous as possible, however for some objects, data are adopted from different sources.
In addition, re-examinations of spectral types and luminosity classes were performed for the sample stars, as described in App.~\ref{sect_st}. Stellar parameters, photometric indices, and their respective uncertainties, spectral types and luminosity classes adopted for the sample are listed in Table~\ref{photo}. 

In the following, we present {\em empirical} relationships of atmospheric parameters and photometric indices based on data of our star sample. Note that the star sample does not cover uniformly the parameter space under consideration, therefore 
the relationships can be refined when adopting a more complete sample. Polynomial fits are chosen to directly compare our results to previous works from \citet{napi93}, \citet{d99} and \citet{l02}. A few stars do not follow tightly the empirical fits for different reasons that need to be further investigated. 
Moreover, the correlation between effective temperature and surface gravity could be treated in an iterative manner in order to find an optimal solution, like the method proposed in \citet{md84}  and \citet{napi93}. We aim at investigating this correlation with a larger sample of stars covering uniformely the parameter space in a further paper.

\subsection{Spectroscopic parameters vs. $Q$}\label{calibrations}

The reddening-free Johnson $Q$-index is particularly interesting for OB stars, as there are many distant objects in the Galactic spiral arms that at present do not 
have measurements in photometric systems other than the Johnson-system. 
Figure~\ref{f1}a shows the relation between our spectroscopic temperatures and $Q$, while Fig.~\ref{f1}b displays the behavior of $T_\mathrm{eff}/\log g$ vs. $Q$. The data from Table~\ref{photo} are indicated by symbols and the polynomial fits by full lines, encoded for different luminosity classes as shown in the legend (in color in the online version). Four objects of luminosity class V had to be excluded from the analysis -- marked with a cross in Fig.~\ref{f1}a (\# 21, 22, 24 and 27). These stars are located in the Orion OB association. Stars \# 21, 22 and 27 are located in a region of dense nebulosity and present anomalous reddening (see Table~\ref{photo}). In these cases, the $Q$-index should be re-evaluated
using an appropriate reddening law, which is, however beyond the scope of the present paper.

The distinction of different luminosity classes reveals tighter trends of $T_\mathrm{eff}$ vs. $Q$ than previously reported. The trends are consistent with those found by \citet{np08} for a sub-set of stars (\#1--6), where the parameter determination was based only on H, He and C. However, that sample was insufficient for a statistical analysis. In the present work, the polynomial fits for luminosity classes III and IV converge into a single curve and the scale for luminosity class V is $\sim$1500 to 2000~K hotter.
The minimisation of uncertainties in our parameter determination and the high accuracy of the photometric data allow clear trends for different luminosity classes to be distinguished. 
In contrast, standard parameter determinations may provide larger systematic uncertainties (see Sect.~\ref{sect_comp}), that would prevent the relations found here to be identified.

The relationship between the temperature and $Q$ for different luminosity classes, i.e., different evolutionary stages, displayed in Fig.~\ref{f1}a can be expected by the fact that stars closer to the ZAMS are more compact and bluer, therefore hotter than the more evolved giants at similar masses. 

In contrast, the ratio $T_\mathrm{eff}/\log g$ vs. $Q$ does not show a dependence on the luminosity class (Fig.~\ref{f1}b), since all three polynomial fits converge approximately to a single solution. Therefore, we adopt one fit to the whole sample (except for the excluded stars), indicated by the thick black curve.

Temperature and gravity scales to the Johnson $Q$-parameter are proposed, based on the curves displayed in Figs.~\ref{f1}a,b.
Polynomial fits result in the following equations for the different luminosity classes:
\begin{eqnarray}\label{vq}
\mathrm{V:}~  T_Q= (22.59 + 48.57\,Q + 70.26\,Q^2)\,10^3\,\mathrm{K}
\end{eqnarray}
\begin{eqnarray}\label{iiiq}
\mathrm{III,\,IV:}~  T_Q= (33.42 + 78.40\,Q + 87.19\,Q^2)\,10^3\,\mathrm{K}
\end{eqnarray}
\begin{eqnarray}\label{gq}
\log g_Q= \frac{T_Q(\mathrm{V,\,IV\,or\,III})}{(6.67 + 13.30\,Q + 16.67
\,Q^2)\,10^3\,\mathrm{K}}
\end{eqnarray}
\normalsize

The range where these equations are valid are delimited by the observational data, as indicated in
 Figs.~\ref{f1}a,b. Note that the parameter range of interest is not uniformly sampled by the observations (Fig.~\ref{f0}), a further addition of few hotter stars with lower surface gravities will allow to refine the fits from the present work. The internal and external uncertainties of $T_Q$ and $\log g_Q$ are listed in Table~\ref{errors}. The internal uncertainty measures the 1$\sigma$-dispersion of the residuals of the fitted values and the spectroscopic parameters and it depends on the quality of the polynomial fits. The external uncertainty is calculated via error propagation of the parameters involved and it depends on the uncertainties of the spectroscopic and photometric parameters, and on the reliability of luminosity classifications. Luminosity classes are the largest sources of external errors , as the relationship between temperature and the $Q$-index for luminosity class V is up to $\sim$2000\,K hotter than that for III,IV. 
The precision of the parameter determination is as indicated in Table~\ref{errors} only when accurate photometric data {\em and} a reliable luminosity class is given for the star\footnote{Luminosity classes from {\sc Simbad} from different sources might differ from each other. See \citet{np08} and \citet{fp12} for corrections of luminosity classes of the {\sc Simbad} database.}. 

When the luminosity class is not known or is not reliable, the uncertainty in the derived temperature increases to $\sim$1500\,K, which is still smaller than the differences between various temperature determinations in the literature (they can amount up to $\sim$2000\,K and in extreme cases up to $\sim$6000\,K; see Sect.~\ref{sect_comp} for more details). 
In the case of the gravity, the lack of precise luminosity classes causes 
an error of $\sim$0.2\,dex, that is also smaller than the maximum differences shown in the literature of $\pm$0.4\,dex. 
Uncertainties in temperature of $\pm$1500\,K and in gravity of $\pm$0.2\,dex are, however, too large for some further applications, such as precise chemical abundance determinations or precise derivation of mass, radius and distances of stars. 
Despite the luminosity classification can compromise the accuracy of temperature and gravity, our calibrations are a useful tool to estimate such parameters for further applications. Therefore, $T_Q$ and $\log g_Q$ can be employed as starting points to be refined using, e.g. spectroscopic analyses.

\subsection{Spectroscopic parameters vs. $[u-b]$}

In this section we investigate a scale based on the narrow-band Str\"omgren colors  for estimating effective temperatures and gravities.
Figures~\ref{f1}c,d show the relationship between $[u-b]$ and our spectroscopic temperatures and $T_\mathrm{eff}/\log g$, respectively. The data are encoded like in Figs.~\ref{f1}a,b. Four objects had to be excluded from the analysis -- marked with a cross in Fig.~\ref{f1}c. Stars \# 21, 22 and 27 have anomalous reddening and star \#30 is apparently too hot for the relationship found for the rest of the main sequence stars\footnote{\citet{napi93} propose an upper limit of 30\,000~K for this relationship.}.
Similarly to the case of the $Q$-parameter, trends of $T_\mathrm{eff}$ vs. $[u-b]$ are found for different luminosity classes. In this case, the polynomial fits for luminosity classes IV and V converge into a single relation, while the scale for luminosity class III is $\sim$2000 to 4000~K cooler. In addition, $T_\mathrm{eff}/\log g$ vs. $[u-b]$ does also not indicate a dependence on the luminosity class (Fig.~\ref{f1}d).
The curves displayed in Figs.~\ref{f1}c,d are employed to propose a temperature and gravity scale to
the Str\"omgren photometric parameter $[u-b]$.
Polynomial fits to the data result in the following equations for the different luminosity classes:
\begin{eqnarray}\label{vu}
\mathrm{IV, V:}~  T_{[u-b]}= (35.82 - 54.86\,[u-b] + 38.07
\,[u-b]^2)\,10^3\,\mathrm{K}
\end{eqnarray}
\begin{eqnarray}\label{iiiu}
\mathrm{III:}~  T_{[u-b]}= (31.45 - 46.00\,[u-b] + 32.96
\,[u-b]^2)\,10^3\,\mathrm{K} 
\end{eqnarray}
\begin{eqnarray}\label{gu}
\log g_{[u-b]}= \frac{T_{[u-b]}\,(\mathrm{V,IV\,or\,III})}{(7.98 - 8.67\,[u-b] + 3.48
\,[u-b]^2)\,10^3\,\mathrm{K}}
\end{eqnarray}
\normalsize

Similarly to the case of the $Q$-parameter, the range where these equations are valid are determined by the observational data.
The internal and external uncertainties of $T_{[u-b]}$ and $\log g_{[u-b]}$ are listed in Table~\ref{errors}. When the luminosity class is not known or not precise, the error in the derived temperature increases to $\sim$2000\,K and for gravity to $\sim0.3$\,dex, both larger than in the case of the calibrations to $Q$.

\subsection{Spectroscopic parameters vs. $[c1]$ and $\beta$}\label{calibrationsc1}

The Str\"omgren parameters $[c1]$ and $\beta$  have been defined as tools to estimate $T_\mathrm{eff}$ and $\log g$ in early-type stars. 
A single trend for the effective temperature vs. $[c1]$ (Fig.~\ref{f1}e) and two curves for $\log g$ vs. $\beta$ (Fig.~\ref{f1}f) are found. For the latter, the relationship for luminosity class V is $\sim$0.3\,dex higher than for III/IV. 
Stars \#4 and 26 have been excluded from the fits because their gravities 
are too low in comparison with the rest of the stars of luminosity class III/IV (they lie slightly above the TAMS in Fig.~\ref{f0}). Star \# 21 is also excluded, as there is a large nebular emission feature centered on H$\beta$, that causes a significantly reduced value of the $\beta$-index. 
As the parameter range of interest is not uniformly sampled by the observations (Fig.~\ref{f0}), a further addition of few hotter stars with low surface gravities and/or cooler stars on the ZAMS might produce a flattening of the slope for luminosity class V in  Fig.~\ref{f1}f, which remains to be investigated.

 Polynomial fits to the present data result in the following equations: 
\begin{eqnarray}\label{tc1}
\mathrm{T}_{[c1]}= (27.57 - 52.92\,[c1] + 61.25\,[c1]^2)\,10^3\,\mathrm{K}
\end{eqnarray}
\begin{eqnarray}
\mathrm{V:}~  \log g= 8.75 - 1.73\,\beta
\end{eqnarray}
\begin{eqnarray}
\mathrm{III, IV:}~  \log g= 12.14 - 3.14\,\beta
\end{eqnarray}
\normalsize

The uncertainties are listed in Table~\ref{errors}. If the luminosity class is not well known, the error in the surface gravity determination increases to $\sim$0.25\,dex.

\begin{figure*}[t!]
\resizebox{0.47\hsize}{!}{\includegraphics[width=12cm,height=5.5cm]{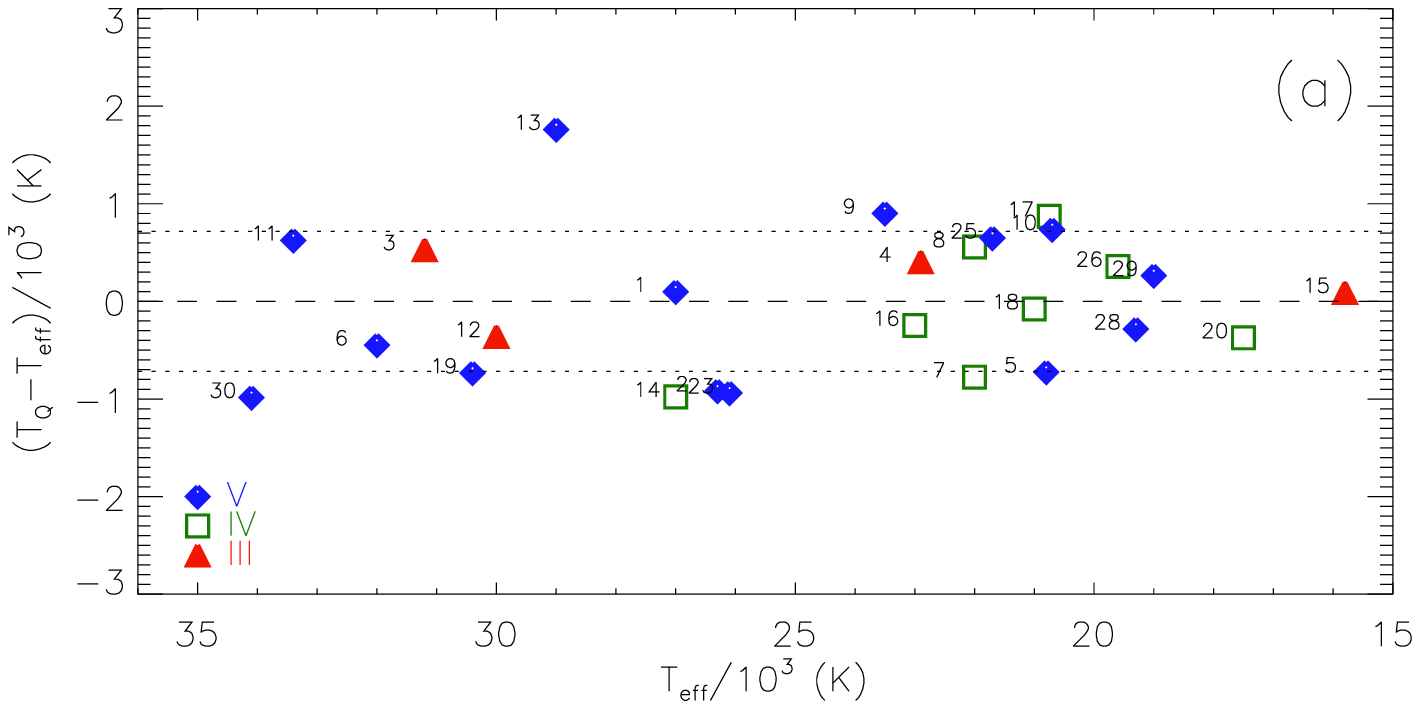}}
\resizebox{0.47\hsize}{!}{\includegraphics[width=12cm,height=5.5cm]{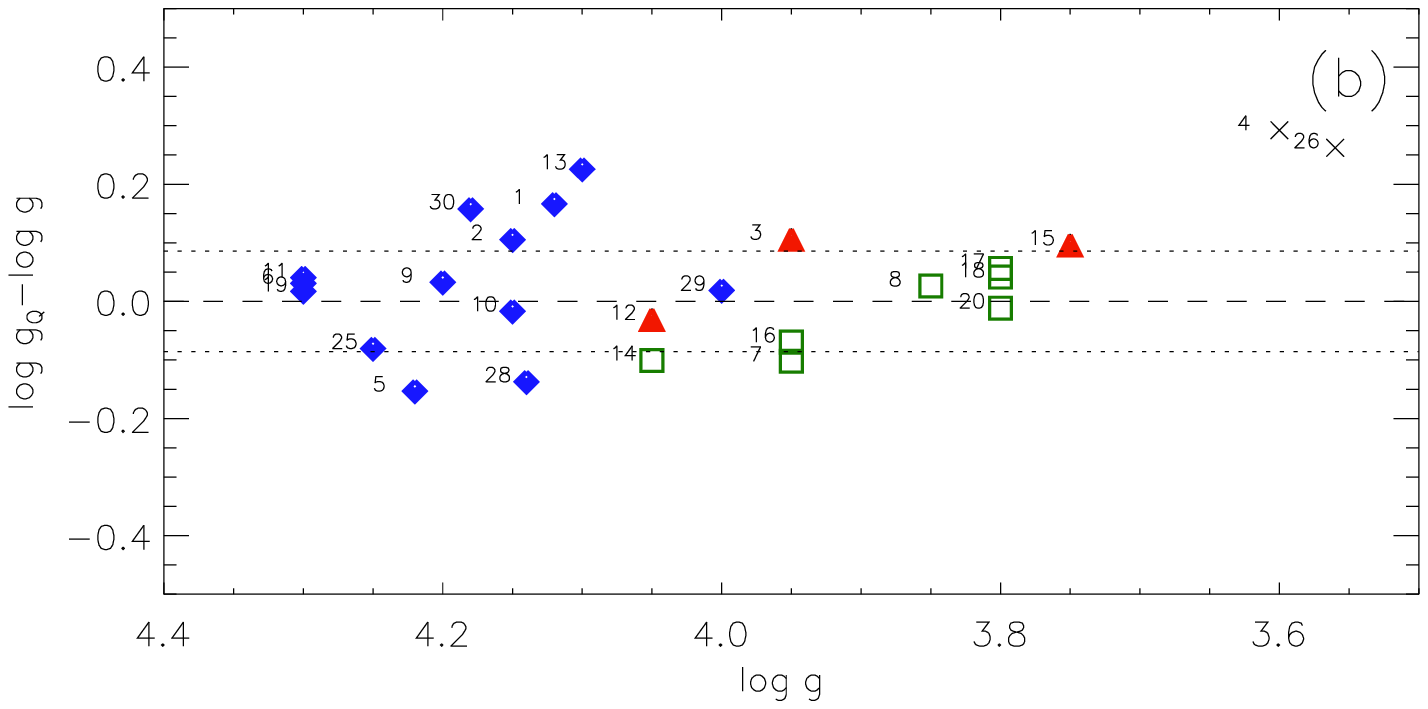}}
\resizebox{0.47\hsize}{!}{\includegraphics[width=12cm,height=5.5cm]{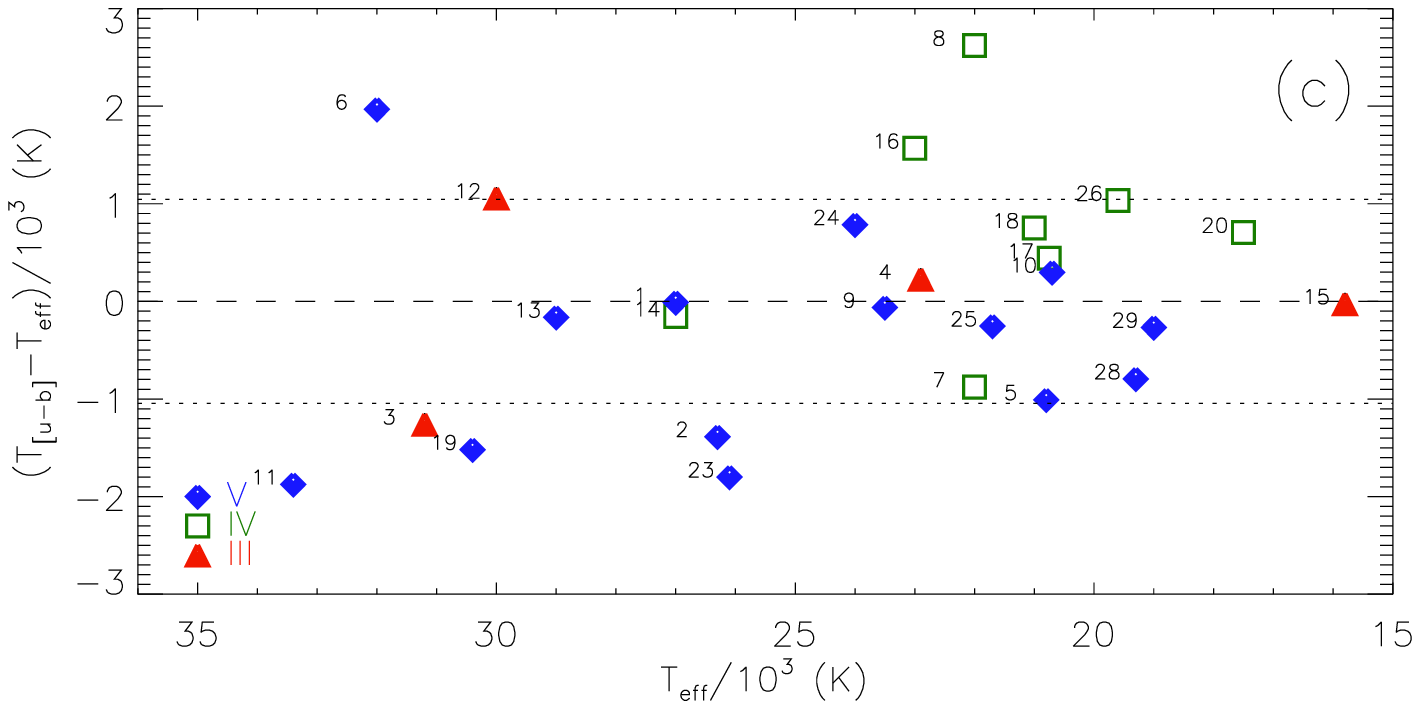}}
\resizebox{0.47\hsize}{!}{\includegraphics[width=12cm,height=5.5cm]{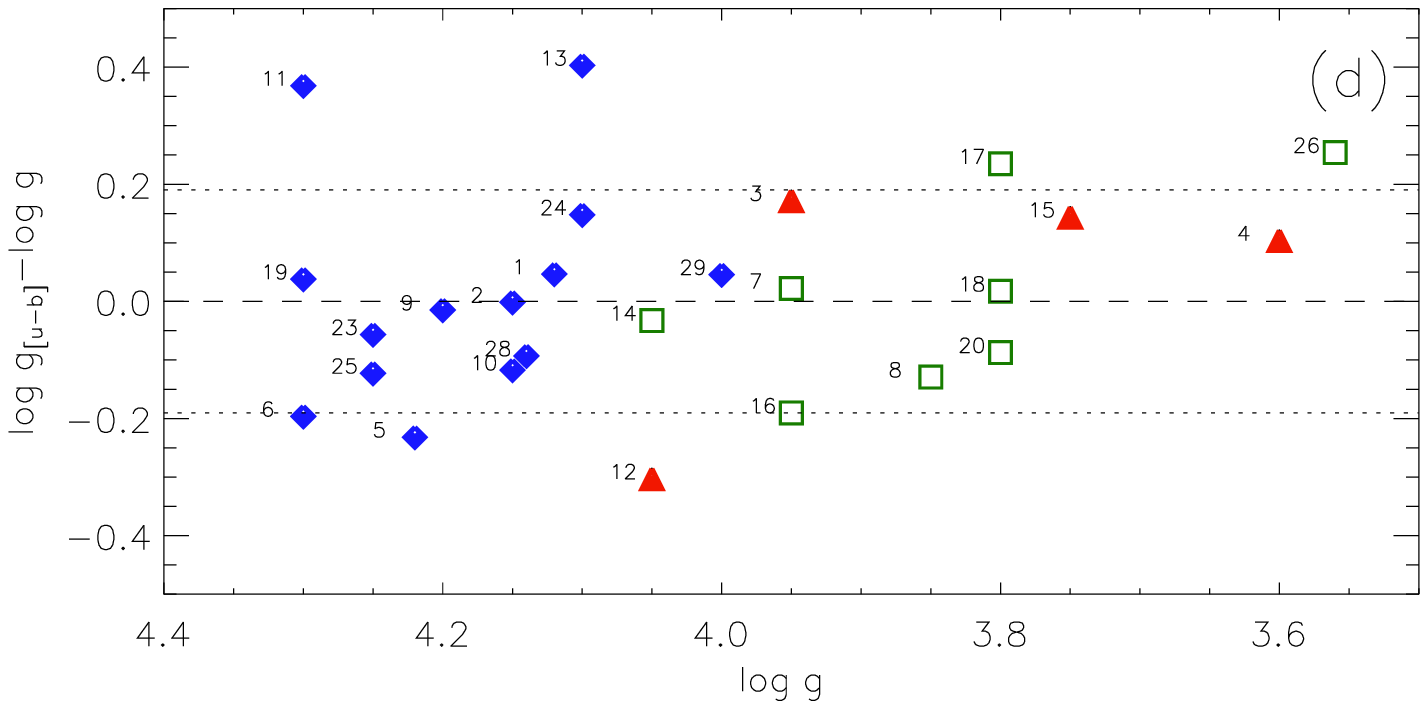}}
\resizebox{0.47\hsize}{!}{\includegraphics[width=12cm,height=5.5cm]{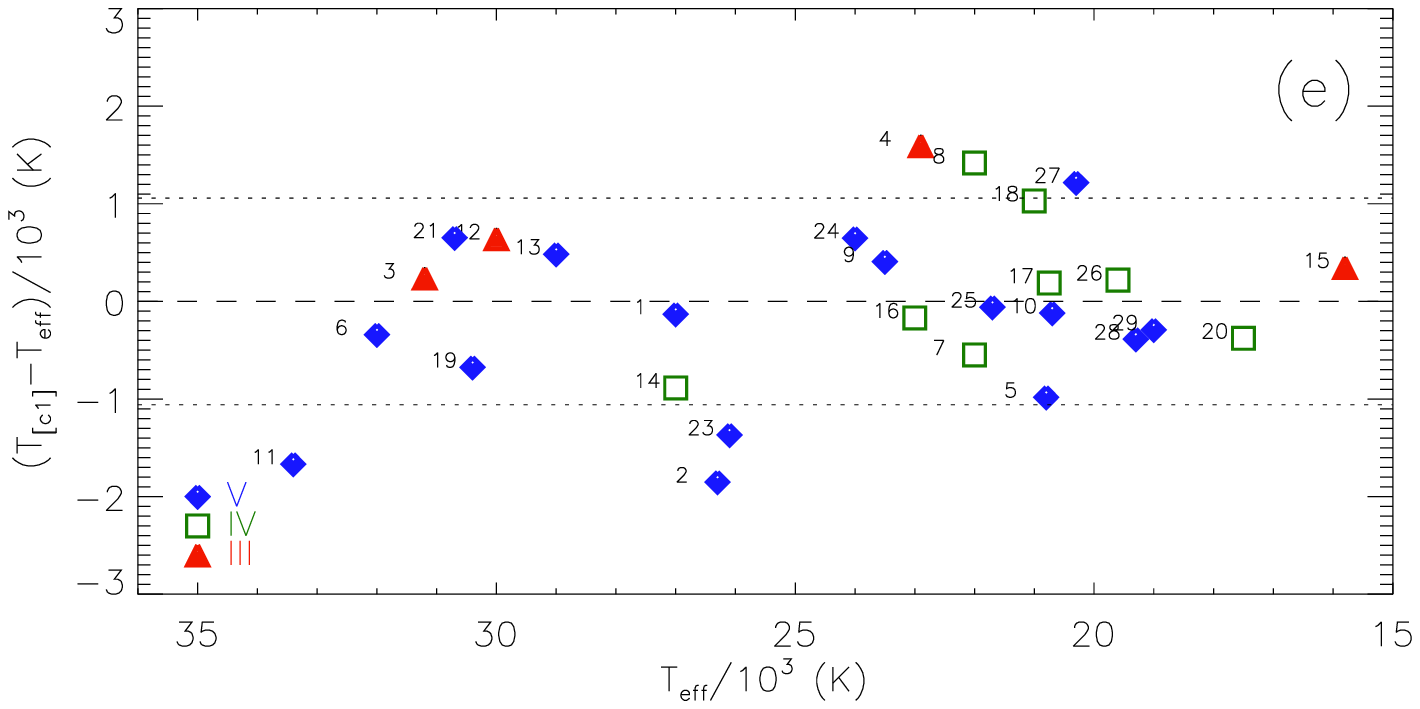}}
\hspace{1cm}
\resizebox{0.47\hsize}{!}{\includegraphics[width=12cm,height=5.5cm]{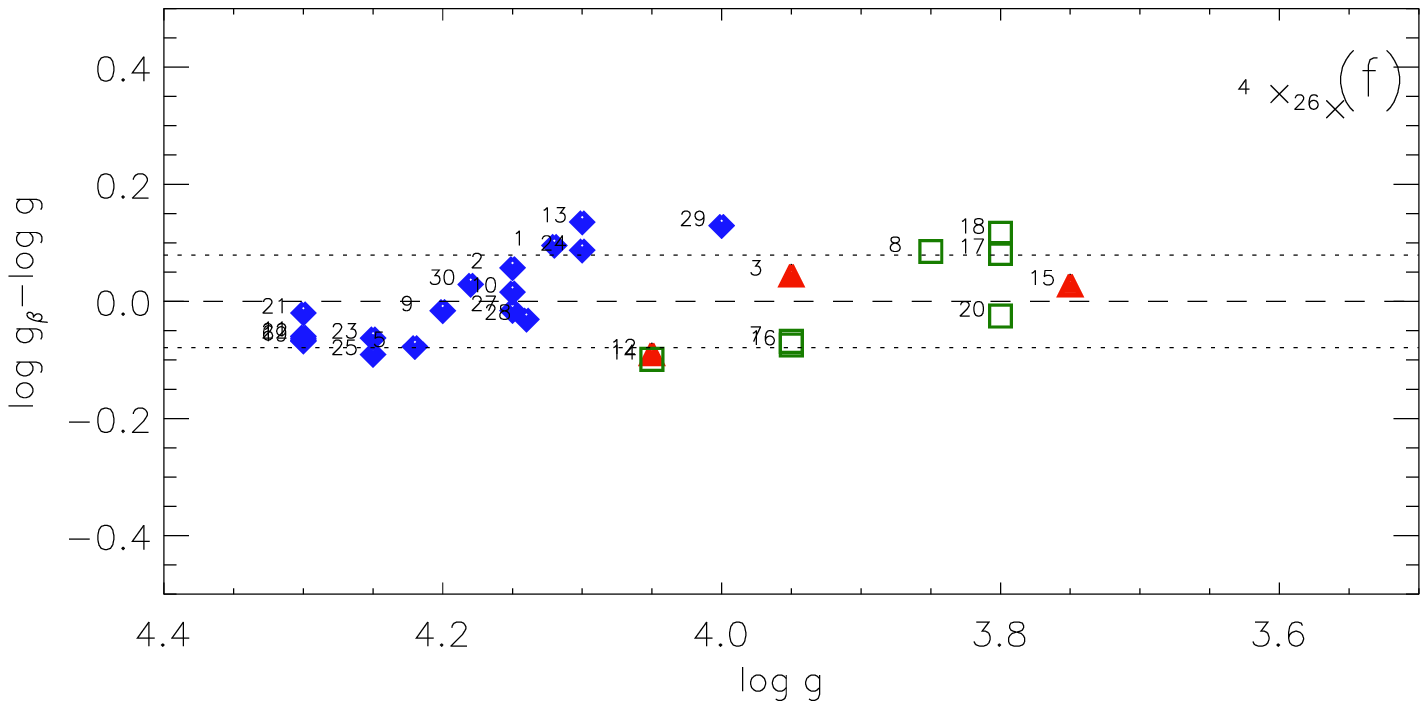}}
\caption[]{Residuals of effective temperatures (left panels) and surface gravities (right panels) determined with our empirical fit formulae to $Q$, $[u-b]$, $[c1]$ and $\beta$ and the spectroscopically derived reference values (Table~\ref{photo}). The dashed lines denote perfect agreement and the dotted lines correspond to the following 1$\sigma$-values: 700\,K (a), 943\,K (c), 1060\,K (e), 0.08\,dex (b), 0.19\,dex (d), and 0.08 (f). See the text for details.}
\label{f2}
\end{figure*}

\subsection{Spectroscopic parameters vs. fitted values}

\begin{table}[t!]
 \centering
\caption[]{
Internal and external uncertainties for temperature and gravity determinations using the calibrations from Sects.~\ref{calibrations}-\ref{calibrationsc1}.\\[-2mm] \label{errors}}
 \setlength{\tabcolsep}{.1cm}
 \begin{tabular}{llrc@{\hspace{3mm}}lcc@{\hspace{1mm}}rrrrrrrrrrrrrrr}
 \noalign{}
\hline
\hline
       &    $\Delta T$            &  Int. &    Ext.         & $\Delta \log g$&      Int. &     Ext.        \\
\hline\\[-0.5mm]
 V    &  $\Delta T_Q$:         &   738\,K      &  800\,K   & $\Delta \log g_Q$:        &  0.08 & 0.23 \\
III,IV&  $\Delta T_Q$:         &   385\,K      &  530\,K   & $\Delta \log g_Q$:        &  0.08 &0.17 \\[1.2mm]
\hline\\
IV,V  &  $\Delta T_{[u-b]}$:     &   1060\,K     &  300\,K   & $\Delta \log g_{[u-b]}$: &  0.19 &0.02  \\
III   &  $\Delta T_{[u-b]}$:     &   960\,K      &  280\,K   & $\Delta \log g_{[u-b]}$: &  0.19 &0.04 \\[1.2mm]
\hline\\
V     &  $\Delta T_{[c1]}$:      &   1060 \,K    &  290 \,K  & $\Delta \log g_{\beta}$:  &  0.07 &0.01  \\
III,IV&  $\Delta T_{[c1]}$:     &   1060 \,K    &  290 \,K  & $\Delta \log g_{\beta}$:  &  0.08 &0.01  \\
\hline\\[-0.5mm]
 \end{tabular}
\end{table}

Internal uncertainties of our calibrations are analysed here via residuals of the fitted values to the spectroscopic parameters, as displayed in Fig.~\ref{f2}. In the left panels $T_Q$ (a), $T_{[u-b]}$ (c) and $T_{[c1]}$ (e) are compared to $T_\mathrm{eff}$ and in the right panels $\log g_Q$ (b), $\log g_{[u-b]}$ (d) and $\log g_{\beta}$ (f) are compared to $\log g$.
Dotted lines indicate internal uncertainties like in Table~\ref{errors} for calibrations independent of the luminosity class ($\Delta \log g_Q$= 0.08\,dex, $\Delta \log g_{[u-b]}$= 0.19\,dex and  $\Delta T_{[c1]}$= 1060\,K). Otherwise, they indicate averaged 1$\sigma$ values for different luminosity classes weighted by star number ($\Delta T_Q$= 700\,K, $\Delta T_{[u-b]}$= 1040\,K and $\Delta \log g_{\beta}$= 0.08\,dex). Dashed lines indicate perfect agreement and symbols are encoded as in the legend. Stars \#4 and 26 were excluded from the fits in Fig.~\ref{f2}f as they are too evolved (i.e., beyond core H-burning) in comparison to the rest of the sample.

As indicated in Table~\ref{errors}, the internal uncertainties based on $Q$ are smaller than in the other cases, but the external uncertainties are larger. 
In further comparisons to other work, we take as reference values the internal uncertainties based on $Q$: for temperature $\Delta T_Q$= 700\,K (averaged value for different luminosity classes weighted by star number) and for gravity $\Delta \log g_Q$= 0.08\,dex, as in Fig.~\ref{f2}. This is because our special interest in the $Q$-based calibrations, as there are more stars with measured UVB indices than with other filters.

\section{Our spectroscopic parameters vs. other works}\label{sect_comp}

\subsection{Photometric $T_\mathrm{eff}$ calibrations}\label{sect_photo}

\begin{figure*}[t!]
\resizebox{0.47\hsize}{!}{\includegraphics[width=12cm]{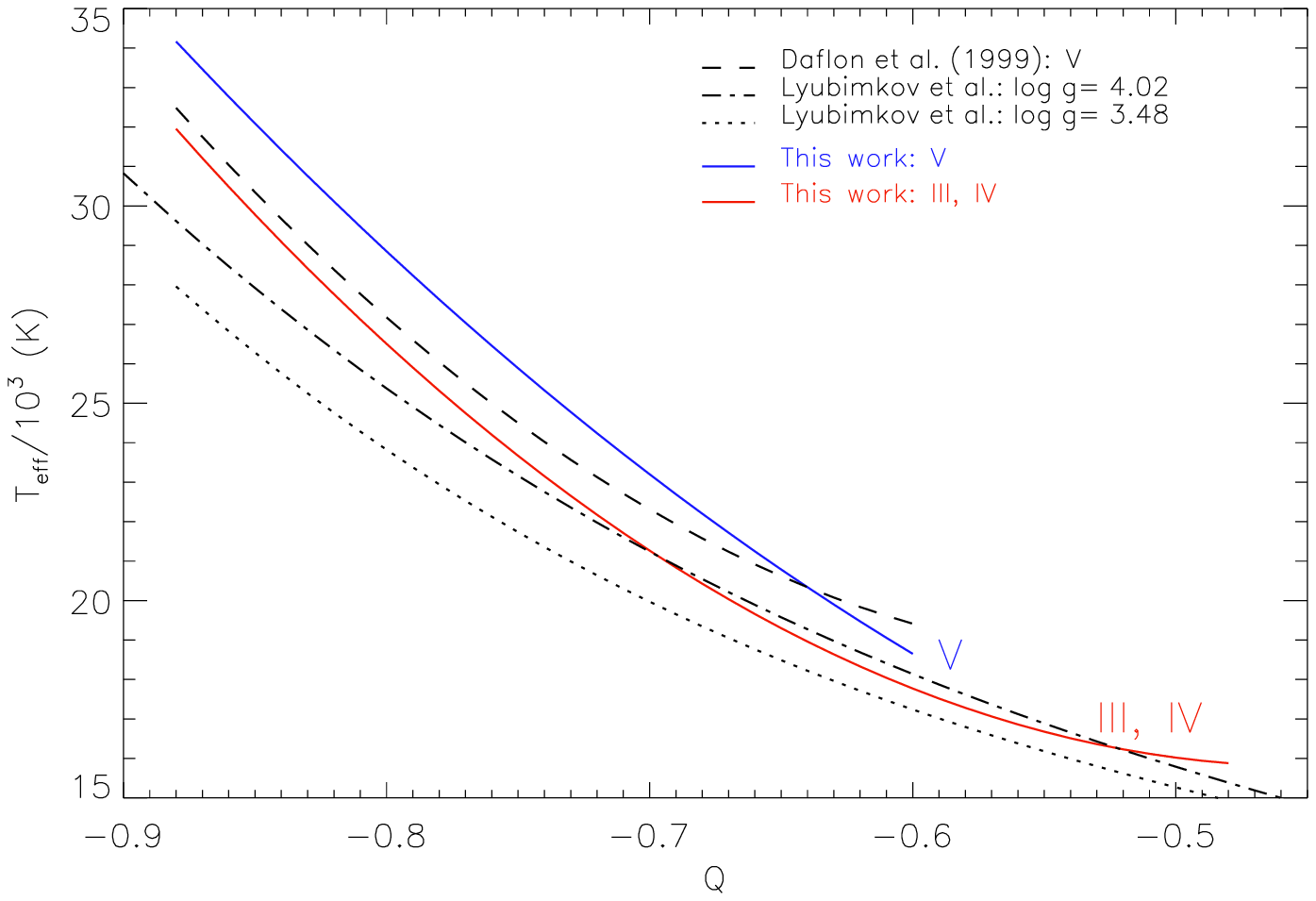}}
\resizebox{0.47\hsize}{!}{\includegraphics[width=12cm]{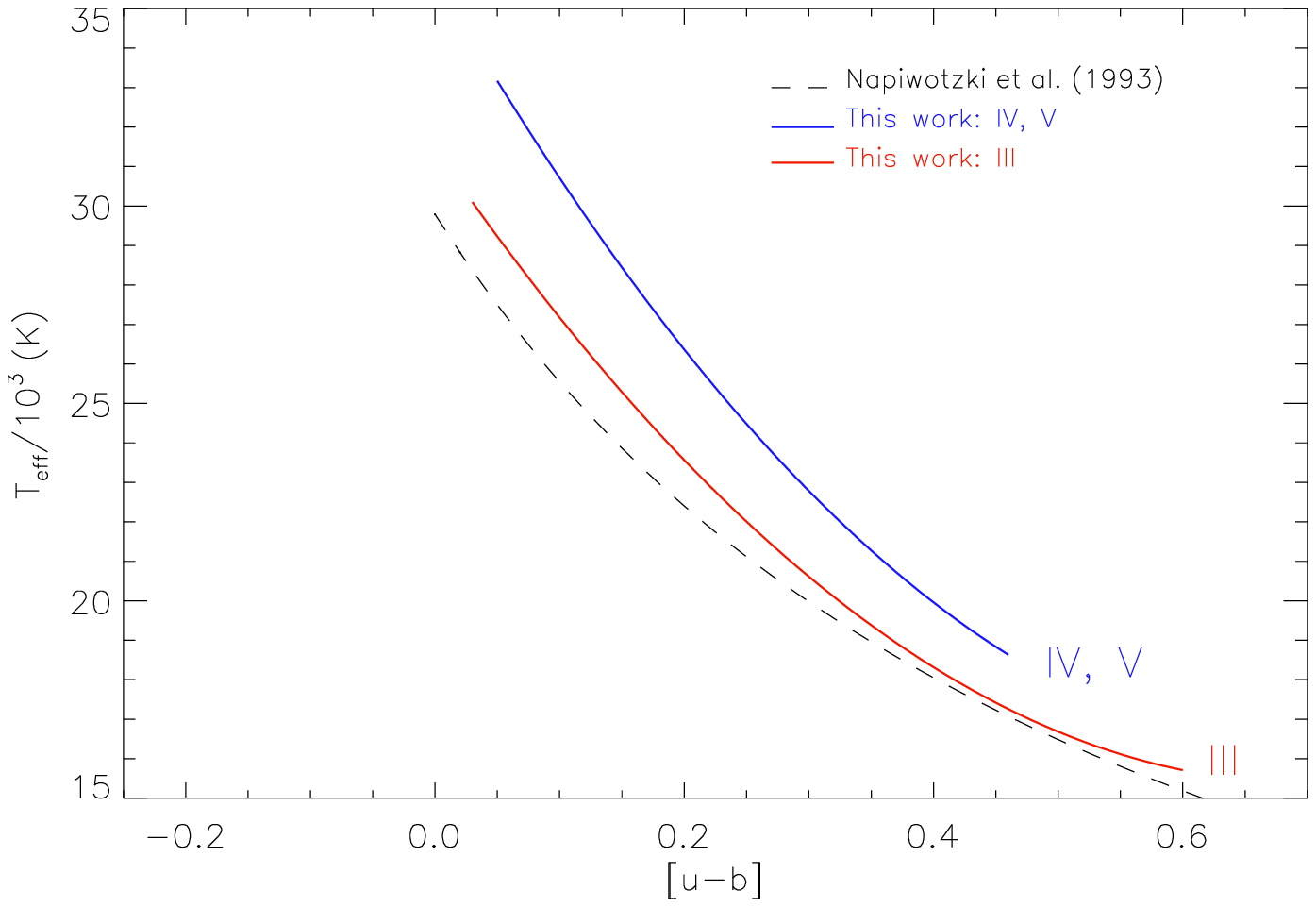}}
\caption[]{Left panel: Comparison of our $Q$-based temperature relation with those proposed 
by \citet{d99} and \citet{l02}. Right panel: comparison of our $[u-b]$-based temperature relation with 
that from \citet{napi93}. See the text for a description.}
\label{f3}
\end{figure*}

Our $T_Q$-scales derived from multiple ionization equilibria are compared in Fig.~\ref{f3}a to photometric calibrations of \citet{d99} for luminosity class V and of \citet{l02} for $\log g=$3.48 and 4.02. In addition, our $T_{[u-b]}$-scales are compared to that of \citet{napi93} in Fig.~\ref{f3}b. Our scales tend to be hotter at higher temperatures than the photometric scales.

Our $T_Q$ for luminosity class V (upper solid curve, blue in the online version) is $\sim$2000\,K hotter than that proposed by \citet{d99} at the highest temperature, the difference decreasing until both scales cross each other at $\sim$21\,000\,K, and remain similar at slightly lower $T_\mathrm{eff}$.
The comparison to \citet{l02} has to be done more cautiously because our sample has a variety of surface gravity values. Only four stars have similar gravities to 4.00: \#3 (III), 7 (IV), 16 (IV) and 29 (V). We compare their upper curve (dashed dotted line) to our lower curve (solid line, red in the online version).   
Our relation is $\sim$2000\,K hotter than theirs at the highest temperatures and both practically coincide below 21\,000\,K. 

Figure~\ref{f3}b shows the comparison of our $T_{[u-b]}$ to the calibration proposed by  \citet{napi93} in their Eqn.~9. 
Their and our curve for luminosity class III practically coincide below 19\,000\,K, and the smallest discrepancy to our IV,V relation is $\sim$1500\,K at $T_\mathrm{eff}$ $\sim$18\,000\,K. 
In contrast, our scales are steeper than their curve, resulting in 
larger differences at higher $T_\mathrm{eff}$, with maximum values of $\sim$1500\,K for luminosity class III and $\sim$6000\,K for IV,V.  Possible reasons for these discrepancies are discussed at the end of next section.

\subsection{Other determinations for individual stars}

\begin{figure*}[t!]
\centering
\resizebox{.33\hsize}{!}{\includegraphics[width=12cm]{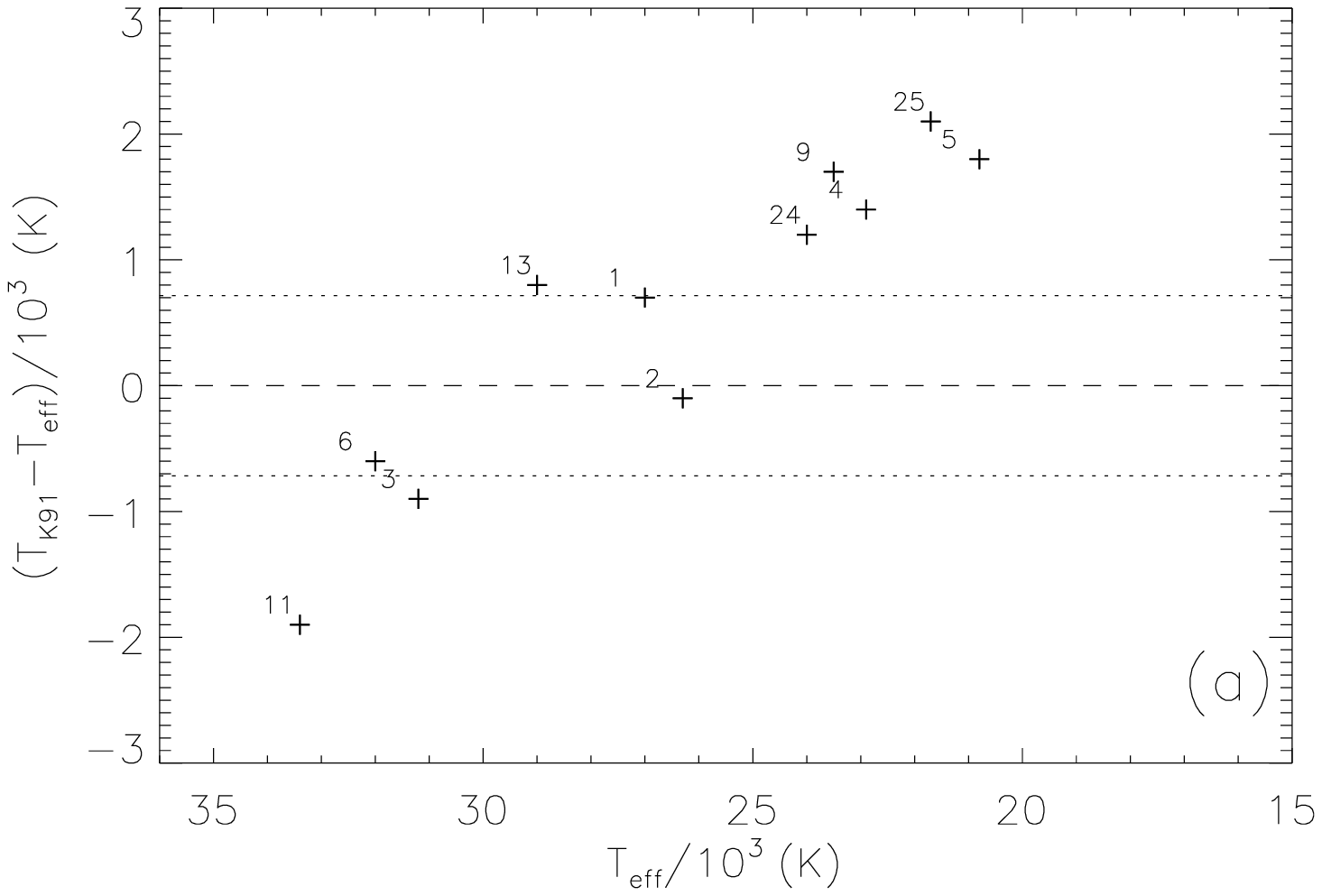}}
\resizebox{.33\hsize}{!}{\includegraphics[width=12cm]{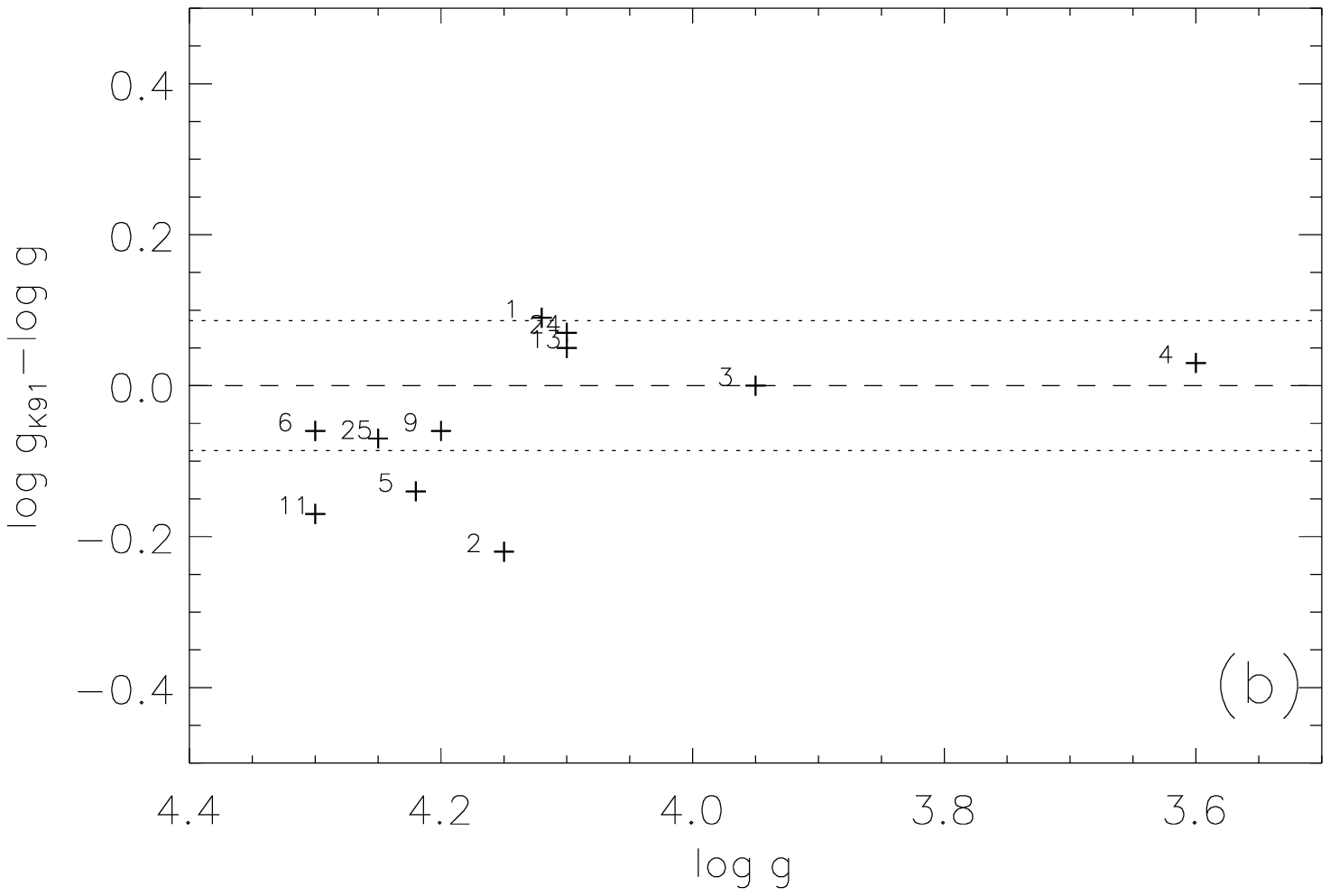}}
\resizebox{.33\hsize}{!}{\includegraphics[width=12cm]{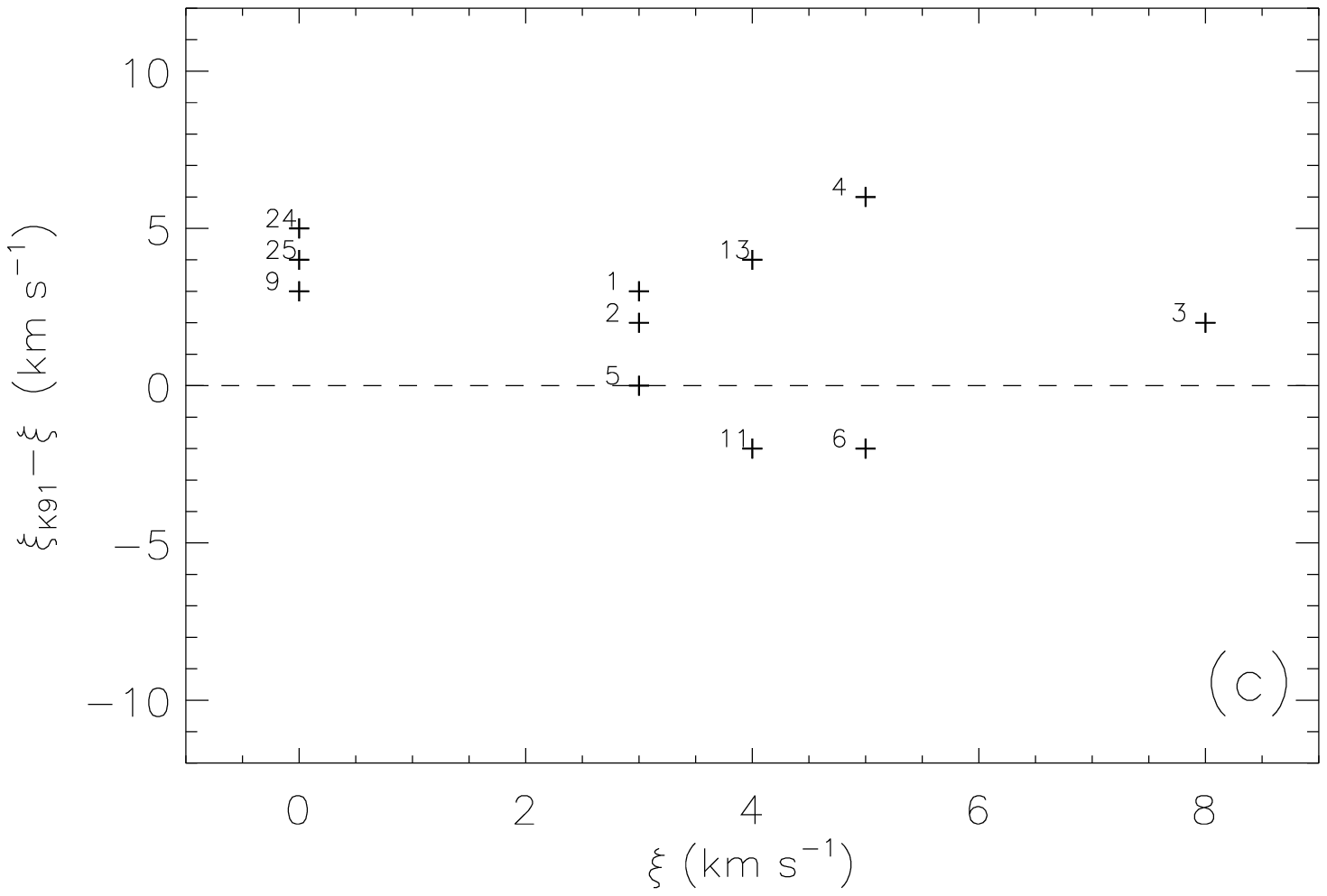}}
\caption[]{Differences between effective temperatures (a), surface gravities (b) and microturbulences (c) determined by \citet{k91} and our spectroscopic parameters (Table~\ref{photo}). 
For comparison, dotted lines indicate to the averaged internal uncertainties of our $T_Q$ and $\log g_Q$-calibrations weighted by star number.}
\label{f4}
\end{figure*}

\begin{figure*}[t!]
\centering
\resizebox{.33\hsize}{!}{\includegraphics[width=12cm]{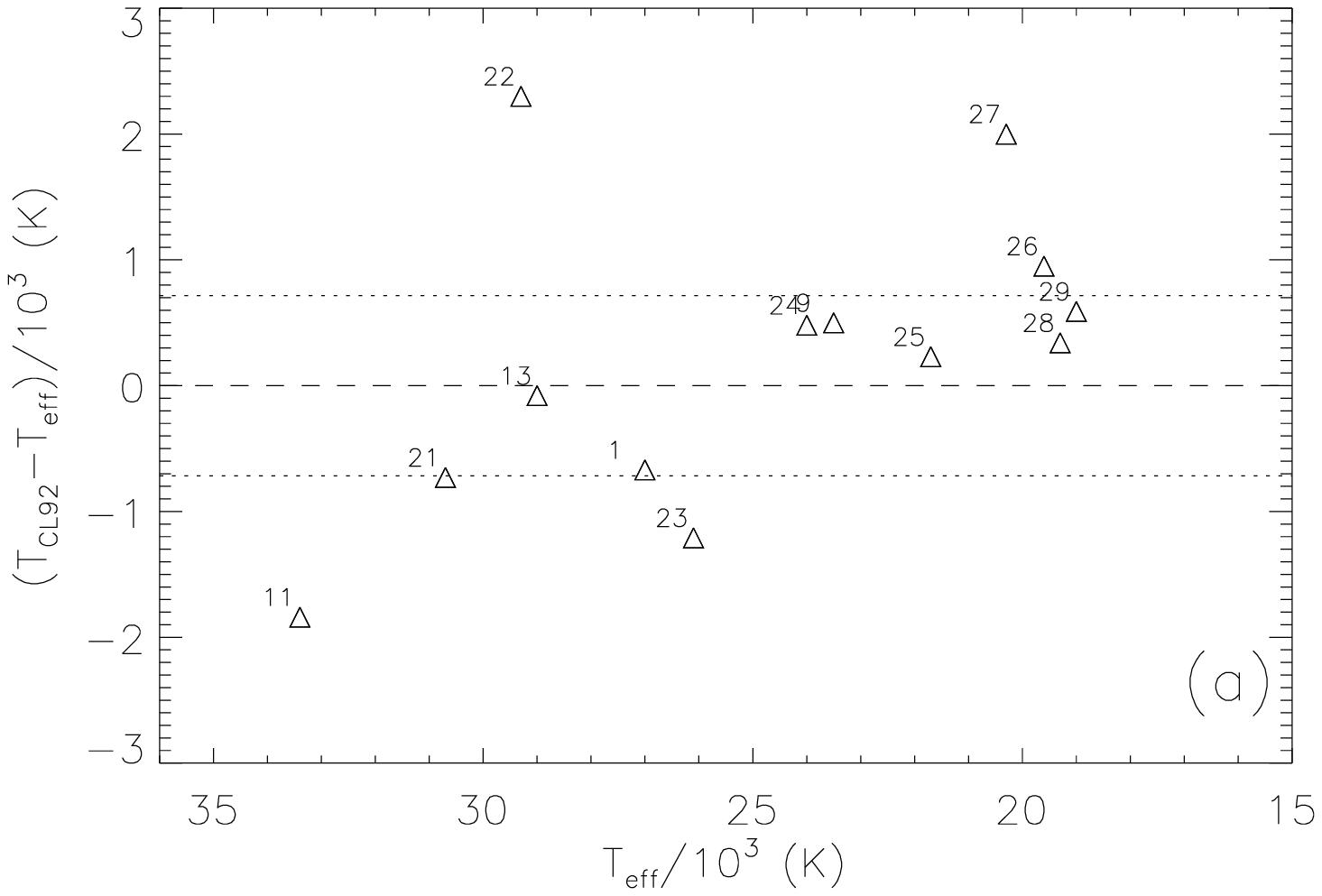}}
\resizebox{.33\hsize}{!}{\includegraphics[width=12cm]{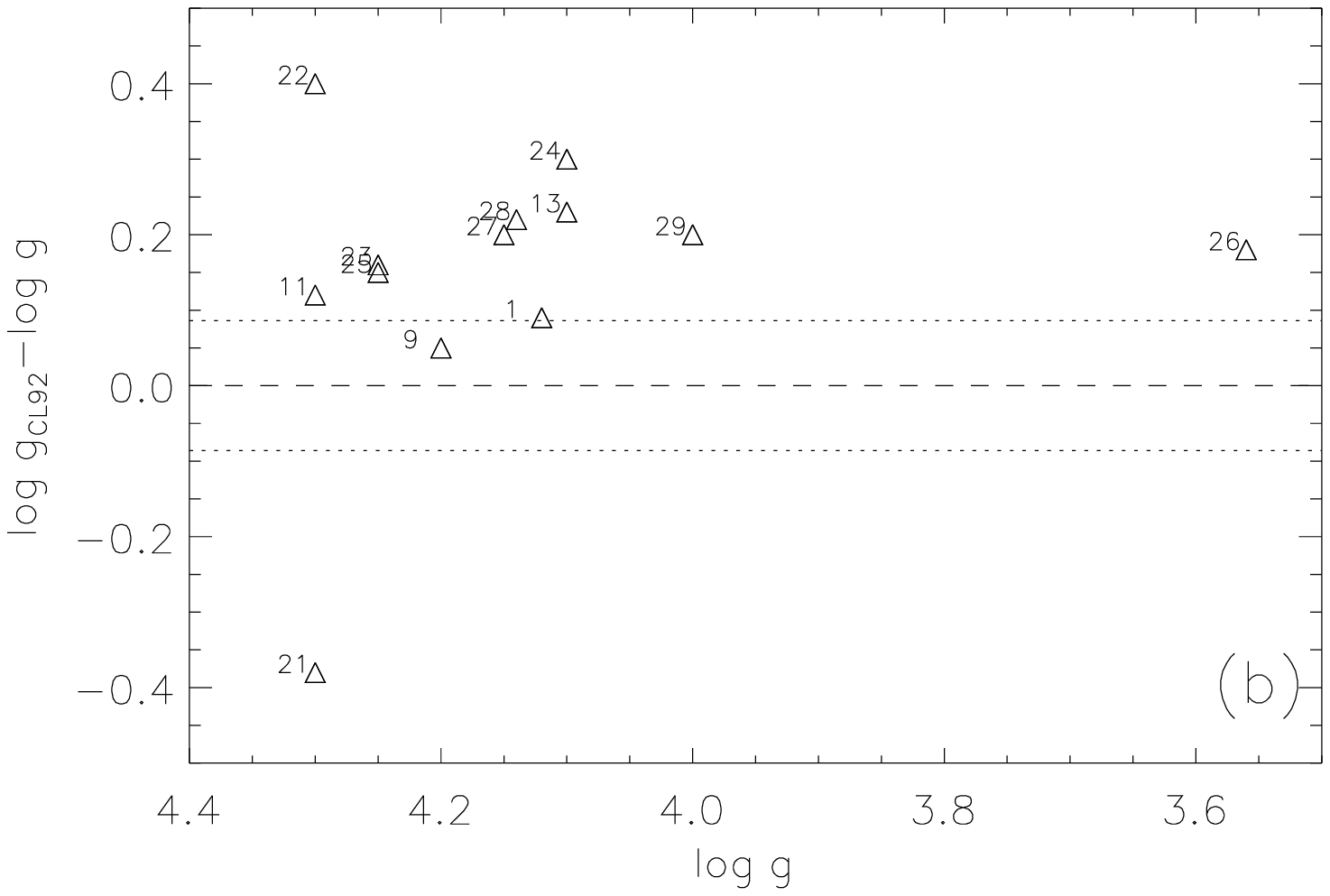}}
\resizebox{.33\hsize}{!}{\includegraphics[width=12cm]{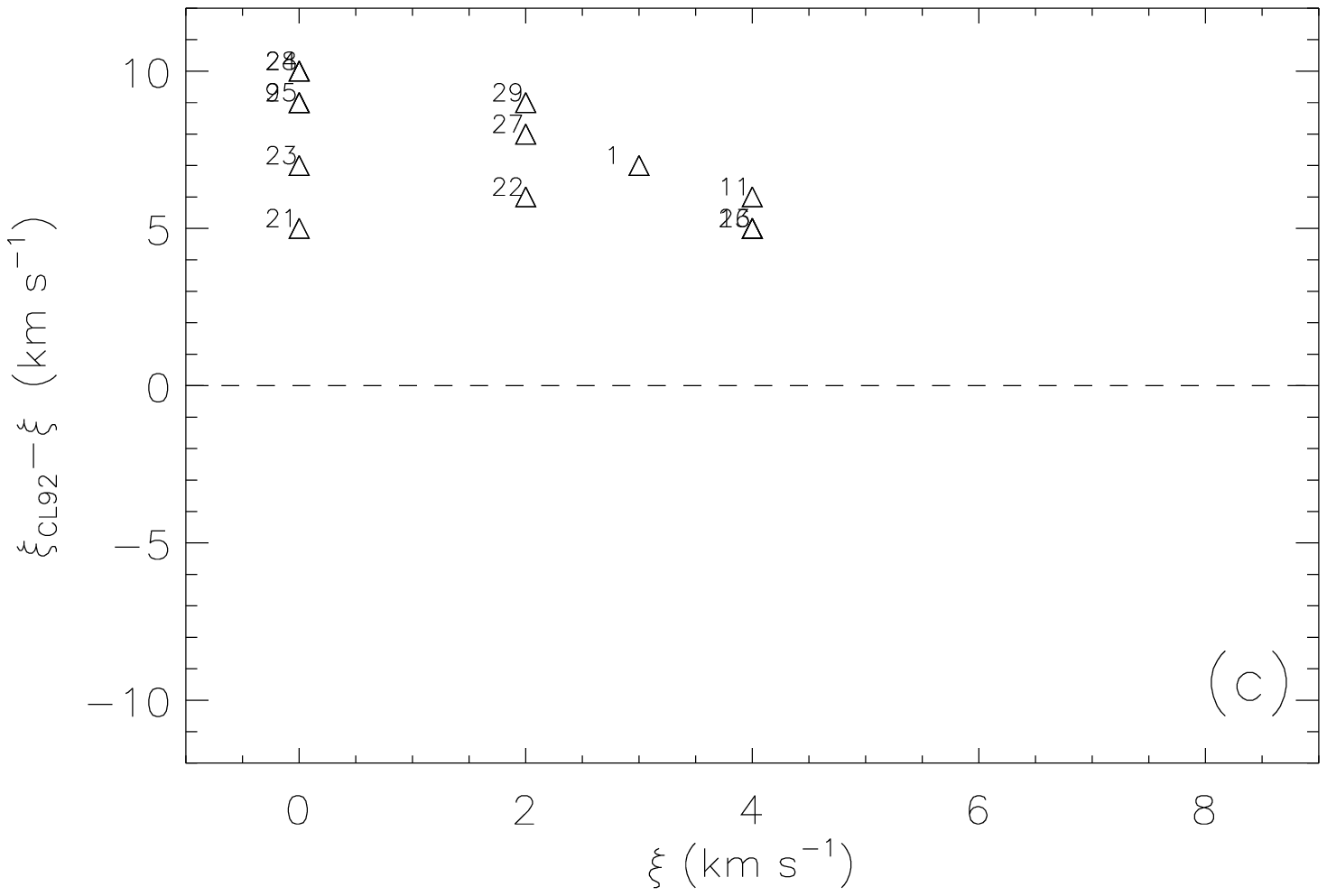}}
\caption[]{Like Fig.~\ref{f4}, but for parameters derived by \citet{cl92} and applied by 
\citet{cl94}, \citet{c06}, \citet{l08} and \citet{d09}. In the latter three studies, stars \#11, 21, 22, 27 and 28 were excluded from the analyses.}
\label{f5}
\end{figure*}

\begin{figure*}[t!]
\centering
\resizebox{.33\hsize}{!}{\includegraphics[width=12cm]{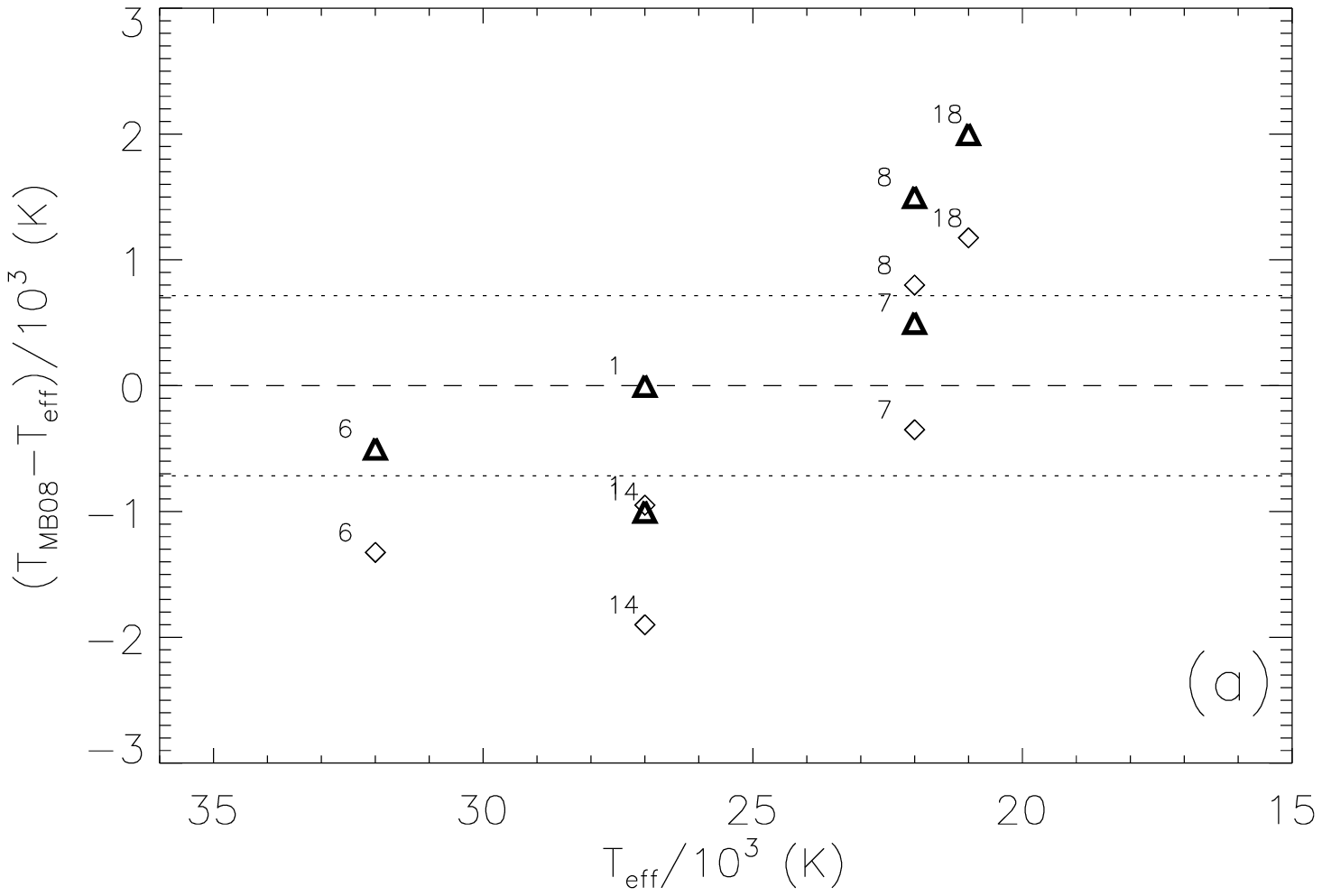}}
\resizebox{.33\hsize}{!}{\includegraphics[width=12cm]{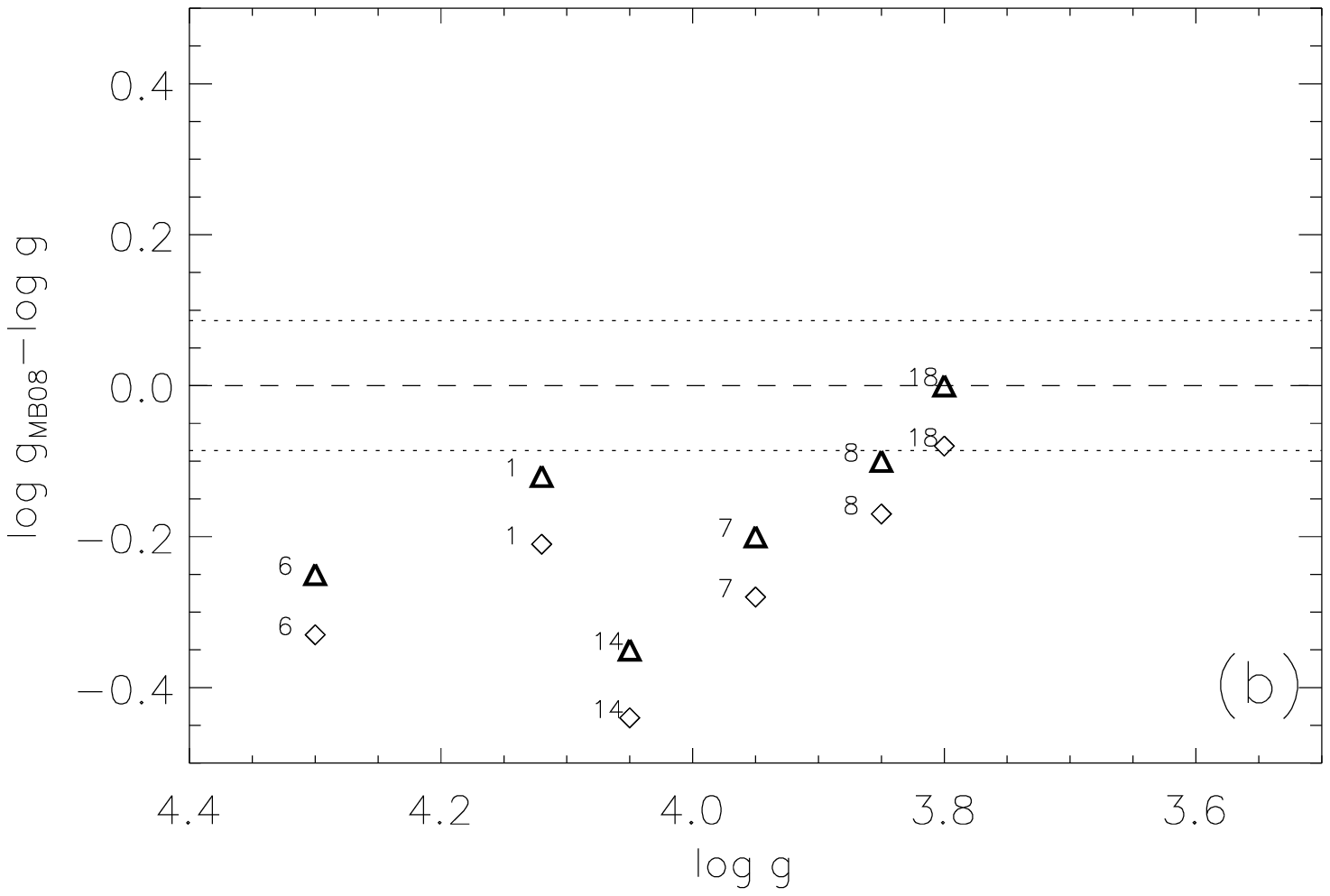}}
\resizebox{.33\hsize}{!}{\includegraphics[width=12cm]{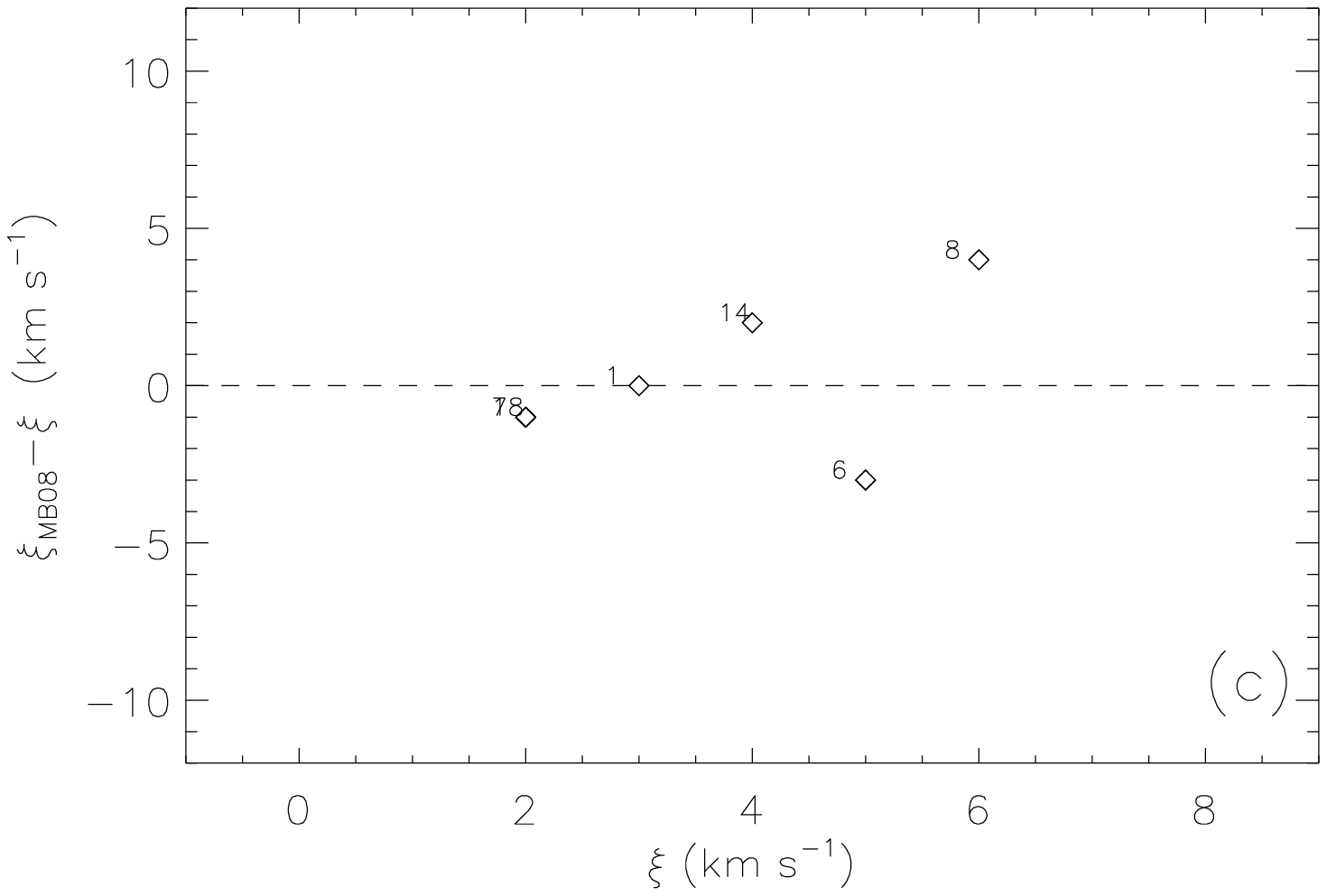}}
\caption[]{Like Fig.~\ref{f4}, but for parameters derived by \citet{mb08} via ionization equilibria of \ion{Ne}{i/ii} (diamonds) and \ion{Si}{ii/iii}-\ion{Si}{iii/iv} (triangles).}
\label{f6}
\end{figure*}

\begin{figure*}[t!]
\centering
\resizebox{.33\hsize}{!}{\includegraphics[width=12cm]{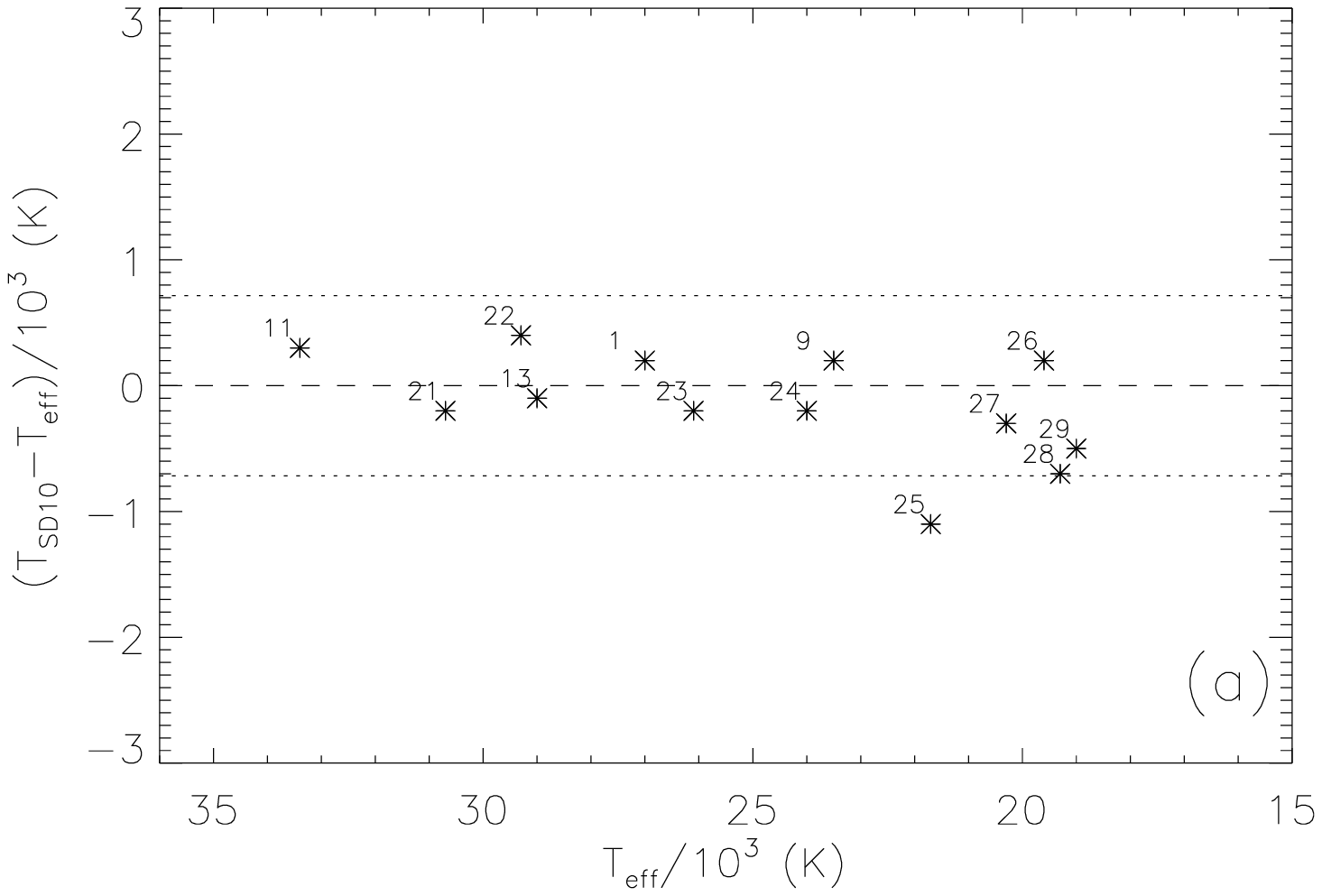}}
\resizebox{.33\hsize}{!}{\includegraphics[width=12cm]{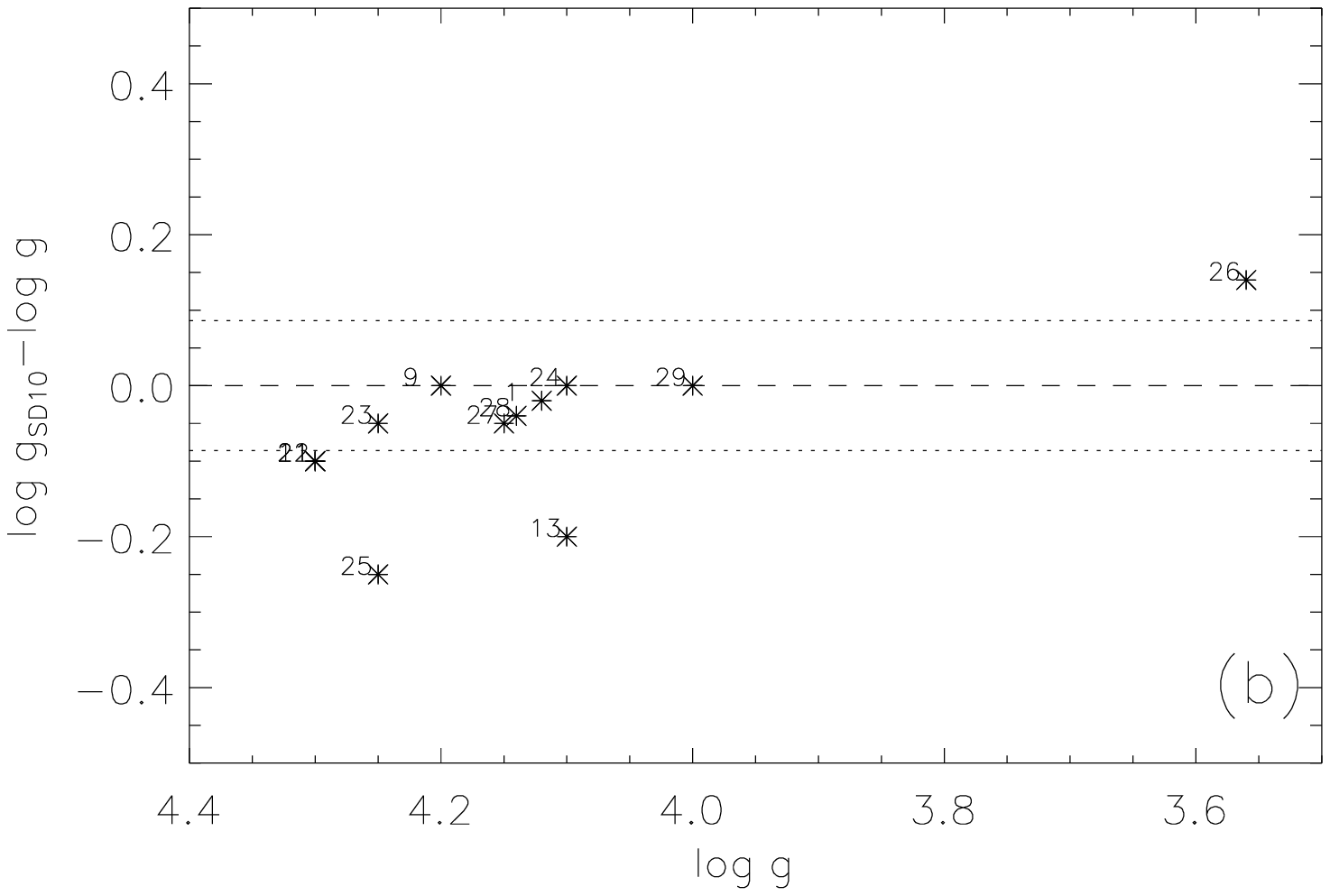}}
\resizebox{.33\hsize}{!}{\includegraphics[width=12cm]{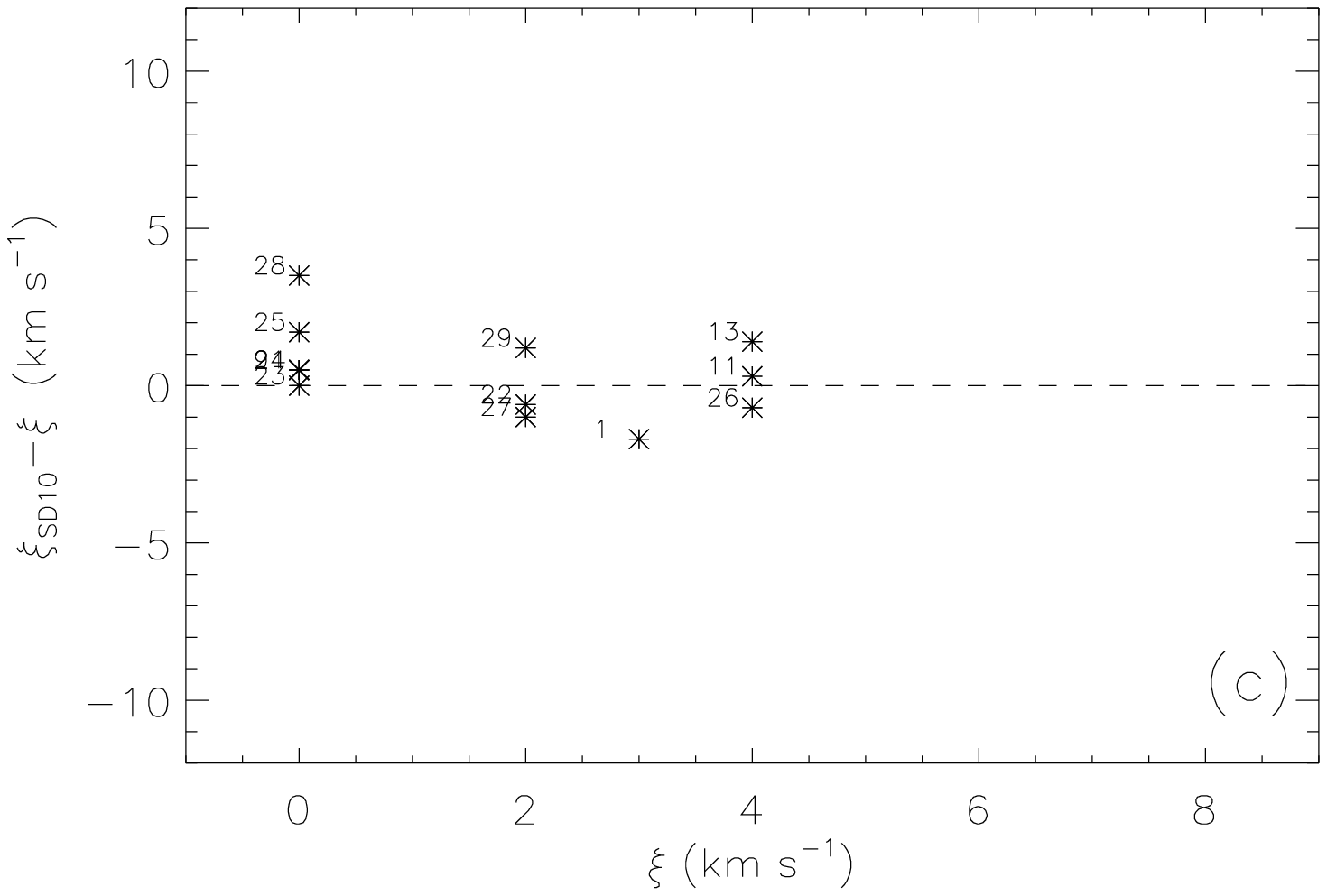}}
\caption[]{Like Fig.~\ref{f4}, but for parameters derived by \citet{s10}}
\label{f7}
\end{figure*}

\begin{figure*}[t!]
\centering
\resizebox{.33\hsize}{!}{\includegraphics[width=12cm]{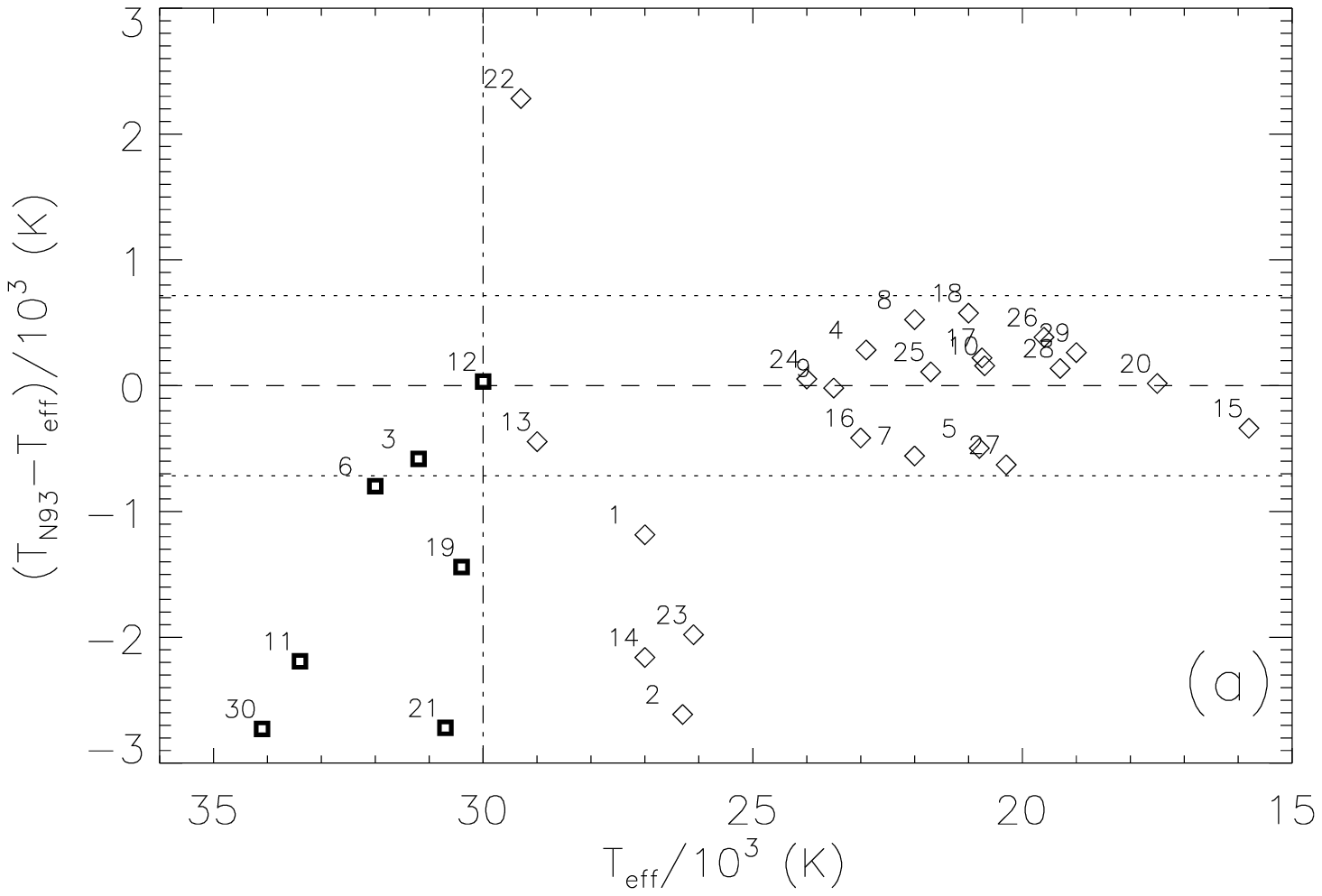}}
\resizebox{.33\hsize}{!}{\includegraphics[width=12cm]{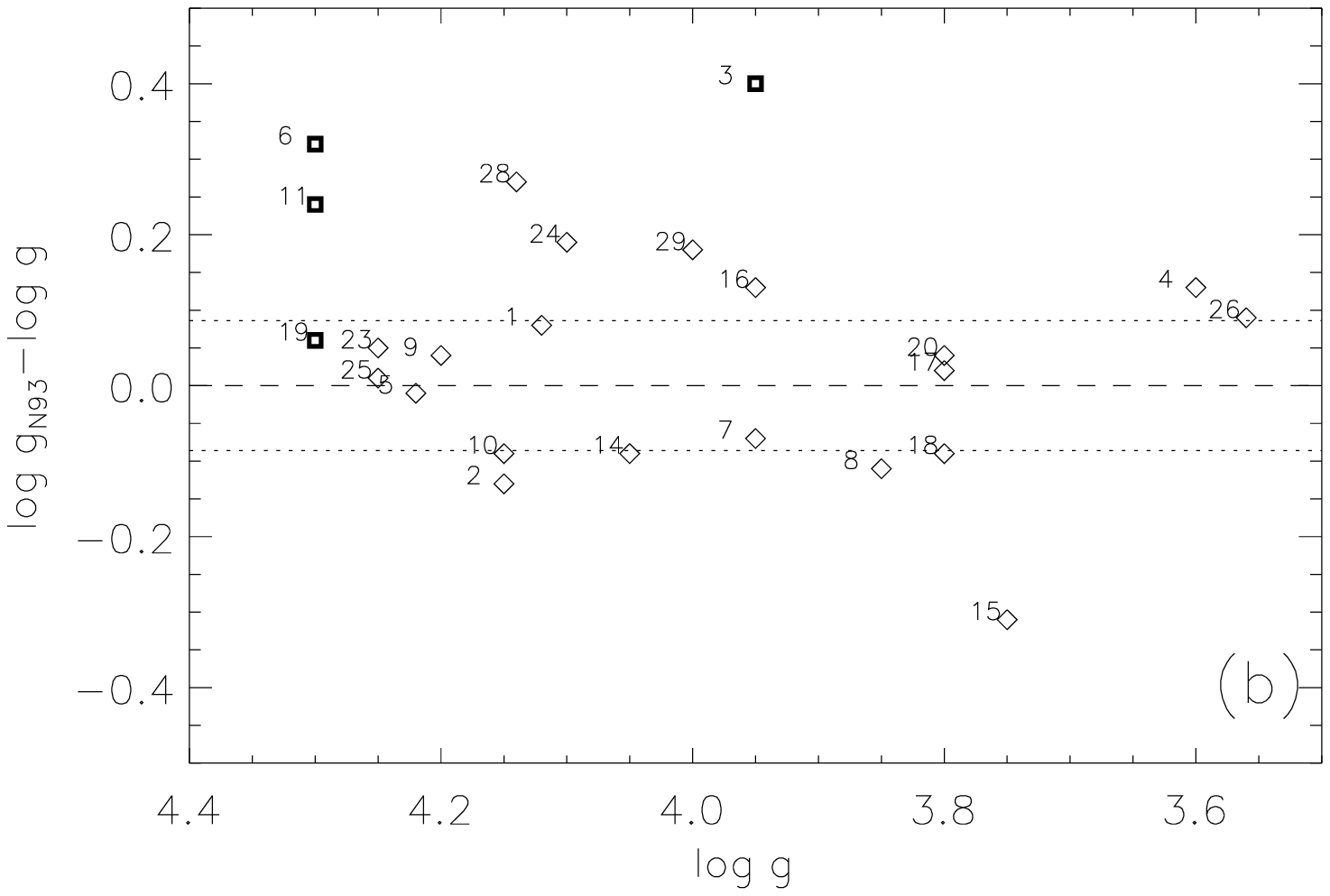}}
\resizebox{.33\hsize}{!}{\includegraphics[width=12cm]{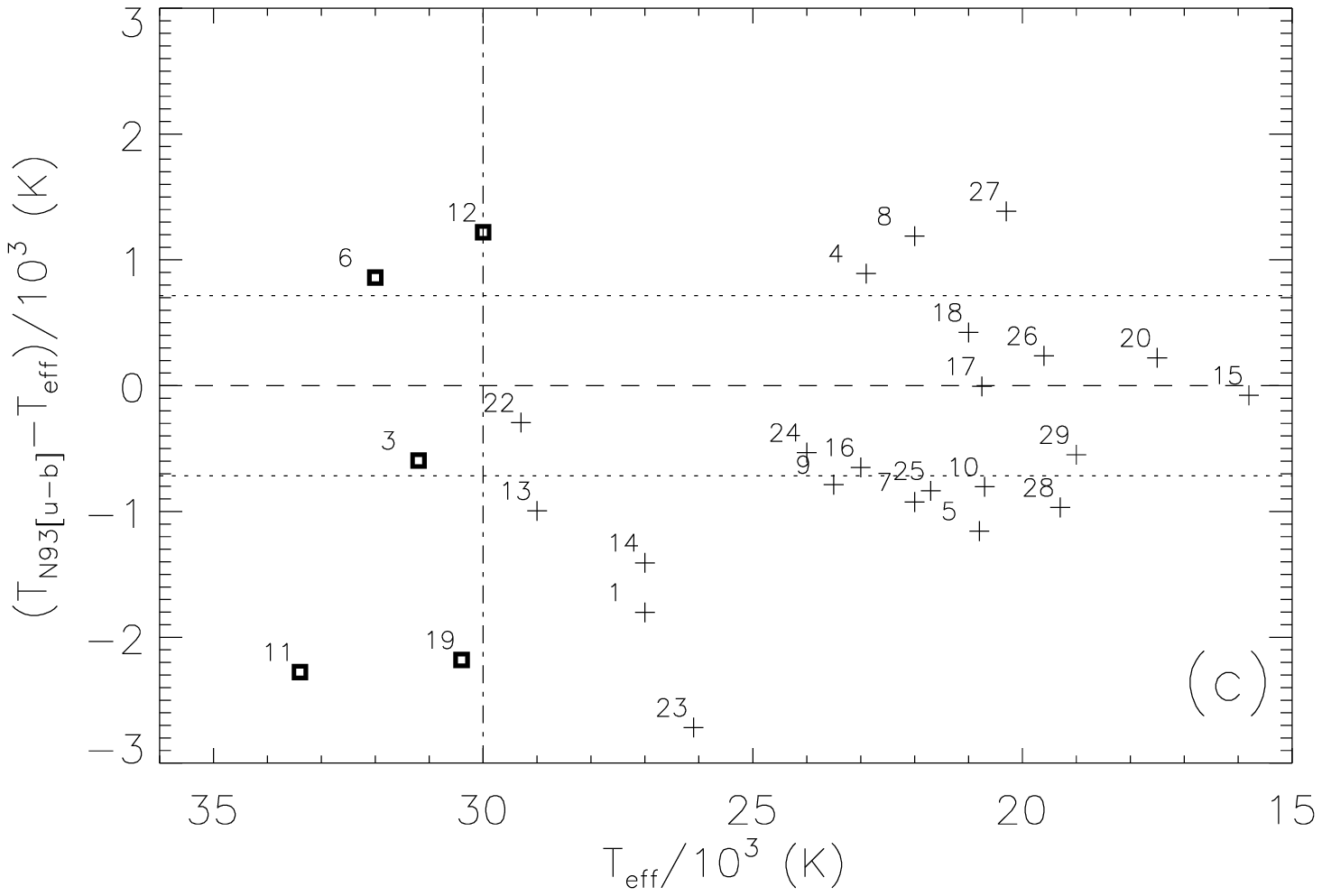}}
\caption[]{Panels (a) and (b) like Fig.~\ref{f4}, but for parameters derived with the calibrations of \citet{md84}, later improved by \citet{napi93}.
Panel (c) shows the difference between the $[u-b]$-based temperature scale proposed by \citet{napi93} 
and our spectroscopic values. Diamonds indicate stars with $T_\mathrm{eff} \le 30\,000$\,K and squares with $T_\mathrm{eff} \ge 30\,000$\,K.}
\label{f8}
\end{figure*}

Besides the photometric temperature scales discussed in Sect.~\ref{sect_photo}, other determinations of temperature and surface gravity are found in the literature for several stars of the sample. 
Whenever possible, we also compare values of microturbulence as all parameters are interrelated.
Figures~\ref{f4}-\ref{f8} illustrate the differences between the atmospheric parameters derived in other works and our spectroscopic $T_\mathrm{eff}$ (a), $\log g$ (b) and $\xi$ (c).
Dashed lines indicate full agreement between other determinations and our values, and dotted lines in 
Figs.~\ref{f4}-\ref{f8}a,b denote $\Delta T_Q$ (700\,K) and $\Delta \log g_Q$ (0.08\,dex), like in Fig.~\ref{f2}, as a reference to our photometric scale. 
 Note that uncertainties in our spectroscopic parameters are even smaller than 
the $1\sigma$-errors for the $Q$-calibrations:  $\sim$300\,K for $T_\mathrm{eff}$ and $\sim$0.05\,dex for $\log g$.

It is beyond the scope of this paper a detailed description of other analysis methods and the causes of discrepancies between their and our results. Based on our previous studies \citep{np07,np08,np10a,np10b}, we propose possible reasons that might cause systematic offsets. However, we should bear in mind that in most cases the discrepancies are likely due to a combination of various effects, which are difficult to be quantified beforehand.

\paragraph{ \citet{k91}.} This work presents a spectroscopic determination of gravity and effective temperature of stars in the solar neighborhood from NLTE hydrogen line profiles (H$\beta$, H$\gamma$, H$\delta$) and the silicon ionization equilibria \ion{Si}{ii/iii} and/or \ion{Si}{iii/iv} (depending on the stellar temperature).
Their procedure is similar to ours, based on an earlier version of the codes {\sc Detail} and {\sc Surface}, however with fewer spectroscopic indicators than those shown in Table~\ref{indicators}. Their silicon abundances are based on the model atom 
by \citet{bb90}\footnote{We do not analyse \ion{Si}{ii} lines because their synthetic spectrum provides erroneous line intensities, as reported in \citet{s10} and \citet{ns11}.}, like in our case, but on model atmospheres by \citet{g84} with much less line blanketing than ATLAS9. Their microturbulences are derived in the standard way from \ion{O}{ii} lines.
Figure~\ref{f4} displays comparisons between their and our stellar parameters for 11 stars in common, identified like in Table~\ref{photo}.
A clear trend is noticed in Fig.~\ref{f4}a for effective temperatures, being their values 
$\sim$2000\,K higher at $T_\mathrm{eff} \sim$20\,000\,K and 
$\sim$2000\,K lower at $T_\mathrm{eff} \sim$34\,000\,K. 
Less blanketed model atmospheres affect the line-blocking and hence the temperature structure of the stellar atmosphere.This could be a reason for discrepancies with our temperatures. Furthermore, for cooler stars, the use of \ion{Si}{ii} in the ionization equilibrium can also cause an offset in the derived temperature.
In contrast, their gravities agree with our spectroscopic values and lie within the 1$\sigma$ uncertainty of $\log g_Q$ for most stars (Fig.~\ref{f4}b). Their microturbulences tend to be larger than ours (Fig.~\ref{f4}c), reaching a maximum difference of $\sim$6\,km\,s$^{-1}$.

\paragraph{ \citet{cl92}.} The parameters determined in this work
were used in a series of subsequent studies of stars in the Orion region by \citet{cl94}, \citet{c06}, \citet{l08}, and \citet{d09}. Their approach for deriving effective temperatures and surface gravities is based
on an iterative scheme using older calibrations of Str\"omgren photometry modified by \citet{gl92} coupled with fits to the wings of H$\gamma$ profiles computed in LTE \citep{k79}. Microturbulences were determined from \ion{O}{ii} lines in NLTE.
Figure~\ref{f5} displays the differences between their and our parameters for 13 stars in common. In spite this sample differs from that in the comparison with \citet{k91}, we notice in Fig.~\ref{f5}a an overall similar trend in temperature than in Fig~\ref{f4}a, except for star \#22. But in this case, there are more stars with temperatures within our 1$\sigma$ uncertainty margins for $T_Q$.
It is difficult to identify the sources of discrepancies in temperature. They could be related to the details of the photometric calibrations, i.e. the use of less blanketed model atmospheres to compute the synthetic photometric indices and to the fits to H$\gamma$ in LTE in the iterative procedure.
Concerning the gravities (Fig.~\ref{f5}b), their values are higher (up to 0.4~dex) 
than our spectroscopic gravities for most stars, lying above the 1$\sigma$ uncertainty of our $\log g_Q$ calibration. 
An exception is the star \#21, that shows a difference of $-$0.4\,dex to our value. 
The discrepancies in surface gravities could be related to the use of LTE synthetic profiles of H$\gamma$, that have been shown to give larger gravities than those computed 
in NLTE \citep{np07}. Since their approach is iterative, this also would affect the determination of temperature. However, differences larger than $\pm$0.2\,dex cannot be explained with 
NLTE effects on the Balmer lines wings. The continuum normalization of H$\gamma$
could also play a role in this respect, since this is an important source of uncertainty in surface gravity determinations \citep{np10a,np10b}.
Their microturbulences are larger -- by 5 to 10\,km\,s$^{-1}$ -- than those derived by us.

In contrast, in \citet{c06}, \citet{l08} and \citet{d09}\footnote {Parameters for 2 stars in common have been revised in \citet{d09} via \ion{S}{ii/iii} ionization equilibrium. For star \#1, offsets of $+$530\,K and $+$0.05\,dex and for \#24, $-$480\,K and $-$0.05\,dex in temperature and surface gravity, respectively, have been applied.} the stars \# 11, 21, 22, 27 and 28 were excluded from the studies, which lead to maximum differences to our work in $T_\mathrm{eff}$ of up to 1000\,K and in $\log g$ of $+$0.05-0.3\,dex. I.e.,
the selected sub-sample of stars analyzed in those works presents a better agreement with our 
results than the original complete sample from \citet{cl92}, however a systematic offset 
in gravities and microturbulences still persists.

\paragraph{ \citet{mb08}.} This work presents an interesting comparison between atmospheric parameters derived in a standard iterative spectroscopic way, i.e. via NLTE \ion{Si}{ii/iii} and/or \ion{Si}{iii/iv} ionization equilibria \citep{m06} and an new \ion{Ne}{i/ii} ionization equilibrium. In addition, the fits of NLTE profiles to H$\epsilon$, H$\delta$, H$\gamma$ and H$\beta$ are used for surface gravity determination.
The computations are performed with similar versions of {\sc Detail} and {\sc Surface} and the same model atmospheres than those used by us.
For Si, a model atom by D. J. Lennon is employed \citep{tl05} while for Ne a new model by K. Butler is applied for the first time. Temperatures derived with Ne are cooler by 825\,K than those derived with Si. In both cases, microturbulences were determined from \ion{O}{ii} lines.
It is worth noticing that they have analyzed 3 stars that were excluded from our 
first sample (Paper~I) because they turned out to be double-lined spectroscopic binaries, namely HD\,35468 (HR\,1790, Bellatrix), HD\,163472 (HR\,6684, V2052\,Oph) and HD\,214993 (HR\,8640, 12\,Lac). Our samples share 6 stars, which parameters are compared in Fig.~\ref{f6}.
Overall, we notice a trend in temperature (Fig.~\ref{f6}a) in a similar way then the two previous cases, the maximum differences being $\sim$1000 at $\sim$21\,000\,K and $\sim$2000\,K at $\sim$27\,000\,K.

It is difficult to evaluate the causes of discrepancies in the results. 
It is not clear whether the new \ion{Si}{ii} model atom provides improved line intensities than the model by \citet{bb90}. If the problem persists, it would affect temperatures lower than $\sim$23\,000\,K (the exact limit depends on the S/N of the spectrum). 
In addition, the Ne model applied in \citet{mb08} showed difficulties to achieve \ion{Ne}{i/ii} ionization equilibrium and to 
match the temperatures derived with \ion{Si}{ii/iii/iv} ionization equilibria. This Ne model atom was subsequently improved 
with more accurate $\log gf$ values in \citet{pnb08} resulting in consistent results between \ion{Ne}{i/ii} and other ionization equilibria: \ion{He}{i/ii}, \ion{C}{ii/iii/iv}, \ion{O}{i/ii}, \ion{Si}{iii/iv}, removing discrepancies found previously; see also Table~\ref{indicators} of the present paper.
Furthermore, their surface gravities (Fig.~\ref{f6}b) are $\sim$0.1-0.45\,dex lower than those found by us. These offsets are unexpected, as their analysis is based on the same codes and a similar H model than those employed in our work. An offset in the temperature determination is not sufficient to explain such large differences in $\log g$. 
The remaining sources of systematics to be investigated in those results is the continuum normalization of the Balmer lines \citep{np10a,np10b}, that might cause differences larger than 0.2\,dex in the derived surface gravities. 
Finally, there is a hint of a trend in microturbulences, except for star \#6 (Fig.~\ref{f6}c).

\paragraph{ \citet{s10}.} Stellar parameters were determined in this work spectroscopically by fitting 
the Balmer lines and matching \ion{He}{i/ii} and/or \ion{Si}{iii/iv} ionization equilibria.
The spectrum synthesis was computed with the unified atmosphere code {\sc Fastwind} \citep{puls05}, that models the stellar atmosphere and wind simultaneously in NLTE. The same observed spectra than  those analyzed in Paper~II have been analyzed in \cite{s10}. Despite the different atmospheric structure, input atomic data and line-formation computations, the temperatures and 
gravities derived in this work agree overall very well with our determinations, as can be seen from Figs.~\ref{f7}a,b, i.e., the differences lie within the 1$\sigma$-uncertainty of the $Q$-calibrations for most stars. Microturbulences also show a good agreement, considering 
the different computations, reaching a maximum difference of $\sim$4\,km\,s$^{-1}$ (Fig.~\ref{f7}c).

\paragraph{ \citet{napi93}.} In Sect.~\ref{sect_comp}, our $T_{[u-b]}$ calibrations were compared with that proposed by \citet{napi93}. 
In Fig.~\ref{f8}a,b, our spectroscopic parameters are compared to temperatures and gravities computed with 
an improved version by \citet{napi93} of a code by \citet{md84}.
Fig.~\ref{f8}c shows a comparison of our spectroscopic temperatures to those derived with the $[u-b]$-temperature scale by \citet{napi93}. The code 
uses as input Str\"omgren $ubvy\beta$ indices and interpolates within a grid of synthetic colors 
iteratively until convergence in effective temperature and surface gravity is reached.
The modification by \citet{napi93} on the gravities is based on fits to Balmer lines.
The code provides a final temperature and gravity, hereafter T$_{N93}$ 
and $\log g_{\rm{N93}}$, respectively. It also returns a temperature from their calibration to the $[u-b]$-parameter, hereafter $T_{\rm{N93}[u-b]}$, that corresponds to their Eqn.~9, discussed in Sect.~\ref{sect_photo} (see also Fig.~\ref{f3}b).

Figure~\ref{f8}a displays the difference between their $T_{\rm{N93}}$ and our $T_\mathrm{eff}$. 
There is a very good agreement for lower temperatures, up to $T_\mathrm{eff}$= 24\,000\,K. 
At higher temperatures there is a larger scatter, their values being smaller than ours from a few hundred up to $\sim$3000\,K. Exceptions are star \#12, with perfect agreement,  and star \#22, $\sim$2500\,K hotter in their approach. We recall that a non-standard reddening law is found for this star, as it is located in a dense nebulous region, therefore a photometric determination of its parameters is dubious.
The vertical dashed-dotted line in Fig.~\ref{f8}a corresponds to the upper temperature limit to which they expect their  determination to be valid. Thick squares denote stars with $T_\mathrm{eff}$$\ge$ 30\,000\,K 
and diamonds the rest of the sample; this corresponds to the upper limit to which they expect their method to be valid. From Fig.~\ref{f8}a we notice, however, that their upper limit is likely $T_\mathrm{eff}$$\approx$ 24\,000\,K. 
Figure~\ref{f8}b shows the differences between their gravities, $\log g_{\rm{N93}}$, and our $\log g$. There is a good agreement between both methods for several stars. A few objects also show larger discrepancies up to $\pm$0.3~dex for stars denoted with diamonds and, as expected,  even larger for those marked with squares, with temperatures beyond their upper limit ($+$0.6~dex for star \#30 and $-$0.5~dex for star \#21, not shown in Fig.~\ref{f8}b because they lie outside the plot range).
Their $T_{\rm{N93}[u-b]}$, extracted from the same code, is compared with our spectroscopic temperature in Fig.~\ref{f8}c. The differences to our work show a similar behaviour than Fig.~\ref{f8}a, but with a larger scatter. Stars \#2, 21 and 30 are not shown because their differences are larger than $-$3000\,K.
Following the comparisons in Fig.~\ref{f8}, we recommend to adopt $T_{\rm{N93}}$ for $T_\mathrm{eff}$$\le$ 24\,000\,K when using 
their approach.

Several reasons could lead to discrepancies between their and our results at higher temperatures. We note that only four stars hotter than 24\,000\,K are present in their calibration sample to derive the $T_{\rm{N93}[u-b]}$-scale. Three of these stars have luminosity class III, therefore the $T_{\rm{N93}[u-b]}$-scale underestimates temperature of stars with luminosity class V (most objects in this work). Two of these four stars are present in our sample: \#3 and \#19. $T_{\rm{N93}[u-b]}$ for star \#3 (HD\,63922, HR\,3055) is 1200\,K lower than our value, and that for star \#19 (HD\,34816, HR\,1756) is 2800\,K than our temperature. 
Furthermore, their calibration is based on temperatures from \citet{m86}, determined from comparisons of observed spectral energy distributions to fluxes computed in LTE by \citet{k79}. Hence, discrepancies with our results could also be caused by less line blanketing in their model atmospheres, neglection of NLTE effects in the fluxes, and also in the computation of synthetic indices and Balmer lines.

\begin{table}[t!]
\caption[]{
Comparison of $T_\mathrm{eff}$- and $BC$-scales from this work to literature data. \\[-4mm] \label{spt}}
 \setlength{\tabcolsep}{.1cm}
 \begin{tabular}{l@{\hspace{3mm}}rrclcc@{\hspace{1mm}}rrrrrrrrrr}
 \noalign{}
\hline
\hline
Sp.\,T    &  \# & $T_\mathrm{eff}$& $\log g$  & $BC$   & $T_{\rm{SK82}}$ &  $BC_{\rm{SK82}}$ & $T_{\rm{GC94}}$ & $T_{\rm{C00}}$ \\
\hline\\[-1.3mm]
O9\,V     &   30         &   34100       &   4.18    & $-$3.23 &  33000   &  $-$3.33    &  \ldots    & 34\,000   \\[1mm]
B0\,V     &   11         &   33400       &   4.30    & $-$3.15 &  30000   &  $-$3.16    &  \ldots    & 30\,000   \\[1mm]
B0.2\,V   &   ~6         &   32000       &   4.30    & $-$3.05 &  \ldots    &  \ldots   &  \ldots    &  \ldots   \\[1mm]
B0.5\,V   &   21         &   30700       &   4.30    & $-$2.98 &  \ldots    &  \ldots   &  \ldots    &  \ldots   \\
          &   19         &   30400       &   4.30    & $-$2.95 &    \\
          &   22         &   29300       &   4.30    & $-$2.88 &    \\[1mm]
B0.7\,V   &   13         &   29000       &   4.10    & $-$2.83 &  \ldots    &  \ldots   &  \ldots &  \ldots \\ [1mm]
B1\,V     &   ~1         &   27000       &   4.12    & $-$2.65 &   25400  &  $-$2.70    &  24620   &  \ldots   \\
          &   ~2         &   26300       &   4.15    & $-$2.59 &            &           &         \\[1mm]
B1.5\,V   &   23         &   26100       &   4.25    & $-$2.63 &  \ldots    &  \ldots   &  \ldots    &  \ldots   \\
          &   24         &   24000       &   4.10    & $-$2.43 &            &           &         \\
          &   ~9         &   23500       &   4.20    & $-$2.38 &     \\[1mm]
B2\,V     &   25         &   21700       &   4.25    & $-$2.19 &  22000   &  $-$2.35    &  19500   & 20900   \\
          &   ~5         &   20800       &   4.22    & $-$2.06 &    \\
          &   10         &   20700       &   4.15    & $-$2.06 &            &           &         \\
          &   27         &   20300       &   4.15    & $-$2.02 &            &           &         \\[1mm]
B2.5\,V   &   28         &   19300       &   4.14    & $-$1.90 & 18700   &   $-$1.94     & 16000    &  \ldots   \\
          &   29         &   19000       &   4.00    & $-$1.85 &   \\[1mm]
 \hline
 \hline\\[-1.3mm]
 B1\,IV   &  14         &  27000        &   4.05    & $-$2.67  &  \ldots    &  \ldots   &  \ldots &  \ldots \\[1mm]
 B1.5\,IV &  16         &  23000        &   3.95    & $-$2.32  &  \ldots    &  \ldots   &  \ldots &  \ldots \\
          &  ~8         &  22000        &   3.85    & $-$2.18 &  \\[1mm]
 B2\,IV   &  ~7         &  22000        &   3.95    & $-$2.20  &  \ldots    &  \ldots   &  \ldots &  \ldots \\
          &  18         &  21250        &   3.80    & $-$2.11    \\
          &  17         &  20750        &   3.80    & $-$2.05    \\[1mm]
 B3\,IV   &  20         &  17500        &   3.80    & $-$1.65  &  \ldots    &  \ldots   &  \ldots &  \ldots \\[1mm]
\hline          &  18         &  21250        &   3.80    & $-$2.11    \\
\hline\\[-1.3mm]
 B0.2\,III  &  ~3         &  31200        &   3.95    &  $-$2.92 &  29000   &  $-$2.88    &  \ldots   &  \ldots  \\[1mm]
 B0.5\,III&  12         &  30000        &   4.05    & $-$2.88  &  \ldots    &  \ldots   &  \ldots &  \ldots \\[1mm]
 B3\,III  &  15         &  15800        &   3.75    &  $-$1.38 &  17100   &  $-$1.60    &  \ldots   &  \ldots  \\[1mm]
\hline
\hline\\[-1.3mm]
\end{tabular}
SK82: \citet{sk82}, GC94: \citet{gc94}, C00: \citet{c00}.\\[1mm]
\end{table}

\subsection{Our recommended parameter determination}

Our scales derived spectroscopically tend to be hotter at higher temperatures than the photometric scales proposed by \citet{d99}, \citet{l02} and \citet{napi93}. 
Furthermore, our values also tend to be higher at $T_\mathrm{eff}$$\ge$ 25\,000\,K and lower at lower temperatures when compared with values derived by \citet{k91}, \citet{cl92} and \citet{mb08} using different analysis methods, synthetic spectra and observations. In contrast, we note that temperatures determined by 
\citet{s10} using the same set of high-quality spectra than those analysed by us,  other state-of-the-art model atmosphere and spectral synthesis in NLTE and simultaneous silicon ionization equilibria and fits to Balmer and helium lines, render values in agreement to our determinations. 
The excellent agreement between our and \citet{s10} temperatures lead us to conclude that our temperature determination based on high quality data, state-of-the-art spectral synthesis in NLTE and several independent temperature indicators are reliable and almost free of large remaining systematic effects.
If adopting significantly different values in the higher temperature domain, like suggested by previous  studies, the ionization equilibrium of no species will be matched any longer, hence the global and detailed observed spectra will not be reproduced by the synthetic model.

We recommend, when possible, to perform a parameter determination like the full spectroscopic method presented by us in previous works (e.g. Paper~I) or by \citet{s10}. If the observed spectra or the spectral synthesis are not available, a good parameter estimation can be provided by our scales presented in Sect.~\ref{sect_tgscales} of this work. For $T_\mathrm{eff}$$\le$ 24\,000\,K, the scale $T_{\rm{N93}}$ by \citet{napi93} also renders reliable temperatures, and for $T_\mathrm{eff}$$\le$ 21\,000\,K, the scales by \citet{d99} and \citet{l02} also give reliable results (note the lower limit indicated by each scale).

\begin{figure*}[t!]
\resizebox{.49\hsize}{!}{\includegraphics{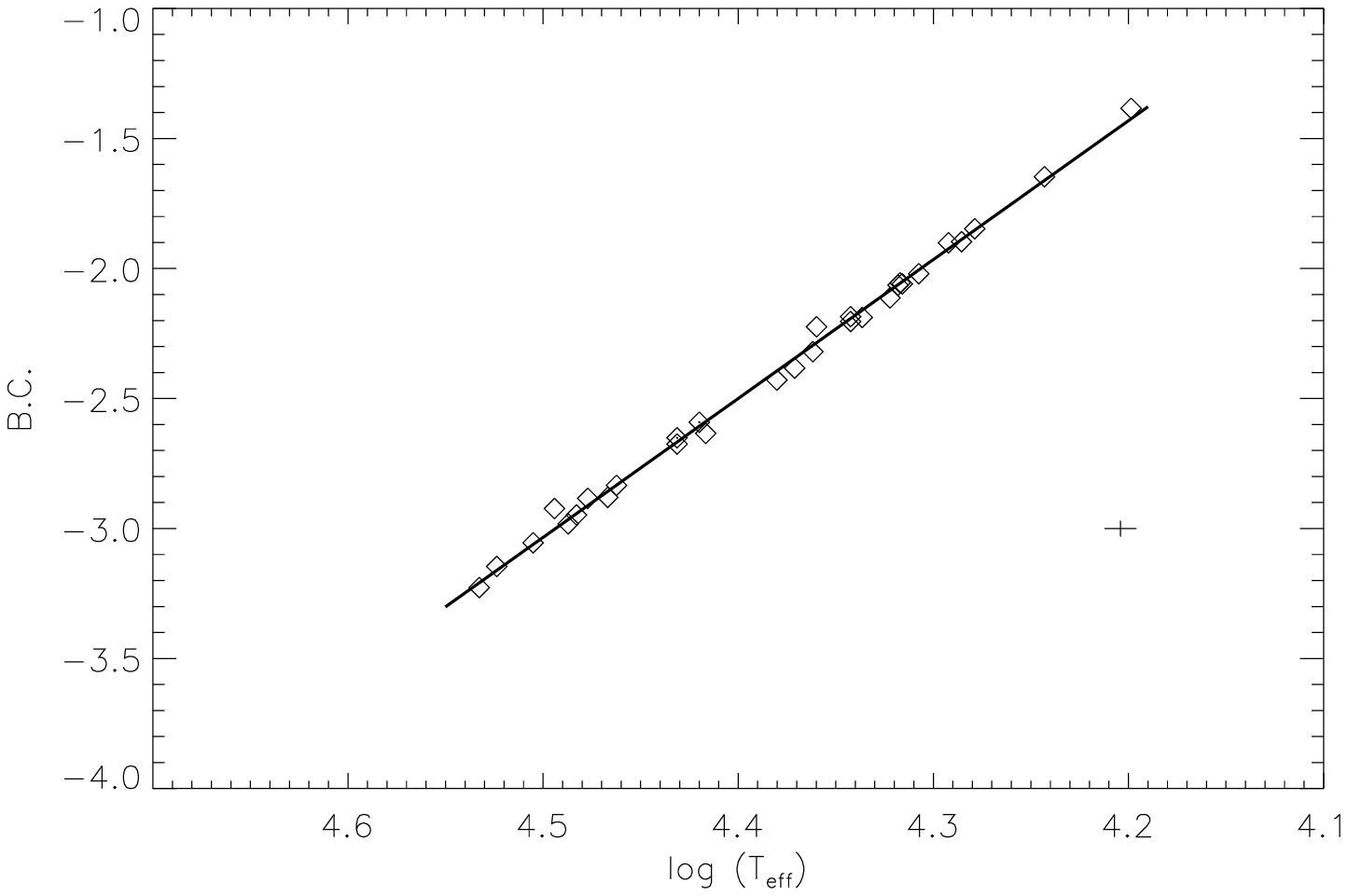}}
\resizebox{.49\hsize}{!}{\includegraphics{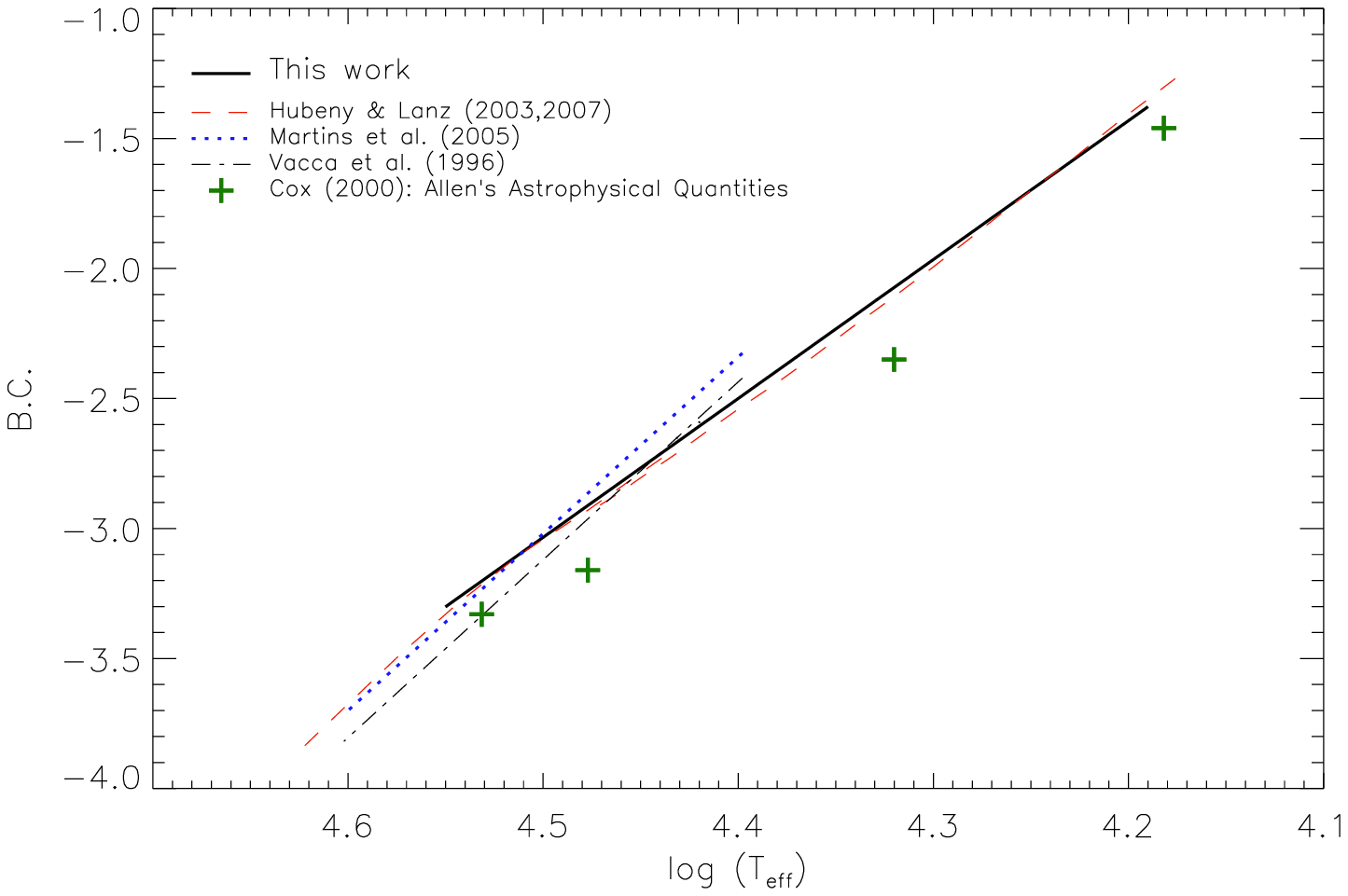}}
\caption[]{Left panel: bolometric corrections vs. effective temperature as derived for the sample stars. A typical error bar is shown, the line indicates a regression curve to the data. Right panel: comparison 
of our $T_\mathrm{eff}$-$BC$ relation with data from the literature. See the text for details.}
\label{bc1}
\end{figure*}

\section{Spectral type calibration}\label{sect_stcali}
In App.~\ref{sect_st}, we re-evaluate the spectral classification of the sample stars. Slight revision of spectral sub-types and luminosity
classes have been performed in several stars, as indicated in Table~\ref{photo}.
The relationship between spectral types and luminosity classes and the stellar parameters $T_\mathrm{eff}$, $\log g$ as well as their corresponding value of bolometric correction
($BC$) is summarized in Table~\ref{spt}. There, also recommended values of temperature and/or $BC$ from \citet{sk82}, \citet{gc94} and \citet{c00} are given for comparison. 

In cases where more than one star belong to a certain spectral/luminosity class, instead of adopting an average value for the parameters, we prefer to indicate the exact parameters of the stars to see how much can they variate. Stars \#4 and 26 have been excluded from the calibration because they seem to be too evolved in comparison to the rest of the objects, as explained above.
Our star sample is not large enough to cover all possible combinations of spectral and luminosity classes within the range O9-B3 V-III, however the carefully derived data allow us to anchor a spectral type calibration to empirical values of temperature for luminosity class V and partially also for luminosity classes III and IV. Our calibration tends to be hotter than  previous reference values in most cases. This is in accordance with our photometric temperature calibrations, that are in general hotter than previous work. This table is also very useful to estimate stellar parameters once the spectral and luminosity classification are well known.

\section{Bolometric corrections vs. temperature}\label{sect_bc}

Bolometric corrections play an important role in the determination of luminosities 
and indirectly also of other fundamental stellar parameters computed from atmospheric parameters $T_\mathrm{eff}$ and $\log g$, e.g. radii, masses. 
Fundamental parameters can be directly compared to predictions from evolution
models in order to learn more about the physics of the stars. 

Taking advantage of our accurate atmospheric parameters determined spectroscopically for the present sample, 
we derive a semiempiric relationship for $BC$ as a function of $T_\mathrm{eff}$.
Bolometric corrections are computed from spectral fluxes calculated with {\sc Atlas9}\footnote{Atmospheric structures and 
fluxes of OB main sequence and giant stars computed in LTE are equivalent to the full NLTE approach. See \citet{np07} and \citet{pnb11} for details.} for each star
of the sample using their corresponding values of $T_\mathrm{eff}$ and $\log g$. 
As expected, a clear relationship is found for $BC$ vs. $T_\mathrm{eff}$ (Fig.~\ref{bc1}, left panel), practically independent from the luminosity class or the surface gravity.
Figure~\ref{bc1}, right panel, illustrates a comparison of our relationship to other work by \citet{v96}, \citet{c00}, \citet{hl03,hl07}, and \citet{m05} derived mainly from synthetic models.
Our slope is less steep than those by \citet{v96} and \citet{m05}, our calibration extending to lower temperatures. In addition, our fit agrees with $BC$ from \citet{hl03,hl07} and for temperatures higher than $\sim$32\,000\,K also with those by \citet{m05}. In contrast, reference values for $BC$ recommended by \citet{c00} are systematically lower than our data.

From the relationship found in Fig.~\ref{bc1} (left panel), we propose a new calibration of the bolometric correction as a function of $T_\mathrm{eff}$:

\begin{eqnarray}
BC= 21.00 - 5.34\,\log (T_\mathrm{eff})\,;~ 1\sigma= 0.01
\end{eqnarray}

The boundaries are limited by the observational data: 15\,800\,$\le T_\mathrm{eff}\le$\,34\,000\,K. For higher temperatures, we recommend to use bolometric corrections from \citet{hl03} or \citet{m05}.

\section{Summary and conclusions}\label{sect_summ}

Based on comprehensive previous studies which lead us to determine effective 
temperatures and surface gravities of 30 OB-type (O9-B3 V-III) stars at high accuracy and precision, 
here we investigate whether such parameters can be used to propose new calibrations to some photometric indices. 
The spectroscopic parameters were derived via multiple ionization equilibria and simultaneous fits to the Balmer lines
based on our most recent spectrum synthesis for NLTE computations and high-quality observed spectra.
We propose very useful relationships between atmospheric parameters and various photometric indices from the Johnson and the Str\"omgren systems. 
Effective temperatures can be estimated at a precision of $\sim$400\,K for luminosity classes III/IV and $\sim$800\,K for luminosity class V. And surface gravities can reach internal uncertainties as low as $\sim$0.08\,dex when using 
our calibration to the Johnson $Q$-parameter. Similar precision is achieved for gravities derived from the $\beta$-index and the precision decreases for both atmospheric parameters when using the Str\"omgren indices $[c1]$ and $[u-b]$. In contrast, external uncertainties are larger for the Johnson than for the Str\"omgren calibrations.
Our uncertainties are smaller than typical differences among other methods in the literature, reaching values up to $\pm$2000\,K for temperature and $\pm$0.25\,dex for gravity, and in extreme cases, $+$6000\,K and $\pm$0.4\,dex, respectively.
We had further re-examined the luminosity and the spectral classes for the star sample and also propose a parameter calibration for sub-spectral types. 
Furthermore, we present a new bolometric correction relation to temperature based on our empirical data, rather than on synthetic grids.

The photometric calibrations presented here are useful tools to estimate effective temperatures and surface gravities of non-supergiant OB stars in a fast manner. Caution has to be taken for undetected double-lined spectroscopic binaries and single objects with anomalous reddening-law, dubious photometric quantities and/or luminosity classes, for which the external uncertainties may increase significantly. We recommend to use these calibrations only as a first step for parameter estimation, with subsequent refinements based on spectroscopy.

\begin{acknowledgements}
The author acknowledges a FFL stipend from the University of Erlangen-Nuremberg and thanks Norbert Przybilla, Eva Ziegerer, Christian Heuser, Ulrich Heber and Sergio Sim\'on-D\'iaz for their suggestions, and the anonymous referee
for his/her comments leading to substantial improvements of the paper. 
\end{acknowledgements}

\appendix

\section{Spectral types and luminosity classes}\label{sect_st}

The photometric calibrations discussed in Sect.~\ref{sect_tgscales} depend on luminosity classifications of the star sample. Here, we re-examine their luminosity and spectral classification.   
As reference, we adopt the classification of one anchor point of the Morgan-Keenan system (star \#11) and five primary standards from N. Walborn (\#6, 7, 11, 19 and 23), all listed in \citet{gc09}. 
In a second step, spectral types and luminosity classes of the sample stars are examined via inspection of Balmer lines and silicon/helium ratios, as recommended by \citet{w71} and \citet{gc09}. Spectral type classification for the hotter stars (O9-B0.7) rely on 
the ratio \ion{Si}{iv}$\lambda$4089/\ion{Si}{iii}$\lambda$4552 and for the cooler stars (B1-B3) on \ion{Si}{ii}$\lambda$4128/\ion{Si}{iii}$\lambda$4552. In addition, luminosity classes are tested via \ion{Si}{iv}$\lambda$4116/\ion{He}{i}$\lambda$4121 for the hotter and \ion{Si}{iii}$\lambda$4552/\ion{He}{i}$\lambda$4387 for the 
cooler stars. 

\begin{figure}[t!]
\centering
\resizebox{0.96\hsize}{!}{\includegraphics[width=12cm]{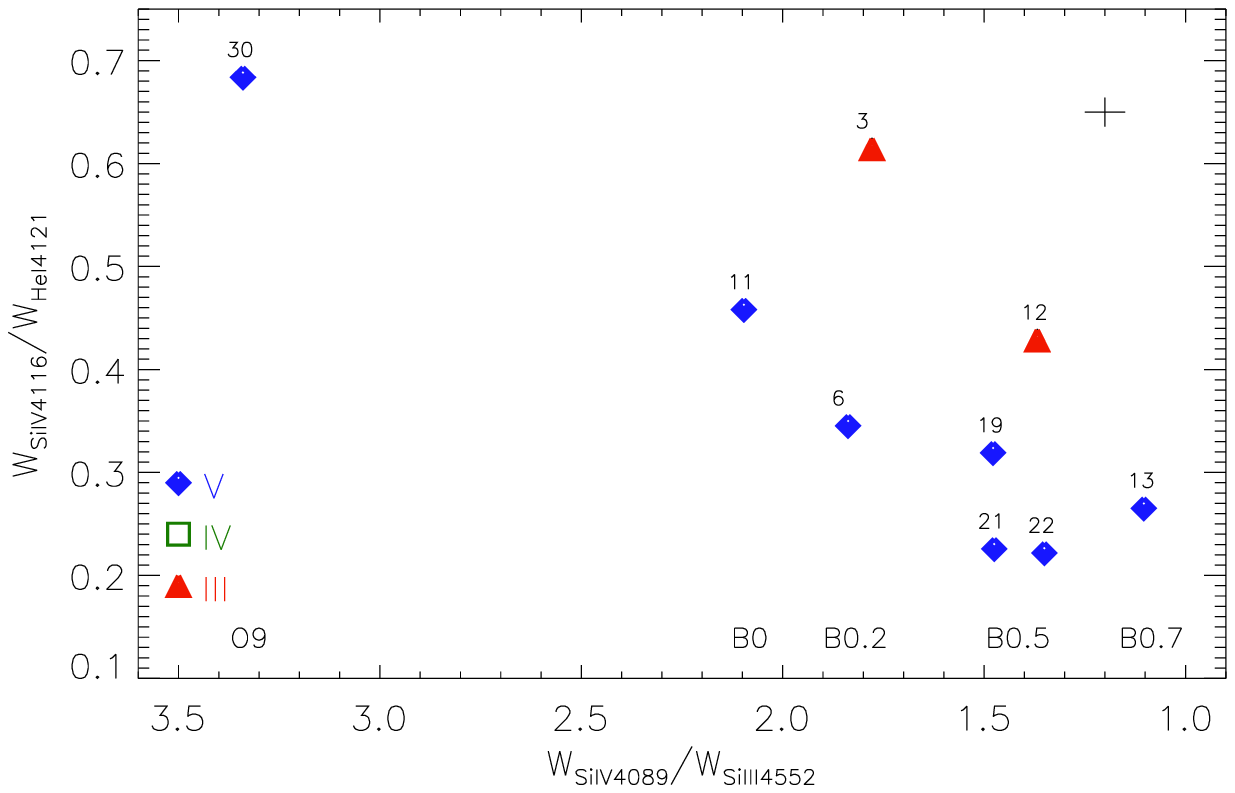}}
\resizebox{0.96\hsize}{!}{\includegraphics[width=12cm]{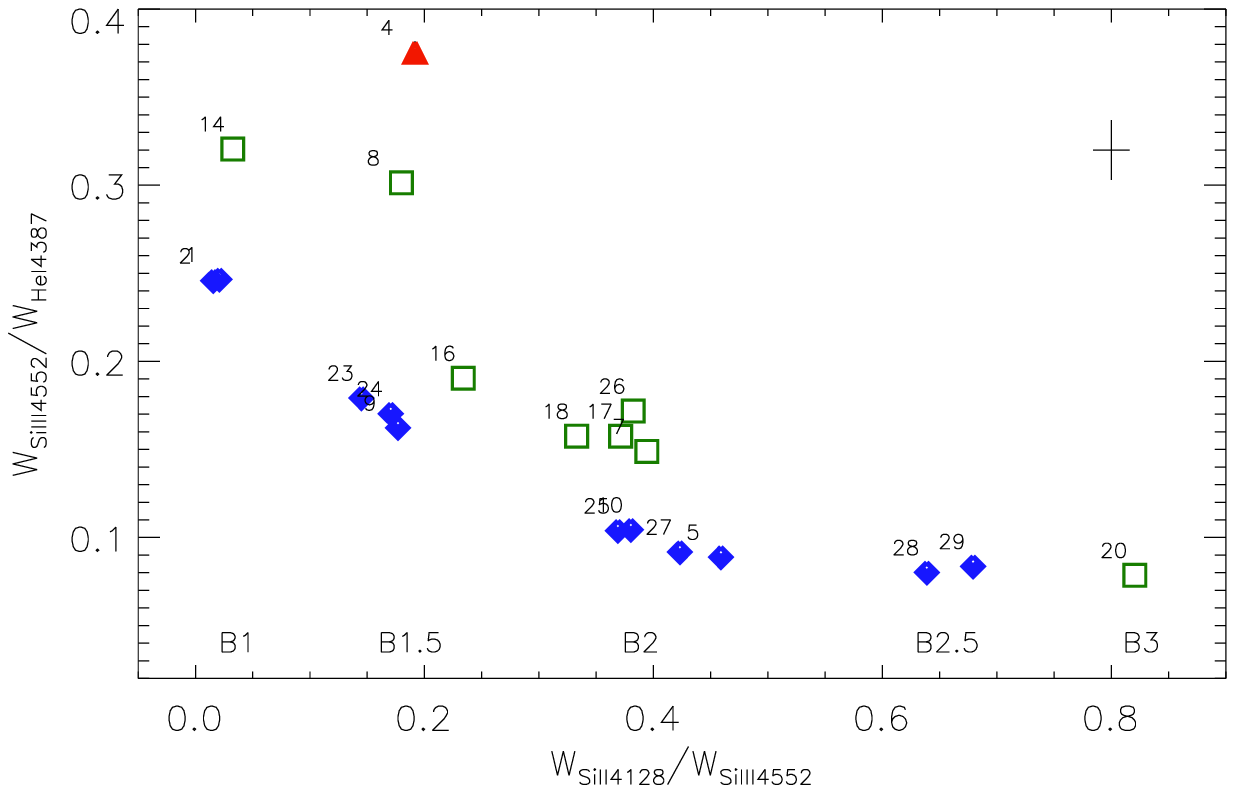}}
\caption[]{Ratios of equivalent widths for temperature (abscissa) and gravity-sensitive (ordinate)
line pairs as basis for the spectral type and luminosity classification of the star sample (Table~\ref{photo}). Upper panel: hotter stars O9-B0.7; lower panel: cooler stars B1-B3.}
\label{fa}
\end{figure}

Figure~\ref{fa} displays the relation of line ratios as spectral type and luminosity class indicators for the hotter (upper panel) and the cooler stars (lower panel). Only few cases -- stars \#2, 8, 10, 19 and 26 -- required a correction in luminosity class.  
Furthermore, slight corrections to (sub-)spectral types for objects \#2, 5, 8, 10, 13, 16, 28, 29, have also been implemented, as listed in Table~\ref{photo}.
The revisions are facilitated by the higher quality of our spectra in comparison to older photographic spectrograms used in the past, affecting in particular the weaker silicon lines. 
Star \#15 is too cool, i.e. has a very low \ion{Si}{ii}$\lambda$4128/\ion{Si}{iii}$\lambda$4552 ratio, therefore is excluded 
from Figure~\ref{fa} (lower panel).

\section{Spectroscopic indicators for temperature and surface determination}\label{sect_indicators}

Table~\ref{indicators} indicates which ionization equilibria in combination with fits to several hydrogen lines, 
have been matched in order to determine effective temperatures and surface gravities for the sample stars.
Dots denote fits to spectral lines and boxes indicate cases where ionization equilibria were matched.
A row denotes the simultaneous fits to different elements/ionization stages using the same set of atmospheric parameters.

\begin{table}[t!]
\centering
\caption[]{Spectroscopic indicators for $T_\mathrm{eff}$ and $\log g$ determination.
Dots indicate spectral fitting and boxes denote ionization equilibria.\\[-4mm]\label{indicators}}
 \setlength{\tabcolsep}{.06cm}
 \tiny
 \begin{tabular}{rrrcc@{\hspace{.1mm}}cc@{\hspace{.1mm}}c@{\hspace{.1mm}}cc@{\hspace{.1mm}}cc@{\hspace{.1mm}}cc@{\hspace{.1mm}}cc@{\hspace{.1mm}}c}
 \noalign{}
\hline
\hline
&HD&$T_\mathrm{eff}$& H & \ion{He}{i} & \ion{He}{ii}                 &
\ion{C}{ii} & \ion{C}{iii}  & \ion{C}{iv} & \ion{O}{i} & \ion{O}{ii} &
\ion{Ne}{i} &  \ion{Ne}{ii} & \ion{Si}{iii}& \ion{Si}{iv} & \ion{Fe}{ii}& \ion{Fe}{iii}\\
 & & 10$^3$\,K\\[-.1mm]
\hline\\[-2mm]
1&36591 & 27.0   & $\bullet$ & \multicolumn{2}{c}{\fbox{$\bullet$~~~~~$\bullet$}}& \multicolumn{2}{c}{\fbox{$\bullet$~~~~~$\bullet$}} &
&                                        & $\bullet$ & \multicolumn{2}{c}{\fbox{$\bullet$~~~~~$\bullet$}} &\multicolumn{2}{c}{\fbox{$\bullet$~~~~~$\bullet$}}& & $\bullet$\\[.5mm]

2&61068 & 26.3   & $\bullet$ & \multicolumn{2}{c}{\fbox{$\bullet$~~~~~$\bullet$}}& \multicolumn{2}{c}{\fbox{$\bullet$~~~~~$\bullet$}} &
&                                        & $\bullet$ & \multicolumn{2}{c}{\fbox{$\bullet$~~~~~$\bullet$}} &\multicolumn{2}{c}{\fbox{$\bullet$~~~~~$\bullet$}}& & $\bullet$\\[.5mm]

3&63922 & 31.2   & $\bullet$ & \multicolumn{2}{c}{\fbox{$\bullet$~~~~~$\bullet$}}& \multicolumn{3}{c}{\fbox{\,$\bullet$~~~~~$\bullet$~~~~~$\bullet$\,}}
&                                        & $\bullet$ &           &                             $\bullet$  &\multicolumn{2}{c}{\fbox{$\bullet$~~~~~$\bullet$}}& & $\bullet$\\[.5mm]

4&74575 & 22.9   & $\bullet$ & $\bullet$ &                                       & \multicolumn{2}{c}{\fbox{$\bullet$~~~~~$\bullet$}} &
& \multicolumn{2}{c}{\fbox{$\bullet$~~~~~$\bullet$}} & \multicolumn{2}{c}{\fbox{$\bullet$~~~~~$\bullet$}} &\multicolumn{2}{c}{\fbox{$\bullet$~~~~~$\bullet$}}& \multicolumn{2}{c}{\fbox{$\bullet$~~~~~$\bullet$}}\\[.5mm]

5&122980& 20.8   & $\bullet$ & $\bullet$ &                                       & \multicolumn{2}{c}{\fbox{$\bullet$~~~~~$\bullet$}} &                 & \multicolumn{2}{c}{\fbox{$\bullet$~~~~~$\bullet$}} & $\bullet$ &                                        &$\bullet$&& \multicolumn{2}{c}{\fbox{$\bullet$~~~~~$\bullet$}}\\[.5mm]

6&149438& 32.0   & $\bullet$ & \multicolumn{2}{c}{\fbox{$\bullet$~~~~~$\bullet$}}& \multicolumn{3}{c}{\fbox{\,$\bullet$~~~~~$\bullet$~~~~~$\bullet$\,}}
&                                        & $\bullet$ & \multicolumn{2}{c}{\fbox{$\bullet$~~~~~$\bullet$}} &\multicolumn{2}{c}{\fbox{$\bullet$~~~~~$\bullet$}}& & $\bullet$\\[.5mm]

7&886   & 22.0   & $\bullet$ & $\bullet$ &                                       & \multicolumn{2}{c}{\fbox{$\bullet$~~~~~$\bullet$}} &
& \multicolumn{2}{c}{\fbox{$\bullet$~~~~~$\bullet$}} & $\bullet$ &                                        &\multicolumn{2}{c}{\fbox{$\bullet$~~~~~$\bullet$}}& \multicolumn{2}{c}{\fbox{$\bullet$~~~~~$\bullet$}}\\[.5mm]

8&29248 & 22.0   & $\bullet$ & $\bullet$ &                                       & \multicolumn{2}{c}{\fbox{$\bullet$~~~~~$\bullet$}} &                 & \multicolumn{2}{c}{\fbox{$\bullet$~~~~~$\bullet$}} & $\bullet$ &                                        &\multicolumn{2}{c}{\fbox{$\bullet$~~~~~$\bullet$}}& \multicolumn{2}{c}{\fbox{$\bullet$~~~~~$\bullet$}}\\[.5mm]

9&35299 & 23.5   & $\bullet$ & $\bullet$ &                                       & \multicolumn{2}{c}{\fbox{$\bullet$~~~~~$\bullet$}} &
& \multicolumn{2}{c}{\fbox{$\bullet$~~~~~$\bullet$}} & \multicolumn{2}{c}{\fbox{$\bullet$~~~~~$\bullet$}} &\multicolumn{2}{c}{\fbox{$\bullet$~~~~~$\bullet$}}& \multicolumn{2}{c}{\fbox{$\bullet$~~~~~$\bullet$}}\\[.5mm]

10&35708 & 20.7   & $\bullet$ & $\bullet$ &                                       & \multicolumn{2}{c}{\fbox{$\bullet$~~~~~$\bullet$}} &                 & \multicolumn{2}{c}{\fbox{$\bullet$~~~~~$\bullet$}} & $\bullet$ &                                        &\multicolumn{2}{c}{\fbox{$\bullet$~~~~~$\bullet$}}& \multicolumn{2}{c}{\fbox{$\bullet$~~~~~$\bullet$}}\\[.5mm]

11&36512 & 33.4   & $\bullet$ & \multicolumn{2}{c}{\fbox{$\bullet$~~~~~$\bullet$}}& \multicolumn{3}{c}{\fbox{\,$\bullet$~~~~~$\bullet$~~~~~$\bullet$\,}}
 &                                        & $\bullet$ &           &                             $\bullet$  &\multicolumn{2}{c}{\fbox{$\bullet$~~~~~$\bullet$}}& & $\bullet$\\[.5mm]

12&36822 & 30.0   & $\bullet$ & \multicolumn{2}{c}{\fbox{$\bullet$~~~~~$\bullet$}}& \multicolumn{3}{c}{\fbox{\,$\bullet$~~~~~$\bullet$~~~~~$\bullet$\,}}
 &                                        & $\bullet$ & \multicolumn{2}{c}{\fbox{$\bullet$~~~~~$\bullet$}} &\multicolumn{2}{c}{\fbox{$\bullet$~~~~~$\bullet$}}& & $\bullet$\\[.5mm]

13&36960 & 29.0   & $\bullet$ & \multicolumn{2}{c}{\fbox{$\bullet$~~~~~$\bullet$}}& \multicolumn{2}{c}{\fbox{$\bullet$~~~~~$\bullet$}} &
 &                                        & $\bullet$ & \multicolumn{2}{c}{\fbox{$\bullet$~~~~~$\bullet$}} &\multicolumn{2}{c}{\fbox{$\bullet$~~~~~$\bullet$}}& & $\bullet$\\[.5mm]

14&205021& 27.0   & $\bullet$ & \multicolumn{2}{c}{\fbox{$\bullet$~~~~~$\bullet$}}& \multicolumn{2}{c}{\fbox{$\bullet$~~~~~$\bullet$}} &
 &                                        & $\bullet$ & \multicolumn{2}{c}{\fbox{$\bullet$~~~~~$\bullet$}} &\multicolumn{2}{c}{\fbox{$\bullet$~~~~~$\bullet$}}& & $\bullet$\\[.5mm]

15&209008& 15.8   & $\bullet$ & $\bullet$ &                                       &  $\bullet$                                       & &                 & \multicolumn{2}{c}{\fbox{$\bullet$~~~~~$\bullet$}} & $\bullet$ &                                    &$\bullet$&& \multicolumn{2}{c}{\fbox{$\bullet$~~~~~$\bullet$}}\\[.5mm]   

16&216916& 23.0   & $\bullet$ & $\bullet$ &                                       & \multicolumn{2}{c}{\fbox{$\bullet$~~~~~$\bullet$}} &
 & \multicolumn{2}{c}{\fbox{$\bullet$~~~~~$\bullet$}} & \multicolumn{2}{c}{\fbox{$\bullet$~~~~~$\bullet$}} &\multicolumn{2}{c}{\fbox{$\bullet$~~~~~$\bullet$}}& \multicolumn{2}{c}{\fbox{$\bullet$~~~~~$\bullet$}}\\[.5mm]

17&  3360& 20.7   & $\bullet$ & $\bullet$ &                                       & \multicolumn{2}{c}{\fbox{$\bullet$~~~~~$\bullet$}} &                 & \multicolumn{2}{c}{\fbox{$\bullet$~~~~~$\bullet$}} & $\bullet$ &                                        &\multicolumn{2}{c}{\fbox{$\bullet$~~~~~$\bullet$}}& \multicolumn{2}{c}{\fbox{$\bullet$~~~~~$\bullet$}}\\[.5mm]

18&16582& 21.2   & $\bullet$ & $\bullet$ &                                       & \multicolumn{2}{c}{\fbox{$\bullet$~~~~~$\bullet$}} &                 & \multicolumn{2}{c}{\fbox{$\bullet$~~~~~$\bullet$}} & $\bullet$ &                                        &\multicolumn{2}{c}{\fbox{$\bullet$~~~~~$\bullet$}}& \multicolumn{2}{c}{\fbox{$\bullet$~~~~~$\bullet$}} \\[.5mm]

19&34816 & 30.4   & $\bullet$ & \multicolumn{2}{c}{\fbox{$\bullet$~~~~~$\bullet$}}& \multicolumn{2}{c}{\fbox{$\bullet$~~~~~$\bullet$}} &
 &                                        & $\bullet$ &  \multicolumn{2}{c}{\fbox{$\bullet$~~~~~$\bullet$}}&\multicolumn{2}{c}{\fbox{$\bullet$~~~~~$\bullet$}}& & $\bullet$\\[.5mm]

20&160762& 17.5   & $\bullet$ & $\bullet$ &                                       &  $\bullet$                                       & &                 & \multicolumn{2}{c}{\fbox{$\bullet$~~~~~$\bullet$}} & $\bullet$ &                                        &$\bullet$&& \multicolumn{2}{c}{\fbox{$\bullet$~~~~~$\bullet$}}\\[.5mm]
21 &  37020  &30.7& $\bullet$ & \multicolumn{2}{c}{\fbox{$\bullet$~~~~~$\bullet$}} & \multicolumn{2}{c}{\fbox{$\bullet$~~~~~$\bullet$}}     &             && $\bullet$  & \multicolumn{2}{c}{\fbox{$\bullet$~~~~~$\bullet$}}& \multicolumn{2}{c}{\fbox{$\bullet$~~~~~$\bullet$}}&& $\bullet$\\[.5mm]
22 &  37042  &29.3  & $\bullet$ & \multicolumn{2}{c}{\fbox{$\bullet$~~~~~$\bullet$}} & \multicolumn{2}{c}{\fbox{$\bullet$~~~~~$\bullet$}}     &             && $\bullet$     & \multicolumn{2}{c}{\fbox{$\bullet$~~~~~$\bullet$}}& \multicolumn{2}{c}{\fbox{$\bullet$~~~~~$\bullet$}}&& $\bullet$\\[.5mm]
 23 &  36959& 26.1 & $\bullet$ & \multicolumn{2}{c}{\fbox{$\bullet$~~~~~$\bullet$}} & \multicolumn{2}{c}{\fbox{$\bullet$~~~~~$\bullet$}}     &             &  \multicolumn{2}{c}{\fbox{$\bullet$~~~~~$\bullet$}}& \multicolumn{2}{c}{\fbox{$\bullet$~~~~~$\bullet$}}& \multicolumn{2}{c}{\fbox{$\bullet$~~~~~$\bullet$}}&& $\bullet$\\[.5mm]
 24 &  37744& 24.0 & $\bullet$ & $\bullet$  &                                       & \multicolumn{2}{c}{\fbox{$\bullet$~~~~~$\bullet$}}     &              & \multicolumn{2}{c}{\fbox{$\bullet$~~~~~$\bullet$}}& \multicolumn{2}{c}{\fbox{$\bullet$~~~~~$\bullet$}}& \multicolumn{2}{c}{\fbox{$\bullet$~~~~~$\bullet$}}&\multicolumn{2}{c}{\fbox{$\bullet$~~~~~$\bullet$}}\\[.5mm]
 25 &  36285& 21.7 & $\bullet$ & $\bullet$  &                                       & \multicolumn{2}{c}{\fbox{$\bullet$~~~~~$\bullet$}}     &              & \multicolumn{2}{c}{\fbox{$\bullet$~~~~~$\bullet$}}& $\bullet$ && \multicolumn{2}{c}{\fbox{$\bullet$~~~~~$\bullet$}}&\multicolumn{2}{c}{\fbox{$\bullet$~~~~~$\bullet$}}\\[.5mm]
 26&  35039& 19.6 & $\bullet$ & $\bullet$  &                                       & \multicolumn{2}{c}{\fbox{$\bullet$~~~~~$\bullet$}}     &              & \multicolumn{2}{c}{\fbox{$\bullet$~~~~~$\bullet$}}& $\bullet$ && \multicolumn{2}{c}{\fbox{$\bullet$~~~~~$\bullet$}}&\multicolumn{2}{c}{\fbox{$\bullet$~~~~~$\bullet$}}\\[.5mm]
 27&  36629& 20.3 & $\bullet$ & $\bullet$  &                                       & \multicolumn{2}{c}{\fbox{$\bullet$~~~~~$\bullet$}}     &              & \multicolumn{2}{c}{\fbox{$\bullet$~~~~~$\bullet$}}& $\bullet$ &&$\bullet$ &&\multicolumn{2}{c}{\fbox{$\bullet$~~~~~$\bullet$}}\\[.5mm]
 28&  36430& 19.3 & $\bullet$ & $\bullet$  &                                       & \multicolumn{2}{c}{\fbox{$\bullet$~~~~~$\bullet$}}     &              & \multicolumn{2}{c}{\fbox{$\bullet$~~~~~$\bullet$}}& $\bullet$ &&$\bullet$ &&\multicolumn{2}{c}{\fbox{$\bullet$~~~~~$\bullet$}}\\[.5mm]
 29&  35912 & 19.0 & $\bullet$ & $\bullet$  &                                       & \multicolumn{2}{c}{\fbox{$\bullet$~~~~~$\bullet$}}     &              & \multicolumn{2}{c}{\fbox{$\bullet$~~~~~$\bullet$}}& $\bullet$ &&$\bullet$ &&\multicolumn{2}{c}{\fbox{$\bullet$~~~~~$\bullet$}}\\[.5mm]
30 &  46202 & 34.1 & $\bullet$ & \multicolumn{2}{c}{\fbox{$\bullet$~~~~~$\bullet$}} & & \multicolumn{2}{c}{\fbox{$\bullet$~~~~~$\bullet$}}& & $\bullet$ & & & \multicolumn{2}{c}{\fbox{$\bullet$~~~~~$\bullet$}}& & $\bullet$\\[1mm]
 \hline
\hline
\end{tabular}
\end{table}


\begin{thebibliography}{}

\bibitem[Anders \& Grevesse(1989)]{ag89}
Anders, E., Grevesse, N.~1989, Geoch. et Cosmoch. Acta, 53, 197

\bibitem[Becker \& Butler(1990)]{bb90}
Becker, S. R., \& Butler, K.~1990, \aap, 235, 326

\bibitem[Butler \& Giddings(1985)]{but_gid85}
Butler, K., \& Giddings, J. R.~1985, in
Newsletter of Analysis of Astronomical Spectra, No. 9 (Univ. London)

\bibitem[Briquet et al.(2011)]{b11}
Briquet, M., Aerts, C., Baglin, A., et al.~2011, \aap, 527, 112 (Paper~III)

\bibitem[Cantiello et al.(2009)]{c09}

Cantiello, M., Langer N., Brott, I. et al.~2009, \aap, 499, 279

\bibitem[Cunha et al.(2006)]{c06}
Cunha, K., Hubeny, I. \& Lanz, T.~2006, \apj, 647, 143

\bibitem[Cunha \& Lambert(1992)]{cl92}
Cunha, K., \& Lambert, D. L.~1992, \apj, 399, 586 (CL92)

\bibitem[Cunha \& Lambert(1994)]{cl94}
Cunha, K., \& Lambert, D. L.~1994, \apj, 426, 170

\bibitem[Cox(2000)]{c00}
Cox, A.~2000, Allen's Astrophysical Quantities (Springer, New York)

\bibitem[Daflon et al.(1999)]{d99}
 Daflon, S., Cunha, K. \& Becker, S.R.~1999, \apj, 522, 950

\bibitem[Daflon et al.(2009)]{d09}
 Daflon, S., Cunha, K., de la Reza, R. et al.~2009, \apj, 138, 1577

\bibitem[Ekstr\"om et al.(2012)]{e12}
Ekstr\"om, S., Georgy, C., Eggenberger, P. et al.~2012, \aap, 537, 146

\bibitem[Firnstein \& Przybilla(2012)]{fp12}
Firnstein, M., \& Przybilla, N.~, A\&A, in press

\bibitem[Conti \& Alschuler(1971)]{conti71}
Conti, P. S., \& Alschuler, W. R.~1971, \apj, 170, 325

\bibitem[Giddings(1981)]{gid81}
Giddings, J. R.~1981, Ph.D. Thesis, London

\bibitem[Gies \& Lambert(1992)]{gl92}
Gies, D. R., \& Lambert, D. L.~1992, \apj, 339, 586

\bibitem[Gold(1984)]{g84}
Gold, M.~1984, Diploma Thesis, Munich

\bibitem[Grevesse \& Sauval(1998)]{gs98}
Grevesse, N., \& Sauval, A.~J.~1998, \ssr, 85, 161

\bibitem[Gray \& Corbally(1994)]{gc94}
Gray, R. O., \& Corbally, C. J.~1994, \aj, 107, 742 (GC94)

\bibitem[Gray \& Corbally(2009)]{gc09}
Gray, R. O., \& Corbally, C. J.~2009, Stellar Spectral Classification (Princeton University Press, Princeton)

\bibitem[Hauck \& Mermilliod(1998)]{hm98}
Hauck, B., \& Mermilliod, M.~1998, \aaps, 129, 431

\bibitem[Hoffleit \& Jaschek(1982)]{hj82}
Hoffleit, D., \& Jaschek, C. 1982, The Bright Star Catalogue, 4th ed. (Yale University Observatory, New Haven)

\bibitem[Hubeny \& Lanz(2003)]{hl03}
Hubeny, I., \& Lanz, T.~2003, \apjs, 146, 417

\bibitem[Hubeny \& Lanz(2007)]{hl07}
Hubeny, I., \& Lanz, T.~2007, \apjs, 169, 83

\bibitem[Jaschek \& Jascheck(1987)]{jj87}
Jascheck, C. \& Jascheck, M.~1987, The classification of stars (Cambridge University Press, Cambridge)

\bibitem[Kilian et al.(1991)]{k91}
Kilian, J., Becker, S. R., Gehren, T. \& Nissen, P. E.~1991, \aap, 244, 419 

\bibitem[Kurucz(1979)]{k79}
Kurucz, R. L.~1979, \apjs, 40, 1

\bibitem[Kurucz(1993a)]{kur93a}
Kurucz, R. L.~1993a, CD-ROM No. 2--12 (SAO, Cambridge, Mass.)

\bibitem[Kurucz(1993b)]{kur93b}
Kurucz, R. L.~1993b, CD-ROM No. 13 (SAO, Cambridge, Mass.)

\bibitem[Lanz et al.(2008)]{l08}
Lanz, T., Cunha, K., Holtzman, J. \& Hubeny, I.~2008, \apj, 678, 1342

\bibitem[Lyubimkov et al.(2002)]{l02}
Lyubimkov, L.S., Rachkovskaya, T.M., Rostopchin, S.I. \& Lambert D.L.~2002, \mnras, 333, 9 

\bibitem[Malagnini et al.(1986)]{m86}
Malagnini, M. L., Morosi, C., Rossi, L. \& Kurucz, R. L.~1986, \aap, 162, 140

\bibitem[Martins et al.(2005)]{m05}
Martins, F., Schaerer, D. \& Hillier, D. J.~2005, \aap, 436, 1049

\bibitem[Mermilliod(1991)]{mer91}
Mermilliod, J. C. 1991, Catalogue of Homogeneous Means in the UBV System (Institut d'Astronomie, Universit\'e de Lausanne)

\bibitem[Meynet \& Maeder(2003)]{mm03}
Meynet, G., \& Maeder, A.~2003, \aap, 411, 543

\bibitem[Moon \& Dworetsky(1984)]{md84}
Moon, T. T., \& Dworetsky, M. M.~1984, Observatory, 104, 273

\bibitem[Morel \& Magnenat(1978)]{mor78}
Morel, M., \& Magnenat, P. 1978, \aaps, 34, 477

\bibitem[Morel et al.(2006)]{m06}
Morel, T., Butler, K., Aerts, C. \& Briquet, M.~2006, \aap, 457, 651

\bibitem[Morel \& Butler(2008)]{mb08}
Morel, T., \& Butler, K.~2008, \aap, 487, 307 (MB08)

\bibitem[Napiwotzki et al.(1993)]{napi93}
Napiwotzki, R., Sch\"onberner, D. \& Wenske, V.~1993, \aap, 268, 653

\bibitem[Nieva(2007)]{n07}
Nieva, M. F.~2007, Ph.D. Thesis, Erlangen-Nuremberg and 
Observat\'orio Nacional, Brazil

\bibitem[Nieva \& Przybilla(2007)]{np07}
Nieva, M. F., \& Przybilla,~N.~2007, \aap, 467, 295

\bibitem[Nieva \& Przybilla(2008)]{np08}
Nieva, M. F., \& Przybilla, N.~2008, \aap, 481, 199

\bibitem[Nieva \& Przybilla(2010a)]{np10a}
Nieva, M. F., \& Przybilla, N.~2010a, in ASP Conf. Ser., 425, 146

\bibitem[Nieva \& Przybilla(2010b)]{np10b}
Nieva, M. F., \& Przybilla, N.~2010b, EAS Publ. Ser., 43, 167

\bibitem[Nieva \& Przybilla(2012)]{np12}
Nieva, M. F., \& Przybilla, N.~2012, \aap, 539, A143 (Paper~I)

\bibitem[Nieva \& Sim\'on-D\'iaz(2011)]{ns11}
Nieva, M. F., \& Sim\'on-D\'iaz~2011, \aap, 532, 2 (Paper~II)

\bibitem[Przybilla et al.(2008)]{pnb08}
Przybilla, N., Nieva, M. F., \& Butler, K.~2008, \apj, 688, L103

\bibitem[Przybilla et al.(2011)]{pnb11}
Przybilla, N., Nieva, M. F., \& Butler, K.~2011, JPhCS, 328, 012015 

\bibitem[Puls et al.(2005)]{puls05}
Puls, J., Urbaneja, M., Venero, R. et al.~2005, \aap, 435, 669

\bibitem[Schaller et al.(1992)]{s92}
Schaller, G., Schaerer, D., Meynet, G., \& Maeder, A. 1992, \aaps, 96, 269

\bibitem[Schmidt-Kaler(1982)]{sk82}
Schmidt-Kaler, T. 1982, in: Landolt-B\"ornstein, Vol. 2, Subvol. b
(Springer-Verlag, Berlin)

\bibitem[Sim\'on-D\'iaz(2010)]{s10} 
Sim\'on-D\'iaz, S. 2010, \aap, 510, A22 (SD10)
 
\bibitem[Trundle \& Lennon(2005)]{tl05}
Trundle, C., \& Lennon, C. J.~2005, \aap, 434, 677

\bibitem[Vacca et al.(1996)]{v96}
Vacca, W., Garmany, C. D., \& Shull, M.~1996, \apj, 460, 87

\bibitem[Walborn(1971)]{w71}
Walborn, N.~1971, \apjs, 198, 257

\end{thebibliography}
\end{document}